\documentclass[lettersize,journal]{IEEEtran}
\usepackage{amsmath,amssymb,amsfonts}
\usepackage{algorithm}
\usepackage{algorithmic}
\usepackage{graphicx}
\usepackage{cite}
\usepackage{multirow}
\usepackage{booktabs}
\usepackage{makecell}
\usepackage[hidelinks]{hyperref}

\newtheorem{lemma}{Lemma}[section]
\newtheorem{proposition}[lemma]{Proposition}
\newtheorem{theorem}[lemma]{Theorem}
\newtheorem{corollary}[lemma]{Corollary}

\def\BibTeX{{\rm B\kern-.05em{\sc i\kern-.025em b}\kern-.08em
T\kern-.1667em\lower.7ex\hbox{E}\kern-.125emX}}

\newcommand{\subparagraph}[1]{%
  \par\noindent\textbf{#1}\par\noindent
}

\begin{document}

\title{FedPLT: Scalable, Resource-Efficient, and Heterogeneity-Aware Federated Learning via Partial Layer Training}

\author{Ahmad Dabaja, and Rachid El-Azouzi
\thanks{Ahmad Dabaja and Rachid El-Azouzi are with the Laboratoire Informatique d'Avignon (LIA), Avignon University, France (email: ahmad.dabaja@univ-avignon.fr, rachid.elazouzi@univ-avignon.fr).}
\thanks{This work was supported in part by the French National Research Agency (ANR) under Grant ANR-22-CE23-0024.}
\thanks{This article substantially extends a conference version published in IEEE PIMRC 2025, doi: \href{https://doi.org/10.1109/PIMRC62392.2025.11274718}{10.1109/PIMRC62392.2025.11274718}.}
\thanks{This work has been submitted to the IEEE for possible publication. Copyright may be transferred without notice, after which this version may no longer be accessible.}
}

\maketitle

\begin{abstract}
Federated Learning (FL) has gained significant attention in distributed machine learning by enabling collaborative model training across decentralized system while preserving data privacy. Although extensive research has addressed statistical data heterogeneity, FL still faces several challenges, including high communication and computation overheads and severe device heterogeneity, which require further investigation. Prior work has addressed these issues through sub-model training and partial parameter training. However, such methods often suffer from inconsistent parameter distributions across clients, inaccurate global loss estimation, and increased bias and variance. Guided by our empirical analysis, we propose FedPLT (Federated Learning with Partial Layer Training), an innovative and structured partial parameter training approach that exhibits training behavior similar to full model training while assigning client-specific portions of the model according to their communication and computational capabilities.
In addition, we evaluate the performance of FedPLT when combined with optimal client sampling under communication constraints. We show that this integration improves FL performance by reducing sampling variance under the same communication budget. Through extensive experiments, we demonstrate that FedPLT achieves performance comparable to, or even surpassing, that of full-model training (i.e., FedAvg), while requiring significantly fewer trainable parameters per client. Moreover, FedPLT outperforms existing methods in highly heterogeneous environments, effectively adapts to client resource constraints, and reduces the number of straggling clients. In particular, FedPLT reduces the number of trainable parameters by 71\%–82\% while achieving performance on par with full-model training.
\end{abstract}


\section{Introduction}

With the increasing demand for privacy-preserving and secure machine learning, particularly in light of regulatory frameworks such as the General Data Protection Regulation (GDPR) \cite{GDPR} and the California Consumer Privacy Act (CCPA)~\cite{CCPA}, federated learning has emerged as a promising paradigm for addressing these concerns.

Federated learning \cite{FedAvg} is a decentralized machine learning approach that allows models to be trained on local edge devices, each with its private data, without the need to share the data itself. Instead, local devices share their model updates with a central server, where they are aggregated to form a global model. This process is repeated over multiple communication rounds, enabling gradual model improvement and enhanced generalization. However, despite its advantages, this approach still faces several significant challenges.  One major challenge is the presence of heterogeneity in local data across clients, which leads to high variance in the trained model, commonly referred to as client drift~\cite{Li2020On, karim20}. This issue arises because devices have different datasets that vary in terms of their source, size, and distribution, causing the global model to struggle to generalize across these diverse datasets \cite{10.1145/3625558}.  Another challenge for federated learning is system heterogeneity, caused by variability in hardware (CPU, memory) and network connectivity \cite{Sahu2018FederatedOI}. Edge devices can range from smartphones and computers to IoT sensors, each with its own computational and communication limitations. This problem is further exacerbated by the growing trend of training large models, such as Transformers~\cite{Transformers}, which are resource-intensive, both in terms of communication and computation, and often exceed the capabilities of many edge devices. The third challenge lies in the communication process itself. Communication overhead remains a significant bottleneck, as frequent model updates between clients and the central server can saturate bandwidth and increase latency, especially in wireless or mobile networks. During each learning round, multiple clients perform local computations before aggregation, collectively contributing to a single global optimization step. In practice, several factors influence the overall communication and computational costs, including the model size, the number of participating clients, and the volume of locally processed data.


To address these issues, three main strategies have emerged in the literature: \textit{sub-model training}, \textit{partial parameter training}, and \textit{client sampling}. Submodeling methods (e.g., FedDrop \cite{FedDrop}, HeteroFL \cite{HeteroFL}, and FedRolex \cite{FedRolex}) reduce the model size assigned to each client according to their communication and computational capabilities. Partial parameter training approaches (e.g., FedPMT \cite{FedPMT}) retain the full model architecture but update only a subset of parameters during each training round. Client sampling strategies, such as Optimal Client Sampling (OCS) \cite{OptimalClientSampling}, aim to reduce communication overhead by selecting only a subset of clients to participate in each round. 

While these methods help mitigate device heterogeneity and reduce communication and computational costs, they also introduce new limitations, such as imbalanced parameter distribution across clients (e.g., in HeteroFL and FedPMT), increased variance (e.g., in FedDrop and HeteroFL), and unstable convergence (e.g., in some client sampling strategies). They may also suffer from limited scalability in federated systems composed mainly of resource-constrained devices, such as IoT devices.


To address the limitations of existing sub-model training and partial parameter training approaches, we propose FedPLT (Federated Learning with Partial Layer Training), a novel framework that provides an effective and structured method for partial training. Unlike previous approaches that introduce noise through random dropout (e.g., FedDrop) or create imbalanced parameter distributions (e.g., HeteroFL, FedRolex, and FedPMT), FedPLT ensures balanced and coordinated training by dividing the layers into equal-sized sub-layers and assigning each device a group of sub-layers based on its communication and computational capacities.

The design of FedPLT is guided by a series of experiments analyzing how training behaves when clients update assigned model partitions instead of the full model. By examining layer-wise convergence through gradient magnitudes and directional changes across training rounds, we identify assignment strategies that promote stronger inter-layer alignment than alternative configurations. These assignment strategies exhibit training dynamics similar to those of full-model training, leading to faster and more stable convergence. Consequently, FedPLT achieves substantial improvements in both convergence speed and model accuracy relative to the number of communication rounds. Another important consideration is that the design of the assigned model partitions depends heavily on the underlying model architecture. For example, convolutional neural networks (CNNs) require careful selection of layers to preserve effective feature extraction, while ResNet-based architectures require partitioning strategies that respect their residual block structure. As a result, the design of the assigned model partitions must be tailored to the specific characteristics of each model type to ensure effective training, stable convergence, and behavior aligned with full-model training. Additionally, FedPLT is flexible, allowing large models to be trained even on low-bandwidth networks and low-resource devices by adapting the proportion of parameters trained per client.

A preliminary version of this work appeared in~\cite{dabaja2025fedplt_conf}. This journal version substantially extends the prior paper with a more detailed formulation of FedPLT, expanded theoretical and efficiency analyses, additional experiments, and a broader discussion of resource-constrained federated systems.

\noindent Our main contributions are as follows:
\begin{itemize}

\item Through empirical experiments, we analyze learning behavior when clients train assigned model partitions instead of the full model. By studying layer-wise convergence in terms of gradient magnitude and directional changes across training cycles, we identify assignment principles that promote inter-layer alignment comparable to that of full-model training.

\item We introduce FedPLT, a structured partial parameter training scheme that addresses key limitations in existing sub-model training and partial parameter training methods in federated learning.

\item We establish convergence guarantees on FedPLT under the standard assumptions, together with an additional positive masked gradient alignment assumption introduced. 

\item We formulate the theoretical efficiency of FedPLT across three factors: communication cost, computation cost, and training time.

\item We extend Optimal Client Sampling (OCS) to the FedPLT setting by incorporating client-specific partial training ratios into the communication budget constraint. This yields a FedPLT-aware sampling rule that accounts for heterogeneous partial training costs while preserving the variance-reduction objective of OCS. This reduces overall communication overhead while maintaining, or even improving, the stability and accuracy of the global model.
    
\item  We conduct extensive experiments to demonstrate that FedPLT achieves performance improvements over full-model training while reducing the number of trained parameters by 84\%, thereby significantly lowering communication costs. Furthermore, FedPLT consistently outperforms other state-of-the-art submodel training and partial learning approaches. The performance gap is particularly notable on the CIFAR-10 dataset trained on an FCN model, where competing methods struggle to generalize. Specifically, FedPLT surpasses FedPMT by 7.49\%, FedRolex by 9\%, HeteroFL by 11.84\%, and FedDrop by 20.05\%. We also evaluate FedPLT under stringent communication constraints using our Optimal Client Sampling (OCS) strategy. Under the same communication budget, the combination of FedPLT and OCS outperforms OCS applied to full-model training, demonstrating the efficiency and effectiveness of integrating partial layer training with optimal client selection.
\end{itemize}

\section{Related Work}
\label{sec:related_work}



In this section, we review existing literature that addresses communication and computation efficiency and system heterogeneity in federated learning (FL). Broadly, these strategies fall into two main categories:

\begin{enumerate}
    \item \textbf{Sub-modeling / Sub-model Training Approach:} This approach reduces the communications and computational burden by shrinking the global model with a factor that depends on the client’s capabilities. This is achieved by omitting certain training parameters, resulting in smaller sub-models that are less resource demanding. Several works follow this approach, including HeteroFL \cite{HeteroFL}, FedDrop \cite{FedDrop}, and FedRolex \cite{FedRolex}. The difference between these methods lies in the way sub-models are created.

    \item \textbf{Partial Parameters Training Approach:} Unlike sub-modeling, this method keeps all the model parameters intact but updates only a subset of them during training. This reduces communication and computation without altering the overall model architecture. FedPMT \cite{FedPMT} is a common method that follows this approach.
\end{enumerate}

Each of the two approaches has distinct advantages and limitations. In all cases, inference is performed using the full global model, which is used to compute both the loss and the evaluation metrics on client data. However, during training, sub-model training approaches rely on reduced models that omit a subset of the trainable parameters. Consequently, the forward pass and the resulting loss are computed on an incomplete architecture that no longer reflects the true local objective associated with the full model. This mismatch leads to inaccurate gradients and, in turn, suboptimal updates being propagated to the server.


Partial parameter training methods address these challenges by maintaining the full model architecture during the forward pass. As a result, the loss is computed on this full architecture, allowing each client to optimize the local objective of the full model. The resulting gradient is therefore not an inaccurate approximation but a projected gradient obtained by masking the coordinates of untrained parameters while preserving the optimization direction over the active subset. This leads to a more consistent optimization trajectory and reduces gradient error. However, these benefits come with slightly higher communication and computation overhead.


Several sub-modeling methods have been proposed to operationalize the idea of training reduced architectures on heterogeneous clients. Although they share the common goal of lowering computational and communication costs, they differ in how sub-models are constructed, assigned to clients, and rotated across training rounds. These design choices influence how much of the global model is trained, how biased the resulting updates are toward certain clients, and how these updates interact during aggregation, leading to different convergence behaviors.

\begin{figure}[!h]
    \centering
    \includegraphics[width=\linewidth]{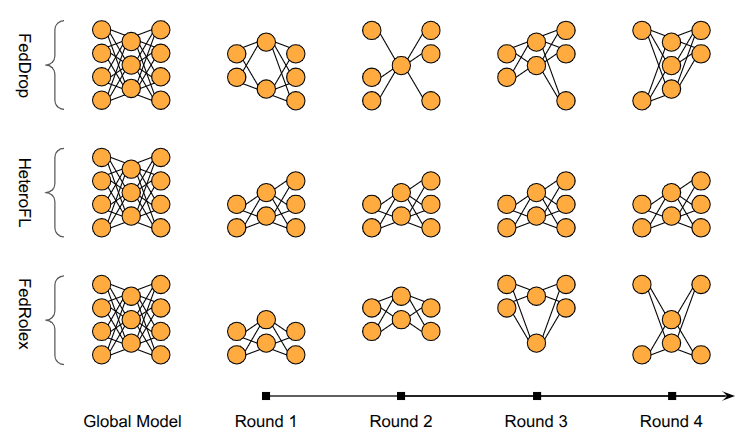}
    \caption{Illustration of sub-model training strategies. FedDrop applies random masks each round, HeteroFL uses a fixed mask, and FedRolex introduces a rolling mask mechanism.}
    \label{fig:submodeling_schemes}
\end{figure}

FedDrop \cite{FedDrop}, HeteroFL \cite{HeteroFL}, and FedRolex \cite{FedRolex} are three existing sub-model training schemes. Figure~\ref{fig:submodeling_schemes} illustrates their differences by showing the local models sent on several round for each method.

FedDrop \cite{FedDrop} adopts a stochastic neuron-dropping mechanism, where each neuron in fully connected layers is independently deactivated according to a Bernoulli distribution at each round during the training. This randomized sparsification reduces computational and communication overhead while allowing different subsets of neurons to be trained across rounds, thereby improving overall parameter coverage over time. This randomness offers simplicity and adaptability but introduces substantial variance in the training process. The fluctuating sub-model structures introduce gradient noise and cause oscillatory loss and unstable global updates. As the random masks change each round, the accumulated variance may slow convergence, especially under non-IID data distributions. HeteroFL \cite{HeteroFL} addresses this instability by adopting a fixed sub-modeling scheme tailored to device capacity. A dropout mask is defined prior to training and remains constant throughout the entire process. Clients with greater computational resources train larger sub-models, while resource-constrained clients train smaller ones formed by truncating neurons on one side of each layer. This eliminates per-round randomness and improves training stability. However, it creates a learning imbalance: Parameters retained across all sub-models are updated by every client and thus benefit from exposure to diverse data. In contrast,  parameters in the truncated region are updated only by high-resource clients.  This uneven update frequency leads to biased parameter learning and can degrade overall model performance, especially when smaller sub-models lack sufficient capacity to generalize well on complex or highly heterogeneous datasets. FedRolex \cite{FedRolex} extends HeteroFL by introducing a rolling or shifting dropout mask. Instead of fixing the truncated region, the mask is gradually shifted across the layer as training progresses, ensuring that different subsets of parameters are activated at different rounds. This mechanism reduces the parameter learning bias observed in HeteroFL, as all neurons eventually receive updates from clients with diverse data.  However, a complete rotation requires as many rounds as the width of the layer. Moreover, the shifting masks reintroduce part of the variance seen in FedDrop, and parameters that are reactivated after long intervals may overwrite earlier learning, leading to a form of catastrophic forgetting. As a result, FedRolex often needs significantly more communication rounds to converge. 

FedPMT~\cite{FedPMT} is a partial parameter training method that generates a limited number of local training configurations by freezing the shallow dense layers of a fully connected network. In this method, high-resource clients with good bandwidth train the full model, whereas low-resource or bandwidth-constrained clients update only the deeper layers, thereby reducing computational load and communication burden without modifying the model architecture. While FedPMT improves stability compared to sub-modeling approaches, it suffers from several structural limitations. First, the flexibility of training proportions is highly limited because FedPMT relies on freezing prefixes of layers. As a result, the number of valid partial-training configurations is bounded by the number of dense layers, and the corresponding training ratios are discrete and often unevenly spaced due to the non-uniform parameter sizes of the layers. Second, FedPMT introduces an inherent imbalance in parameter exposure across clients. The deeper layers updated by all clients benefit from more diverse data distributions, whereas the shallower layers trained only by high-resource clients receive updates from a much narrower subset of clients. This issue becomes more pronounced in architectures such as FCNs, where the early layers, which contain the majority of parameters, are trained by only a few clients and on less diverse data. This imbalance can bias training, degrade generalization, and slow global convergence, particularly under strong data heterogeneity.

A common limitation of existing sub-model training and partial parameter training methods is their reliance on the availability of sufficiently powerful clients. In fully resource-constrained systems, such as IoT-dominated environments, this assumption may not hold, making large-model training impractical since some model components may remain insufficiently trained or never updated.

\section{Background and  Preliminary Analysis of Partial Parameter Training}

\subsection{Federated Learning (FedAvg)}

We consider a federated system composed of a set of clients $\mathcal{K}=\{1,\dotsc,K\}$ and a central server. Each client $k$ holds a private local dataset $D_k=\{(x_i,y_i)\}_{i=1}^{n_k}$ of size $n_k$. The global training objective is defined as
\begin{equation}
    \min_{W\in \mathbb{R}^d} F(W) := \sum_{k=1}^{K} \frac{n_k}{n} F_k(W),
\end{equation}
where $F_k(W)=\mathbb{E}_{(x,y)\sim D_k}[\ell(W;x,y)]$ denotes the local expected loss of client $k$, $W$ denotes the model parameters, and $n=\sum_{k=1}^K n_k$. We recall the standard local update rule of FedAvg. At each communication round $t$, the server broadcasts to each client $k$ the current global model $W^t$. Each client initializes its local model as $W_k^{t,0}=W^t$ and performs $\tau$ steps of local stochastic gradient descent (SGD):
\[
W_k^{t,s+1} = W_k^{t,s} - \eta_k \nabla F_k(W_k^{t,s}),
\qquad s=0,\dots,\tau-1.
\]
where $s$ is the current local epoch and $\eta_k$ is the learning rate of client $k$. After local training, each client $k$ returns its final model $W_k^{t,\tau}$ to the server which aggregates them using a weighted average:
\[
W^{t+1} = \sum_{k=1}^K \frac{n_k}{n}\, W_k^{t,\tau}. 
\]

\begin{figure}[!h]
    \centering
    \includegraphics[width=\linewidth]{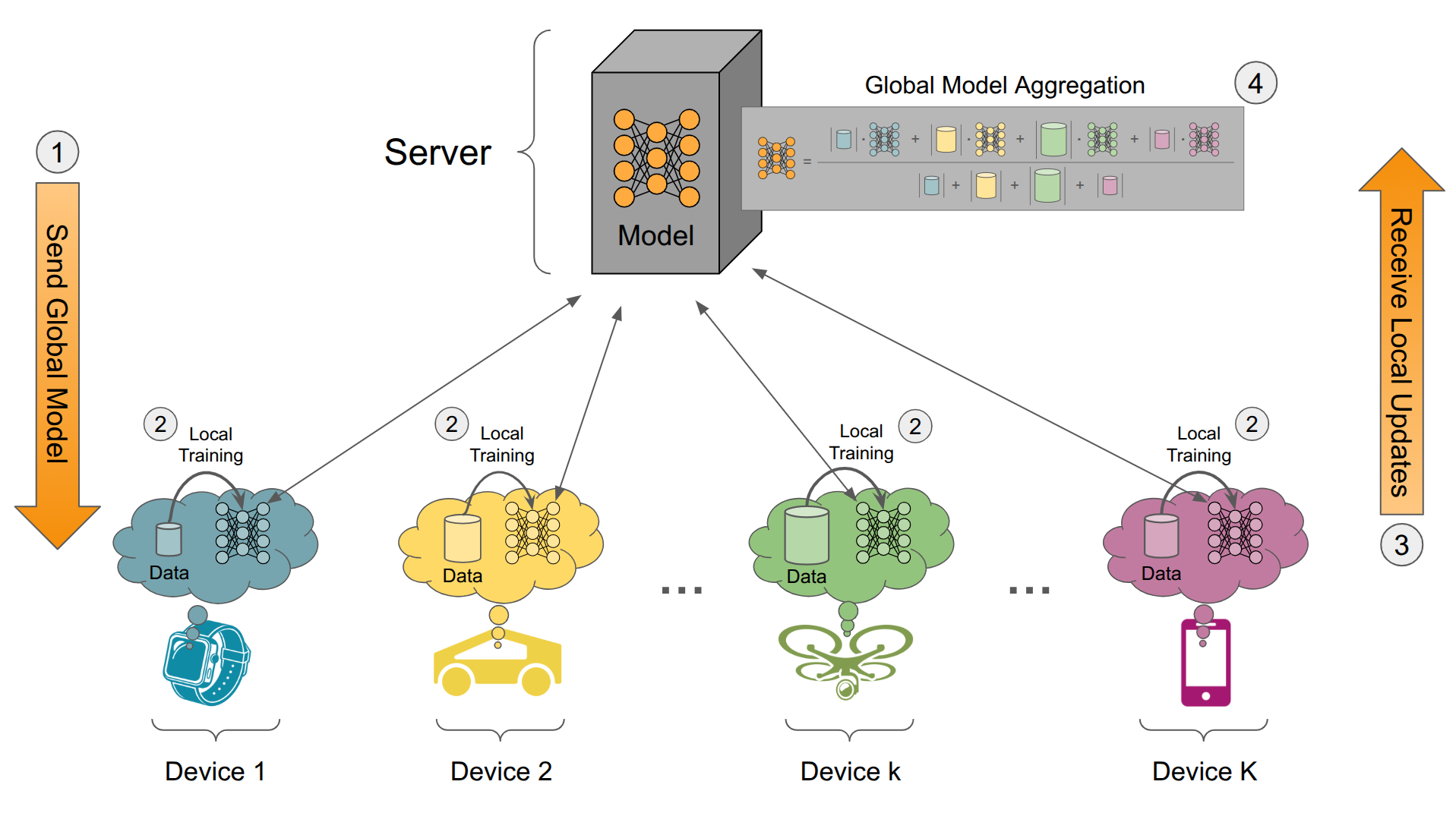}
    \caption{Overview of the standard Federated Learning (FedAvg) workflow.
    (1) The server broadcasts the global model to participating clients;
    (2) each client performs local training on its private dataset;
    (3) clients send their local model updates back to the server;
    (4) the server aggregates these updates to obtain the new global model.}
    \label{fig:fedavg_workflow}
\end{figure}

This process is repeated for multiple rounds $R$ until convergence. Figure~\ref{fig:fedavg_workflow} provides an overview of the four stages of a standard FedAvg round, illustrating model broadcast, local training, update transmission, and server aggregation.


Although FedAvg is widely used, training and communicating the entire global model presents significant challenges in realistic heterogeneous environments. These challenges arise from device resource constraints, high communication costs, and the presence of straggler clients with limited computational capabilities or unstable connectivity. To address these issues, prior work has explored resource-aware local training strategies, including sub-model training~\cite{FedDrop, HeteroFL, FedRolex} and partial parameter training~\cite{FedPMT}. However, even with such solutions, designing effective client-specific assignments of trainable model partitions remains challenging, particularly when the goal is to preserve training behavior and performance comparable to full-model training.

\subsection{Formalizing Partial Parameter Training}
To formalize partial-training configurations adapted to device capacities, we represent the configuration of client \(k\) by a \emph{layer-wise allocation vector}
\[
Q_k = (q_{k,1}, q_{k,2}, \dots, q_{k,L}),
\]
where \(q_{k,l} \in [0,1]\) denotes the proportion of parameters in layer \(l\) that are trainable on client \(k\). This layer-wise allocation induces an overall \emph{training ratio} for client \(k\), defined as the fraction of model parameters that the client is allowed to train:
\[
    r_k = \frac{\sum_{l=1}^L q_{k,l} \, h_l}{\sum_{l=1}^L h_l},
\]
where \(h_l\) denotes the number of trainable parameters in layer \(l\).

While \(Q_k\) specifies the trainable proportion within each layer, it does not directly describe how the overall trained portion is distributed across layers. To capture this distribution, we introduce the \emph{contribution vector}
\[
    X_k = (x_{k,1}, x_{k,2}, \dots, x_{k,L}),
\]
where
\[
    x_{k,l} = \frac{q_{k,l} \, h_l}{r_k \cdot \sum_{j=1}^{L} h_j}.
\]
By construction, \(x_{k,l} \in [0,1]\) and \(\sum_{l=1}^{L} x_{k,l} = 1\).

\subsection{Exploring Parameter-Efficient Training}



In this subsection, we empirically investigate how different trainable parameter allocations affect training dynamics across model architectures (FCNs and CNNs), with the goal of identifying an effective allocation strategy. To study these dynamics, we use diagnostic metrics such as Magnitude Gradient (MG) and Effective Perturbation (EP).


\begin{table*}[t]
    \centering
    \caption{Partial parameter training configurations for the MNIST FCN model. Each configuration is defined by the vector $Q_k$, the proportion of trained parameters $r_k$, the per-layer contribution vector $X_k$, and the standard deviation of $X_k$.}
    \label{tab:Q_k4DenseMNIST}
    \begin{tabular}{l c c c c}
        \toprule
        Config
        & $Q_k$
        & $r_k (\%)$
        & $X_k$
        & $\mathrm{Std}(X_k)$ \\
        \midrule
        Full model 
        & $(1.00,1.00,1.00,1.00)$ 
        & $100.0$ 
        & $(0.708,\ 0.231,\ 0.058,\ 0.002)$ 
        & $0.278$ \\
        
        Deep-heavy 
        & $(0.08,0.68,0.68,1.00)$ 
        & $25.6$ 
        & $(0.222,\ 0.615,\ 0.154,\ 0.009)$ 
        & $\mathbf{0.224}$ \\
        
        Shallow-dominant
        & $(0.30,0.10,0.10,1.00)$ 
        & $24.4$ 
        & $(0.872,\ 0.095,\ 0.024,\ 0.009)$ 
        & $0.361$ \\
        \bottomrule
    \end{tabular}
\end{table*}

\begin{figure*}[t]
  \centering
  \begin{minipage}[t]{0.32\linewidth}
    \centering
    \includegraphics[width=\linewidth]{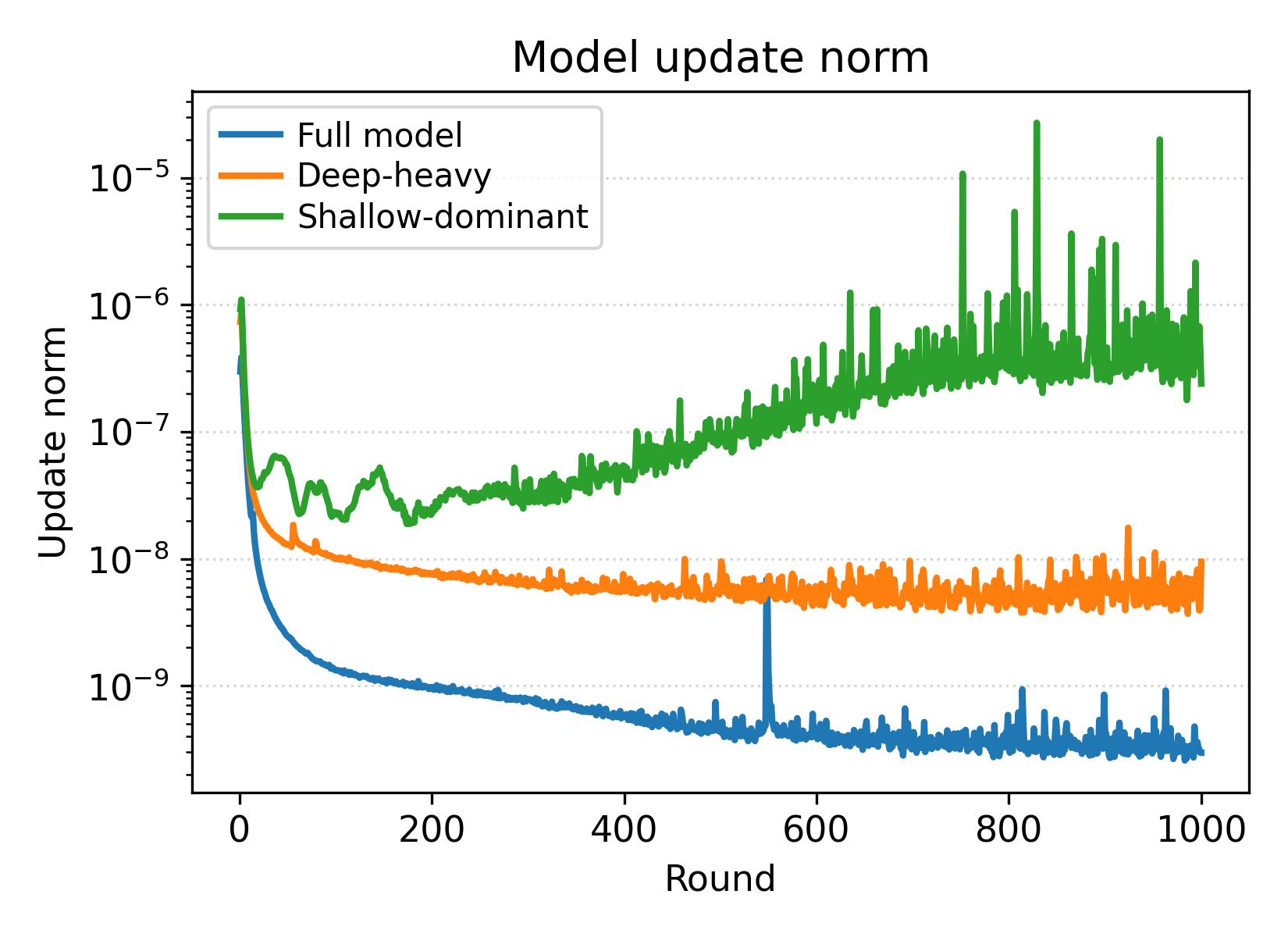}
  \end{minipage}
  \hfill
  \begin{minipage}[t]{0.32\linewidth}
    \centering
    \includegraphics[width=\linewidth]{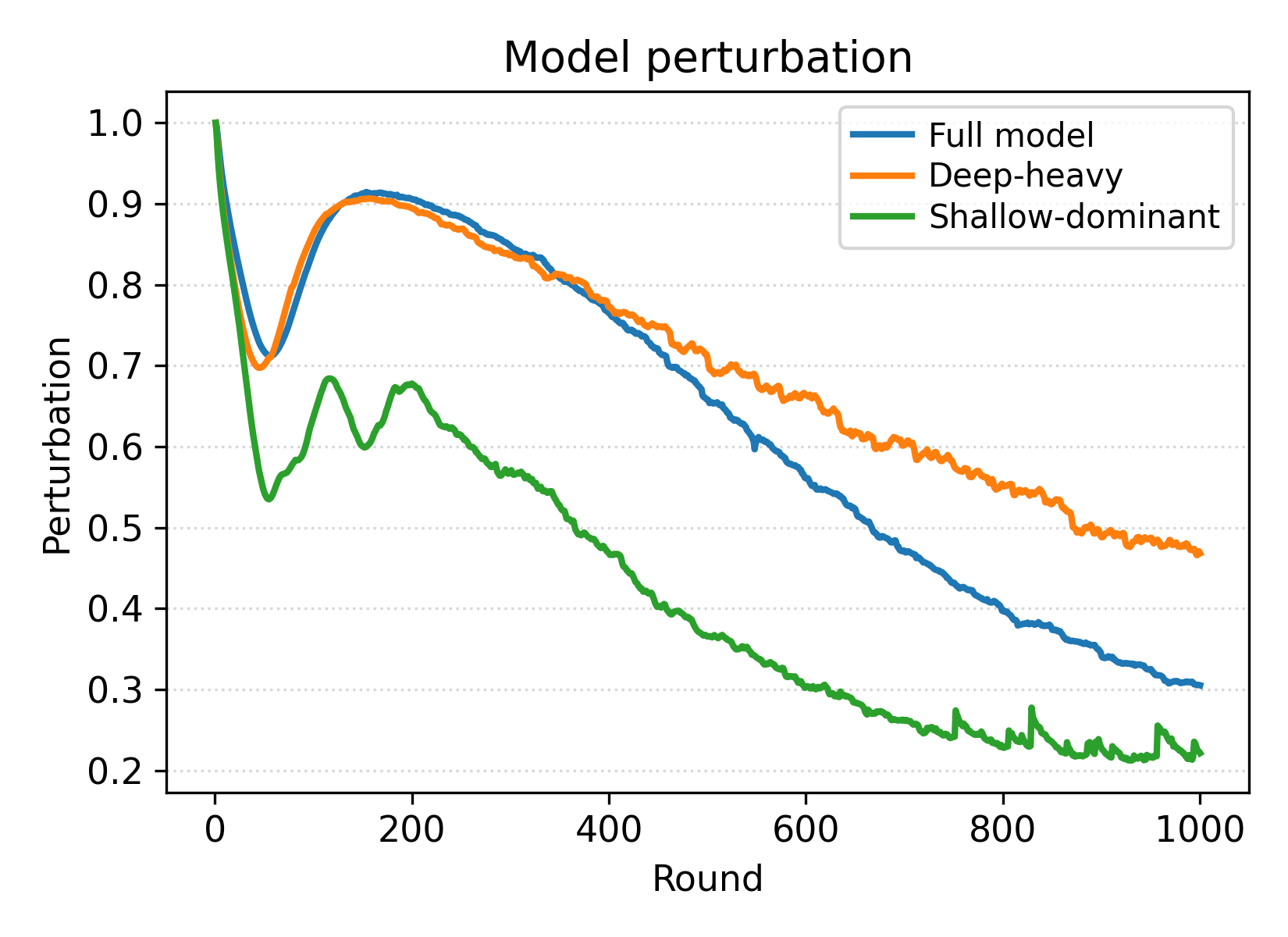}
  \end{minipage}
  \hfill
  \begin{minipage}[t]{0.32\linewidth}
    \centering
    \includegraphics[width=\linewidth]{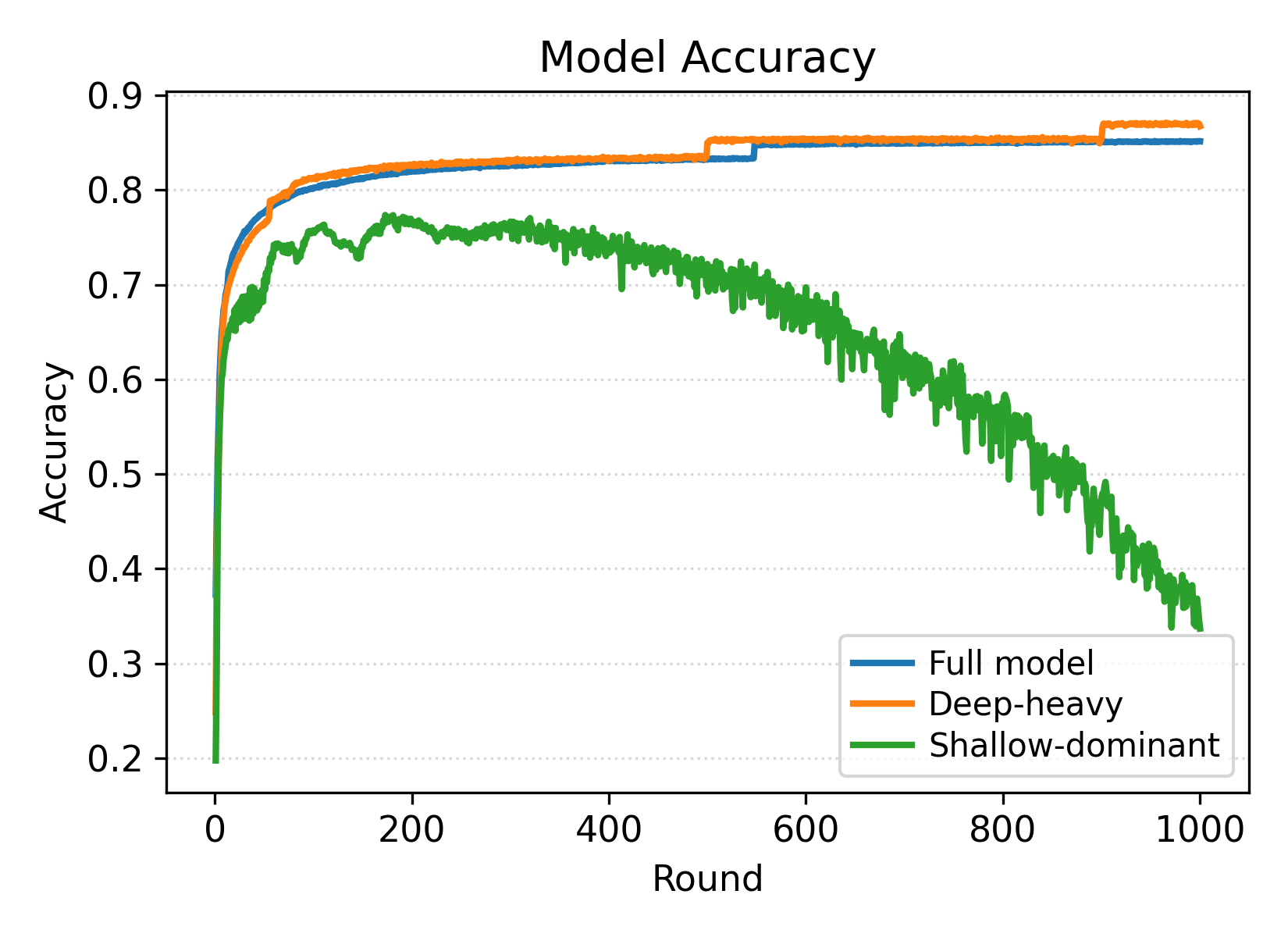}
  \end{minipage}
  \caption{The global update norm (left) and the effective perturbation (mid) at the model level and the model accuracy (right) for different partial training configurations.}
  \label{fig:mnist_model_metrics}
\end{figure*}

\begin{figure*}[t]
    \centering

    \centering
    \includegraphics[width=\linewidth]{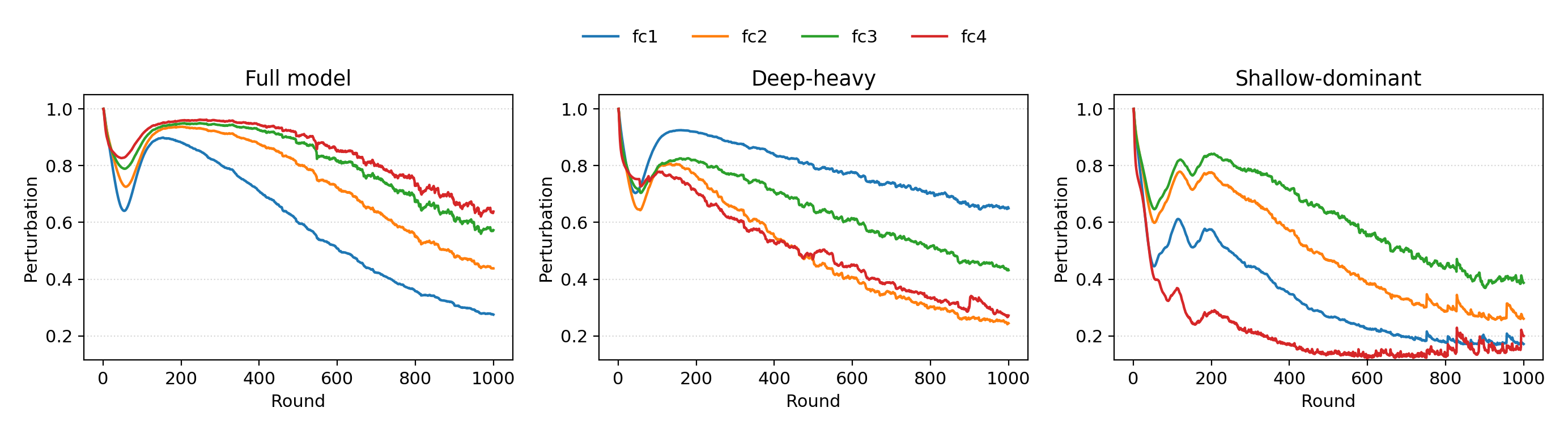}
    \caption{Layer-wise effective perturbation for fc1-fc4 weights under different partial training configurations.}
    \label{fig:mnist_perturb_layers}
    
\end{figure*}

\subsubsection{Diagnostic Metrics for Training Dynamics}
\paragraph{{\bf Magnitude Gradient (MG)}} 
To characterize the strength of parameter updates, we track the Magnitude Gradient (MG) at each round:
\[
MG^{t}=\left\| \Delta w^{t} \right\|^2, \qquad \text{where } \Delta w^{t}=w^{t}-w^{t-1}.
\]
Here, $w^{t}$ denotes the parameter vector at round $t$, which may represent either the full model or a specific layer, and $\Delta w^{t}$ is the corresponding update between two successive rounds. The Magnitude Gradient reflects how much the parameters change during training. Larger values correspond to stronger parameter changes, while smaller values indicate weaker or more stable updates.

\paragraph{\bf Effective Perturbation}
We quantify the temporal alignment of parameter updates using Effective Perturbation (EP)~\cite{chen2023synchronize}, defined as
\[
\mathrm{EP}^{t}=
\frac{
\left\| \sum_{t'=t-\tau+1}^{t} \Delta w^{t'} \right\|
}{
\sum_{t'=t-\tau+1}^{t} \left\| \Delta w^{t'} \right\|
}.
\]
$EP^{t}$ can be computed at the model level or the layer level. It takes values in $[0,1]$ and measures the directional consistency of parameter updates over time. Values close to one indicate that successive updates are well aligned and optimization proceeds coherently, whereas values close to zero reflect oscillatory or inconsistent update directions caused by noise or conflicting gradients. We note that $EP^t$ does not indicate proximity to the optimum. A learning trajectory may exhibit high directional alignment while moving away from the optimum, or low alignment while oscillating near a solution. Therefore, this metric should be interpreted together with amplitude-based measures, such as the update norm, to assess the quality of the optimization process.

\subsubsection{Empirical Analysis of Parameter-Efficient Training}
\label{subsec:empirical_analysis}


To better understand how partial parameter training affects optimization dynamics and to identify an effective layer-wise allocation strategy, we conduct a series of experiments on two model families: fully connected networks (FCNs) and convolutional neural networks (CNNs). This analysis allows us to examine the optimization behavior induced by different trainable parameter allocations and to compare it with that of full-model training within each model family.


\begin{table*}[t]
\centering
\caption{the different partial training configurations for CIfAR-10 ResNet-8 Model.
Here $Q_k$ is the contribution vector, $r_k$ the training ratio (in \%), $X_k$ the induced contribution vector, and $\mathrm{Std}(X_k)$ the standard deviation of $X_k$.}
\label{tab:Q_k4ResnetCifar}
\resizebox{\textwidth}{!}{
\begin{tabular}{lcccc}
\toprule
Config & $Q_k$ & $r_k$ (\%) & $X_k$ & $\mathrm{Std}(X_k)$ \\
\midrule
Full Model
& $(1.00,\ 1.00,\ 1.00,\ 1.00,\ 1.00)$
& $100.00$
& $(0.00598,\ 0.06193,\ 0.18537,\ 0.73805,\ 0.00868)$
& $0.27677$
\\
Deep-heavy
& $(0.06,\ 0.06,\ 0.12,\ 0.42,\ 1.00)$
& $34.75$
& $(0.00108,\ 0.01114,\ 0.06669,\ 0.89613,\ 0.02497)$
& $0.34878$
\\
Shallow-dominant
& $(1.00,\ 1.00,\ 0.50,\ 0.25,\ 1.00)$
& $35.38$
& $(0.01690,\ 0.17505,\ 0.26198,\ 0.52154,\ 0.02452)$
& $\mathbf{0.18556}$
\\
\bottomrule
\end{tabular}
}
\end{table*}

\begin{figure*}[t]
  \centering
  \begin{minipage}[t]{0.32\linewidth}
    \centering
    \includegraphics[width=\linewidth]{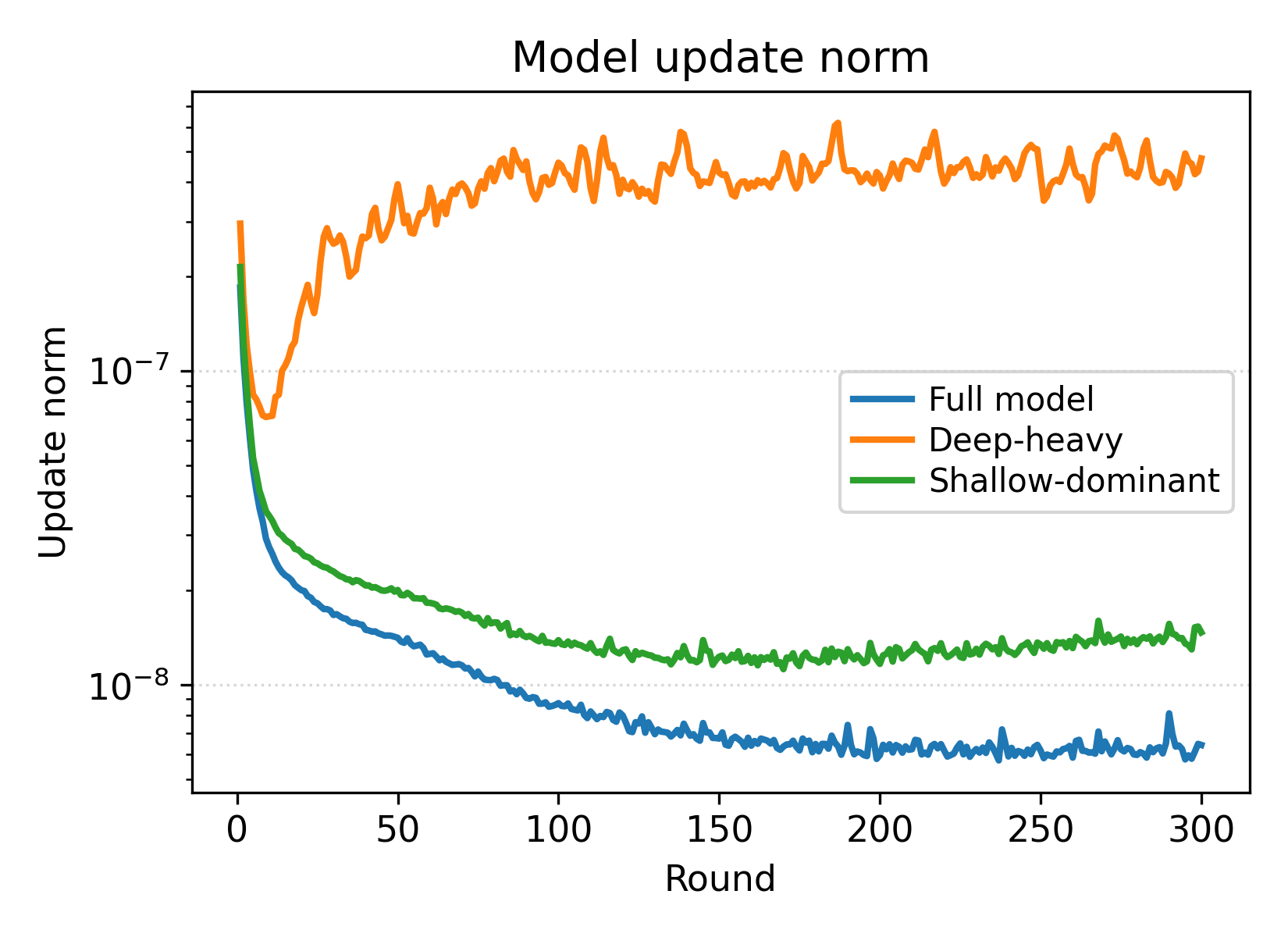}
  \end{minipage}
  \hfill
  \begin{minipage}[t]{0.32\linewidth}
    \centering
    \includegraphics[width=\linewidth]{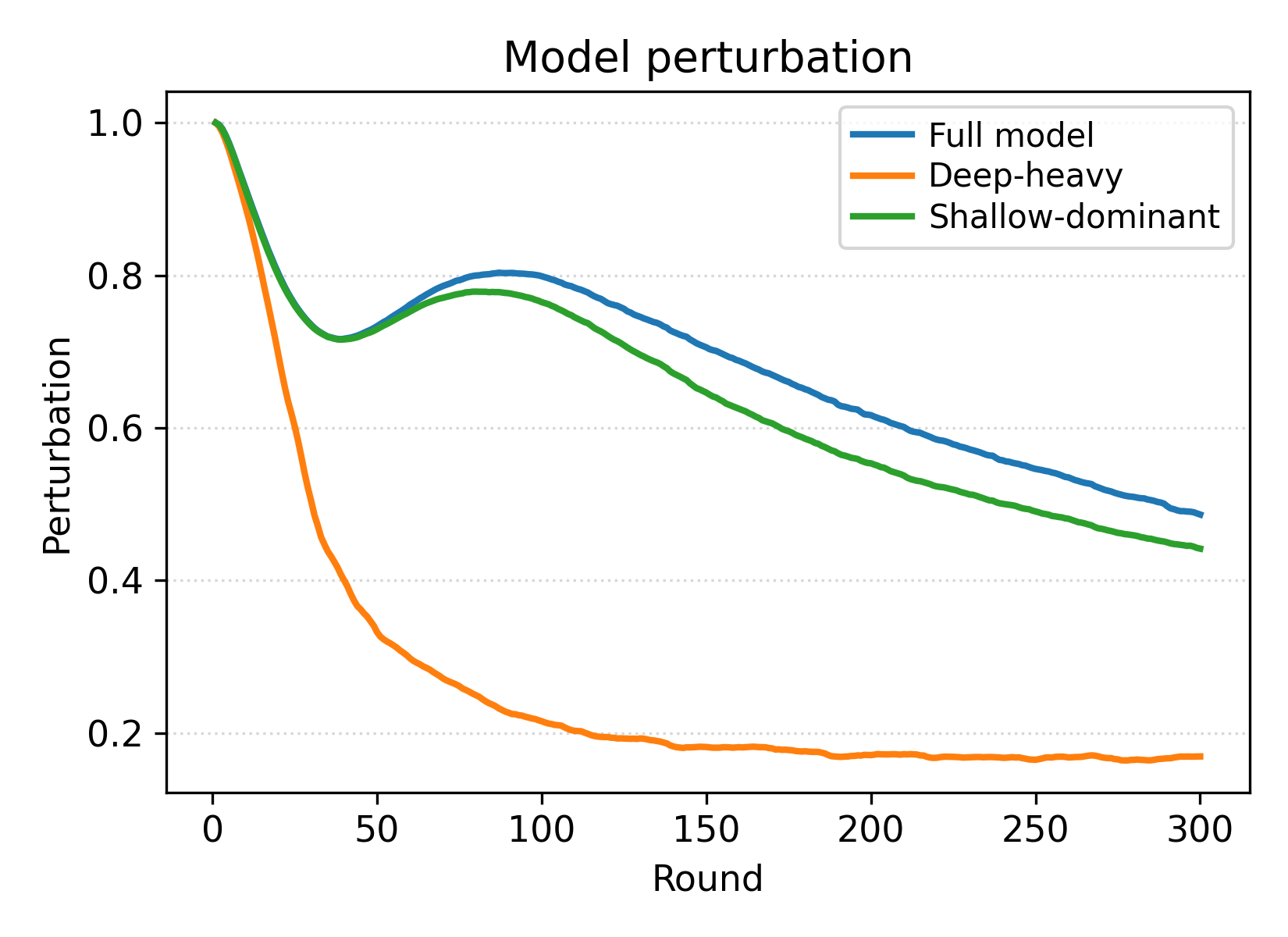}
  \end{minipage}
  \hfill
  \begin{minipage}[t]{0.32\linewidth}
    \centering
    \includegraphics[width=\linewidth]{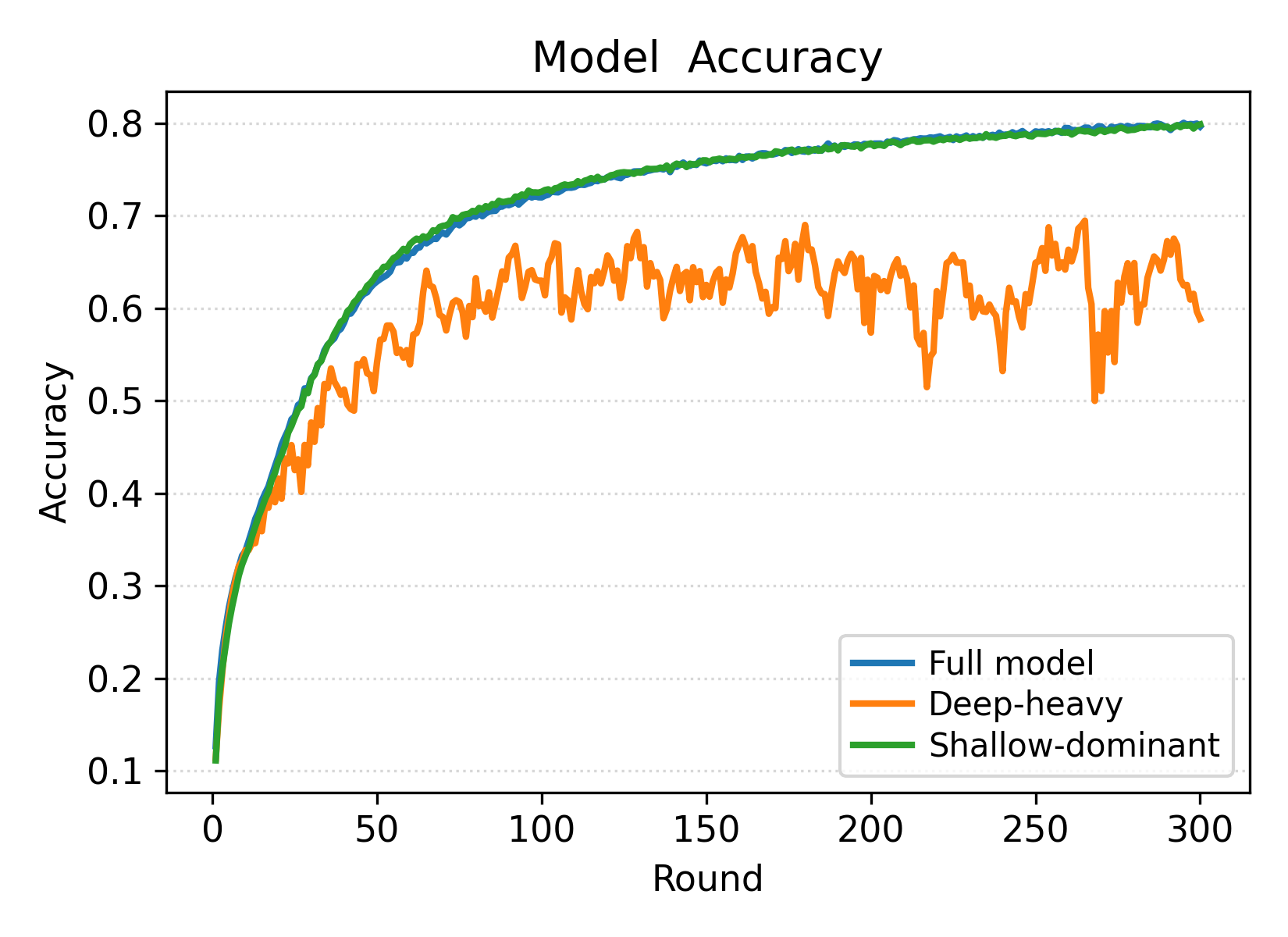}
  \end{minipage}
  \caption{The global update norm (left) and the effective perturbation (mid) at the model level and the model accuracy (right) for different partial training configurations.}
  \label{fig:cifar_model_metrics}
\end{figure*}

\begin{figure*}[htbp]
    \centering

    \begin{minipage}{\textwidth}
        \centering
        \includegraphics[width=\linewidth]{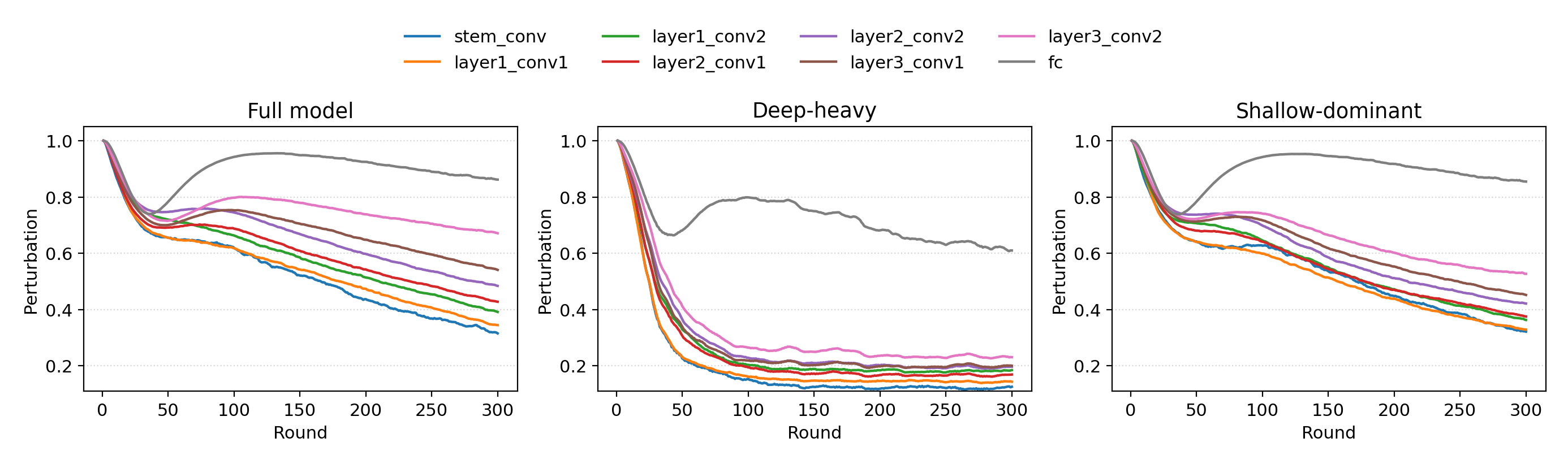}
        \caption{Effective perturbation for each ResNet-8 layer under different partial training configurations.}
        \label{fig:cifar_perturb_layers}
    \end{minipage}

\end{figure*}

\paragraph{\bf Fully Connected Networks (FCN)}
We consider a four-layer fully connected neural network for Fashion-MNIST, which takes grayscale images of size $28 \times 28$ as input and consists of three hidden fully connected layers of widths $(512,256,128)$ followed by a $10$-class output layer. We evaluate this model in a homogeneous setting, where all clients use the same partial-training configuration. Table~\ref{tab:Q_k4DenseMNIST} reports the three studied configurations obtained from different layer-wise allocations $Q_k$: (i) Full Model, where all layers are trained; (ii) Deep-Heavy, where larger values of $q_{k,l}$ are assigned to deeper layers; and (iii) Shallow-Dominant, where larger values of $q_{k,l}$ are assigned to earlier layers.


We first analyze the accuracy and model-level training dynamics induced by different partial-training configurations to determine which one remains closest to full-model training in terms of model-level behavior and performance. Figure~\ref{fig:mnist_model_metrics} reports these metrics for the different configurations.


At the model level, both the deep-heavy and full-model configurations exhibit a decreasing update norm during the early rounds, followed by smooth convergence toward low values, whereas the shallow-dominant configuration presented a noisier pattern that diverges since early rounds. This suggests that the deep-heavy configuration preserves training dynamics closer to those of full-model training and that partial training has not adversely affected performance, whereas the shallow-dominant configuration departs from the full model behavior by drifting into a suboptimal region of the optimization landscape. A similar trend appears in the perturbation metric. The deep-heavy configuration remains closely aligned with full-model training until around round 400, maintaining values above 0.7, and then diverges only slightly while still reaching a perturbation value of about 0.5 by round 1000. In contrast, the shallow-dominant configuration exhibits much lower perturbation values from the beginning of training, decreasing to around 0.2 by round 1000. This suggests weaker coordination across layers, which may lead to less stable optimization and poorer generalization. To further support this observation, we analyze the layer-level perturbation induced by these configurations and examine how closely their perturbation profiles align across layers and match those of full-model training. These profiles are shown in Figure~\ref{fig:mnist_perturb_layers}. We observe that, in both the full-model and deep-heavy configurations, the layers remain strongly aligned, leading to more coordinated and effective learning. In contrast, the shallow-dominant configuration exhibits greater divergence across layers, which makes training more difficult and may lead to suboptimal convergence. Overall, the shallow-dominant configuration departs from full-model training behavior and underperforms, whereas the deep-heavy configuration remains closer to full-model dynamics and achieves stronger performance, as also reflected in the accuracy curves in Figure~\ref{fig:mnist_model_metrics}. This makes the latter the more suitable choice for FCN training.


\paragraph{\bf Convolutional Neural Networks (CNN)}

CNNs differ from fully connected networks in how parameters are distributed across depth. In particular, early layers generally contain fewer parameters, while deeper layers account for a larger proportion of the trainable weights due to increasing channel dimensionality. This raises the question of whether the allocation strategy identified for FCNs also applies to CNNs, or whether it must be adapted to this structural difference. To investigate this, we train a ResNet-8 model~\cite{ResNet} on CIFAR-10 using different layer-wise training configurations $Q_k$, analogous to those examined in the FCN study: full model, deep-heavy, and shallow-dominant. These configurations are summarized in Table~\ref{tab:Q_k4ResnetCifar}. Note that ResNet-8 contains 8 layers, but we group each block of two convolutional layers into a single unit, yielding $H=[448,4640,13888,55296,650]$, corresponding to (step\_conv, block\_1, block\_2, block\_3, fc).



We first report the model-level update norm, perturbation metric, and model accuracy for the different configurations in Figure~\ref{fig:cifar_model_metrics}. In the CNN setting, we observe the opposite trend: the shallow-dominant configuration achieves better performance, whereas the deep-heavy configuration underperforms. As shown in Figure~\ref{fig:cifar_perturb_layers}, this behavior may be related to stronger inter-layer divergence in the deep-heavy setting, which can hinder representation learning and negatively affect convergence.


These empirical findings suggest a simple principle: layers with higher parameter density should receive lower training ratios, whereas layers with lower parameter density should receive larger allocations. Consequently, the best-performing configurations yield a more balanced contribution vector $X_k$, corresponding to a more uniform distribution of trained parameters across layers. Tables~\ref{tab:Q_k4DenseMNIST} and~\ref{tab:Q_k4ResnetCifar} confirm that these configurations consistently achieve the lowest standard deviation of $X_k$ (marked in bold). In the next section, we describe how to determine the allocation vector $Q_k$ from $r_k$ o that the resulting contribution vector $X_k$ is balanced.

\subsection{The Optimal Allocation Vector for Parameter Training}

For a fixed client training ratio $r_k$, different layer-wise configurations $Q_k$ can induce markedly different optimization behaviors. In particular, configurations whose induced contribution vectors $X_k$ are relatively balanced exhibit coherent layer-wise training dynamics and strong performance across both MLP and CNN architectures. However, we emphasize that although configurations with balanced contribution vectors consistently achieve strong accuracy, this study does not guarantee optimal performance among all feasible choices of $Q_k$.

We formulate below the optimization problem to identify the most balanced feasible contribution vector:
\begin{align}
\min_{X \in \mathbb{R}^L} \quad & J(X)
\label{eq:proj_obj} \\
\text{s.t.} \quad
& \sum_{l=1}^L x_l = 1, \label{eq:proj_sum} \\
& 0 \le x_l \le \bar{x}_l, \quad \forall l.
\label{eq:proj_bounds}
\end{align}
where
\[
J(X):=\frac{1}{2}\left\| X - \frac{1}{L}\mathbf{1} \right\|_2^2 \quad
\text{and} \quad 
\bar{x}_l = \frac{h_l}{r_k \sum_{j=1}^L h_j}
\]
denotes the maximum feasible contribution of layer $l$. Since the objective function is strictly convex and the feasible set is convex, a unique global minimizer exists. By applying the KKT conditions, the optimal contribution vector $X^*$ is given by
\begin{equation}
x_l^* = \min(\bar{x}_l,\tau), \quad l = 1,\dots,L,
\end{equation}
where the scalar $\tau$ is chosen such that
\begin{equation}
\sum_{l=1}^L \min(\bar{x}_l , \tau) = 1.
\end{equation}

The optimal contribution vector $X_k^*$ can then be mapped to a corresponding layer-wise allocation vector $Q_k$, yielding a balanced training configuration for the client $k$ under the fixed ratio $r_k$. This strategy provides a principled way to distribute each client’s training budget across layers within the proposed framework.

\section{FedPLT: Federated Learning with Partial Layer Training}

Motivated by the limitations of sub-model and partial parameter training strategies discussed in Section~\ref{sec:related_work}, and guided by the empirical analysis in Section~\ref{subsec:empirical_analysis}, we propose a new federated learning framework, FedPLT (Federated Learning with Partial Layer Training). Its core principle is a fine-grained allocation of training ratios across layers, achieved by decomposing each layer into smaller sub-layers and assigning subsets of these to clients. In this section, we introduce the FedPLT framework and describe its main design components.

\subsection{Partial Layer Training and Sublayer Assignment}

After determining the layer-wise allocation vector $Q_k$ for each client, FedPLT divides every trainable layer into multiple equal-sized sub-layers (blocks), and instantiates each allocation by assigning to client $k$ a subset of these blocks according to $Q_k$. This defines the client-specific assignment:
\[
\mathcal{A}_k = \{\mathcal{A}_{k,1}, \dots, \mathcal{A}_{k,L}\},
\]
where $\mathcal{A}_{k,l} \subseteq \{1,\dots,\mathcal{H}_l\}$ denotes the set of indices of sub-layers in layer $l$ assigned to client $k$, and $\mathcal{H}_l$ is the number of sub-layers in layer $l$.

\begin{figure}[!h]
  \centering
  \includegraphics[width=0.36\textwidth]{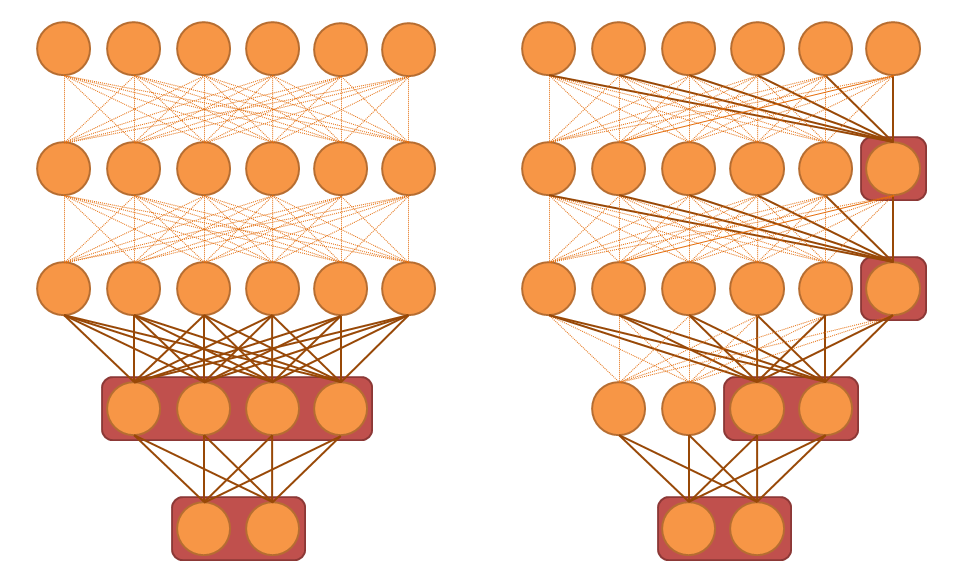}
  \caption{Comparison of two partial-parameter training configurations generated by FedPMT (left) and FedPLT (right) under the same training ratio. Brown blocks indicate the sub-layers assigned to the client, and brown solid lines indicate the corresponding trainable weights.}
  \label{fig:PMT/PLT}
\end{figure}

To illustrate a client assignment, figure~\ref{fig:PMT/PLT} compares two partial-training configurations with identical training ratios generated by FedPMT and FedPLT. For a client with $r_k\approx 30\%$, FedPMT applies $Q_k=\{0,0,1,1\}$ and concentrates training in deeper layers, whereas FedPLT realizes $Q_k=\{1/6,1/6,1/2,1\}$ by distributing updates across layers. In this FCN illustration, the assigned sub-layers in $\mathcal{A}_k$ are shown as brown blocks, and the corresponding trainable parameters are the incoming weights to these selected blocks, shown as brown solid lines, whereas the frozen parameters are shown in dotted lines. Unlike FedPMT, which restricts training to contiguous layers, FedPLT enables fine-grained block-wise allocation, allowing balanced participation of all layers in the learning process. Thus, it distributes the training depth and width rather than activating or freezing entire layers. This enables flexible training ratios.

\begin{figure}[!h]
  \centering
  \includegraphics[width=0.42\textwidth]{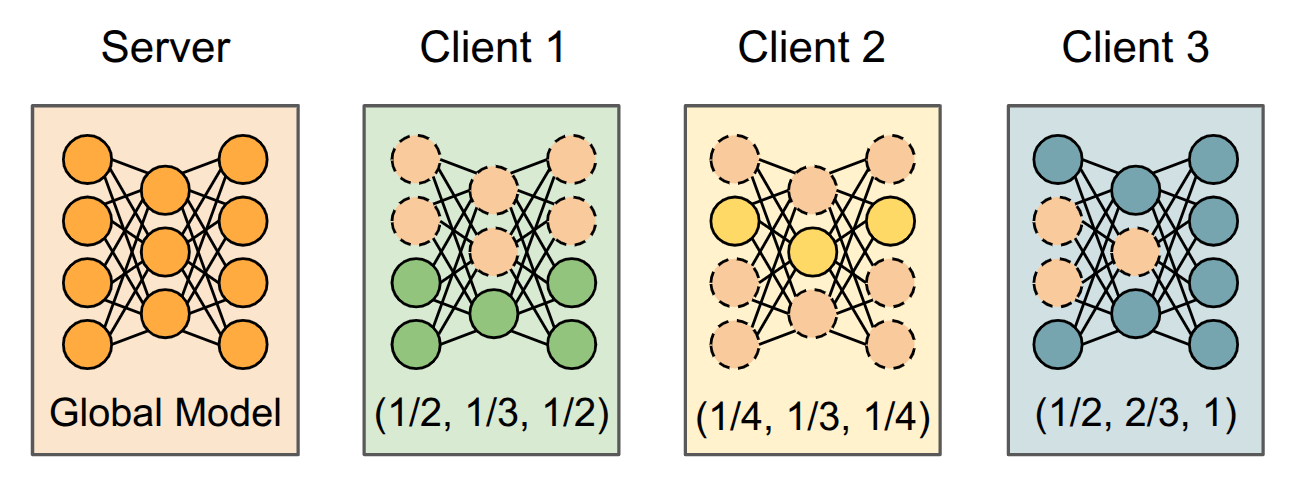}
    \caption{Illustration of FedPLT sub-layer allocation. Sub-layers are assigned to clients in a rotational manner so that each sub-layer is trained by approximately the same number of clients.}
  \label{fig:FedPLT_layer_assignment}
\end{figure}

Moreover, these assignments are generated in a structured, rotating manner across clients, such that over all clients each sub-layer is trained by approximately the same number of clients (Figure~\ref{fig:FedPLT_layer_assignment}). This prevents persistent under-training of any parameter group and ensures uniform exposure of all model parts to heterogeneous data. It also ensures that all parts of the global model are updated across clients, even in fully resource-constrained systems. Importantly, these assignments are computed once before training begins and remain fixed throughout the entire federated process.

During training, each client receives the full global model but updates only the parameters associated with its assigned sub-layers, freezing all others. More precisely, if client $k$ is assigned sub-layer index $h \in \mathcal{A}_{k,l}$ in layer $l$, then it updates the corresponding parameter block $W_{l,h}$.

\subsection{Aggregation}

During each round, each client transmits only the updated parameters of the sub-layers it trained. The server aggregates these updated sub-layer weights independently using weighted averaging over the clients that trained them, with weights proportional to their local dataset sizes. Clients that did not train a given sub-layer are excluded from its aggregation.

Let $W_{l,h}$ denote the parameter block associated with the $h$-th sub-layer of layer $l$, and let $W_{k,l,h}^{t,\tau}$ denote its locally updated version returned by client $k$ at round $t$. The global parameter block for sub-layer $h$ in layer $l$ is computed as:
\[
W_{l,h}^{t+1} =
\sum_{k \in \mathcal{S}_{l,h}} c_{k,l,h} \, W_{k,l,h}^{t,\tau},
\qquad
c_{k,l,h} = \frac{n_k}{\sum_{j \in \mathcal{S}_{l,h}} n_j},
\]

where $\mathcal{S}_{l,h}$ is the set of clients that trained sub-layer $h$ in layer $l$, and $n_k$ is the dataset size of client $k$.

The formulation above is written at the sub-layer level to make the aggregation mechanism explicit. However, it can also be expressed compactly at the model level.

Let
\[
W=\{W_{l,h}\}_{l=1,\dots,L;\,h=1,\dots,\mathcal H_l}
\]
denote the full model obtained by stacking all sub-layer parameter blocks. For each client $k$, define a binary mask
\[
m_k=\{m_{k,l,h}\}_{l,h},
\qquad
m_{k,l,h}=
\begin{cases}
1, & \text{if } k\in \mathcal S_{l,h},\\
0, & \text{otherwise},
\end{cases}
\]
where each $m_{k,l,h}$ is understood to act on the whole block $W_{l,h}$. We also define the corresponding block-wise aggregation weight vector
\[
\boldsymbol{c}_k=\{c_{k,l,h}\}_{l,h},
\qquad
c_{k,l,h}=
\begin{cases}
\dfrac{n_k}{\sum_{j \in \mathcal{S}_{l,h}} n_j}, & \text{if } k\in \mathcal S_{l,h},\\[1.2ex]
0, & \text{otherwise}.
\end{cases}
\]
Then the aggregation can be written equivalently as
\begin{equation}
    \label{eq:masked_aggregation}
    W^{t+1}=\sum_{k=1}^K \boldsymbol{c}_k \odot m_k \odot W_k^{t,\tau},
\end{equation}
where $\odot$ is applied element-wise.

\subsection{FedPLT scheme}
Given each client’s training ratio $r_k$, the server first determines a layer-wise allocation vector $Q_k$ using optimal solution of~(\ref{eq:proj_obj}-\ref{eq:proj_bounds}), and partitions each layer into equal-sized sub-layers. Based on $Q_k$, sub-layers are assigned to clients in a fixed rotational manner such that, across the population, each sub-layer is trained by approximately the same number of clients.

During federated training, each client receives the full global model, updates only the parameters associated with its assigned sub-layers while freezing the remaining parameters, and sends back only these updated parameters to the server. The server then aggregates each sub-layer independently using weighted averaging over the clients that trained it. This process is repeated for multiple rounds until convergence.









The complete FedPLT procedure is summarized in Algorithm~\ref{alg:FedPLT}.
\begin{algorithm}[!h]
\caption{FedPLT Scheme}
\label{alg:FedPLT}
\begin{algorithmic}[1]
\STATE \textbf{Input:} Initial global model $W^0$, number of layers $L$, client set $\mathcal{K}=\{1,\dots,K\}$, client training ratios $\{r_k\}_{k=1}^K$, rounds count $R$, local iterations count $\tau$, learning rates $\{\eta_k\}_{k=1}^K$ 

\STATE \textbf{1. Client Initialization}
\FOR{each client $k \in \mathcal{K}$}
    \STATE Compute $Q_k=\{q_{k,1},\dots,q_{k,L}\}$ from $r_k$
\ENDFOR
\FOR{each layer $l=1,\dots,L$}
    \STATE Partition layer $l$ into $\mathcal{H}_l$ equal-sized sub-layers
\ENDFOR
\FOR{each client $k \in \mathcal{K}$}
    \STATE Generate $\mathcal{A}_k=\{\mathcal{A}_{k,1},\dots,\mathcal{A}_{k,L}\}$ according to $Q_k$
\ENDFOR

\STATE \textbf{2. Federated Training}
\FOR{each round $t=0,\dots,R-1$}
    \FOR{each client $k \in \mathcal{K}$ in parallel}
        \STATE Download the global model $W^t$
        \STATE Initialize the local model $W_k^{t,0} \leftarrow W^t$
        \FOR{each local iteration $s=0,\dots,\tau-1$}
            \STATE Sample a mini-batch $\xi_k^{t,s}$ from $D_k$
            \FOR{each layer $l=1,\dots,L$}
                \FOR{each sub-layer $h=1,\dots,\mathcal{H}_l$}
                    \IF{$h \in \mathcal{A}_{k,l}$}
                        \STATE Update the assigned block using SGD:
                        \[
                        W_{k,l,h}^{t,s+1}
                        =
                        W_{k,l,h}^{t,s}
                        -
                        \eta_k \nabla_{W_{l,h}} \ell(W_k^{t,s};\xi_k^{t,s})
                        \]
                    \ELSE
                        \STATE Freeze the block: \( W_{k,l,h}^{t,s+1}=W_{k,l,h}^{t,s}\)
                    \ENDIF
                \ENDFOR
            \ENDFOR
        \ENDFOR
        \STATE Upload the updated blocks $W_{k,l,h}^{t,\tau}; h \in \mathcal{A}_{k,l}$
    \ENDFOR

    \FOR{each layer $l=1,\dots,L$}
        \FOR{each sub-layer $h=1,\dots,\mathcal{H}_l$}
            \STATE Define $\mathcal{S}_{l,h}=\{k \in \mathcal{K} : h \in \mathcal{A}_{k,l}\}$
            \STATE Aggregate the updated block:
            \[
            W_{l,h}^{t+1}
            =
            \sum_{k \in \mathcal{S}_{l,h}} c_{k,l,h}\, W_{k,l,h}^{t,\tau}, \quad
            c_{k,l,h} = \frac{n_k}{\sum_{j \in \mathcal{S}_{l,h}} n_j}
            \]
        \ENDFOR
    \ENDFOR
\ENDFOR

\STATE \textbf{Output:} Final global model $W^R$
\end{algorithmic}
\end{algorithm}

\section{Convergence Analysis}
\label{sec:convergence}

In this section, we provide a theoretical analysis of the convergence behavior of FedPLT. Our goal is to establish that, despite the use of fixed binary masks that restrict each client to updating only a subset of the model parameters, the global model converges under standard optimization assumptions.

For the convergence analysis, we use the simplified client weight $c_k=\frac{n_k}{n}$ instead of the coordinate-dependent aggregation weights defined in~\eqref{eq:masked_aggregation}. This simplification is introduced only to ease the proof and does not affect the main convergence argument.

We make the following standard assumptions commonly used in the convergence analysis of federated learning.

\begin{itemize}
    \item \textbf{Assumption 1 (Smoothness):} Each local objective function \( F_k \) is \( L \)-smooth. That is, for all \( W, W' \in \mathbb{R}^d \),
    \begin{equation}
    \|\nabla F_k(W) - \nabla F_k(W')\| \leq L \|W - W'\|.
    \end{equation}
    This implies that the gradient of \( F_k \) does not change too rapidly. Equivalently, by the Descent Lemma, we have:
    \begin{equation}
    F_k(W') \leq F_k(W) + \nabla F_k(W)^\top (W' - W) + \frac{L}{2} \|W' - W\|^2.
    \end{equation}

    \item \textbf{Assumption 2 (Strong Convexity):} Each local objective function \( F_k \) is \( \mu \)-strongly convex. That is, for all \( W, W' \in \mathbb{R}^d \),
    \begin{equation}
    F_k(W') \geq F_k(W) + \nabla F_k(W)^\top (W' - W) + \frac{\mu}{2} \|W' - W\|^2.
    \end{equation}

    \item \textbf{Assumption 3 (Unbiased Gradient Estimates):} The stochastic gradients computed from local mini-batches are unbiased estimators of the true local gradient:
    \begin{equation}
    \mathbb{E}_{\xi_k}[\nabla \ell(W; \xi_k)] = \nabla F_k(W),
    \end{equation}
    where the expectation is taken over the random mini-batch $\xi_k$ sampled at client $k$.

    \item \textbf{Assumption 4 (Bounded Variance):} The variance of the stochastic gradients is uniformly bounded:
    \begin{equation}
    \mathbb{E}_{\xi_k}\left[\left\|\nabla \ell(W; \xi_k) - \nabla F_k(W)\right\|^2\right] \leq \sigma^2,
    \end{equation}
    for all \( W \in \mathbb{R}^d \) and all clients \( k \).

    \item \textbf{Assumption 5 (Bounded Global Gradient Norm):} The norm of the gradient of the global objective function $F(W)$ is uniformly bounded:
    \begin{equation}
    \|\nabla F(W)\|^2 \leq G^2, \quad \forall W \in \mathbb{R}^d.
    \end{equation}

    \item \textbf{Assumption 6 (Positive Masked Gradient Alignment):} 
    We assume that for all clients \(k\), the masked gradient retains positive alignment with the descent direction:
    \[
    \rho_k := \frac{\left\langle m_k \odot \nabla F_k(W),\ W - W^* \right\rangle}{\left\langle \nabla F_k(W),\ W - W^* \right\rangle} > 0,
    \]
    where \(W^*\) is the global minimizer. This ensures that masking does not reverse the gradient direction, preserving progress toward optimality. A full geometric interpretation of this assumption is provided in Appendix~\ref{app:masked_alignment}.
\end{itemize}

Let \( W^t \in \mathbb{R}^d \) denote the global model at communication round \( t \), and let \( W^* \in \mathbb{R}^d \) be an optimal solution minimizing the global objective \( F(W) \). We define the expected squared distance to optimality as:
\[
D^t := \mathbb{E}[\|W^t - W^*\|^2].
\]

To simplify notation, define
\[
z := \mu \,\underline{\rho},
\]
and
\[
B := 2\tau^2(G^2+\sigma^2)\Gamma + 2L\nu\Lambda + \sigma^2 \Gamma,
\]
where \( \underline{\rho} := \min_{k,t} \rho_k^t \), \( \Gamma \) captures the aggregate masking effect, \( \Lambda \) is the aggregate heterogeneity gap, \(\tau\) is the local training steps, and \( \nu \) is a constant depending on the masking and smoothness terms. All these quantities are defined explicitly and derived in Appendix~\ref{app:convergence}.

\begin{proposition}[One-step recursive bound] \\
\label{prop:one_step_recursion}
Under Assumptions 1--6, the sequence \( \{D^t\} \) satisfies, for every round \( t \),
\[
D^{t+1} \leq (1 - z \eta^t)\, D^t + (\eta^t)^2 B.
\]
\end{proposition}

Proposition~\ref{prop:one_step_recursion} shows that the convergence behavior is governed by two competing terms: a contraction term \( (1-z\eta^t)D^t \), which drives the iterates toward the optimum, and a second-order error term \( (\eta^t)^2 B \), which captures the effect of stochasticity, heterogeneity, and masking.

The above recursion immediately yields the following finite-horizon convergence bound.

\begin{theorem}[General finite-horizon convergence bound] \\
\label{thm:convergence}
Under Assumptions 1--6, for any horizon \( T \geq 1 \), the iterates generated by FedPLT satisfy
\[
D^{T}
\leq
\left(\prod_{t=0}^{T-1}(1-z\eta^t)\right) D^0
+
B \sum_{t=0}^{T-1} (\eta^t)^2
\prod_{s=t+1}^{T-1}(1-z\eta^s).
\]
\end{theorem}

Theorem~\ref{thm:convergence} provides a unified upper bound that explicitly captures the influence of all algorithmic and problem-dependent parameters through the constants \( z \) and \( B \). In particular, \( z \) measures the effective contraction induced by strong convexity and masked gradient alignment, while \( B \) aggregates the impact of stochastic gradient noise, local update drift, smoothness, and objective heterogeneity.

We next specialize this result to two standard learning-rate schedules.

\begin{corollary}[Constant step size] \\
\label{cor:constant_stepsize}
If \( \eta^t = \eta \) for all \( t \), with \( 0 < \eta < 1/z \), then
\[
D^t \leq (1-z\eta)^t D^0 + \frac{\eta B}{z}.
\]
Thus, the method converges geometrically to a neighborhood of \( W^* \), with asymptotic error floor \( \frac{\eta B}{z} \). A smaller learning rate yields a tighter neighborhood, at the cost of slower convergence.
\end{corollary}

\begin{corollary}[Decaying step size] \\
\label{cor:decaying_stepsize}
If \( \eta^t = \frac{1}{z(t+1)} \), then
\[
D^t = \mathcal{O}\!\left(\frac{1}{t}\right).
\]
Hence, the method converges sublinearly to the exact optimum, with vanishing error over time.
\end{corollary}

These results show that FedPLT preserves standard convergence guarantees despite relying on static partial updates and masked gradients. The complete derivation and all intermediate steps are provided in Appendix~\ref{app:convergence}.

\section{Optimal Client Sampling with Partial Layer Training}

\subsection{Optimal Client Sampling}

\textit{Client Sampling} methods are widely studied approaches where only a subset of clients participates in each FL round. This family of approaches reduces the overall bandwidth consumption and lowers system-wide resource usage, which is particularly beneficial for limiting computation and communication costs. However, client sampling also introduces gradient noise and oscillations, especially when clients' data are heterogeneous. The randomness of client selection, combined with the variability in local data distributions, can cause the aggregated global update to drift away from the true global objective. This drift manifests over rounds as an oscillatory training trajectory, which may slow convergence and degrade model stability.

A key distinction between client sampling methods lies in the criteria used to assign sampling probabilities, whether based on dataset size, computational capability, or training time. A principled strategy is to choose these probabilities so as to directly minimize the variance introduced by partial participation.

One method that follows this principle is \textit{Optimal Client Sampling (OCS)} \cite{OptimalClientSampling}, which selects clients to minimize the variance of the aggregated update under a fixed communication budget~$\kappa$. Formally, OCS solves the optimization problem
\begin{equation}
\min_{\{p_k^t\}} \; \mathbb{E}_{A^t} \left[ \left\| \sum_{k \in A^t} \frac{n_k}{p_k^t} U_k^t - \sum_{k=1}^{K} n_k U_k^t \right\|^2 \right],
\end{equation}
subject to the communication constraint
\begin{equation}
\sum_{k=1}^{K} p_k^t = \kappa,
\end{equation}
where $n_k$ is the number of local samples on client~$k$, $U_k^t$ denotes its model update (or gradient), $p_k^t$ is its sampling probability, and $A^t$ is the set of selected clients at round~$t$. The constraint ensures that, in expectation, only $\kappa$ clients participate per round.

The optimal solution assigns higher sampling probabilities to clients with larger gradient norms and larger local datasets, thereby prioritizing those with stronger contributions to the global update. OCS has demonstrated significant improvements in communication efficiency and convergence speed compared to uniform sampling.

However, OCS requires estimating gradient information for \textit{all} clients at each round, which may be computationally expensive. More importantly, its formulation assumes that every client trains the full global model. As a result, OCS does not naturally extend to settings where clients train only fractions of the model, as is the case with sub-model training or partial parameter training methods.

Once clients no longer train the full set of parameters, as in sub-modeling or partial parameter training, the communication cost assumed in OCS fundamentally changes. This is because each client now contributes updates on only a fraction of the parameters and incurs a proportionally smaller communication cost. This shift naturally raises the question of how the OCS formulation should be modified to account for partial training ratios. The following subsection addresses this by extending OCS to the FedPLT setting.

\subsection{Extending OCS to Partial Layer Training}

We now extend the OCS formulation to the FedPLT setting, where each client trains only a fraction of the global model determined by its partial training ratio $r_k$. Under this regime, the communication cost of selecting client $k$ scales proportionally with $r_k$, rather than being uniform across clients as assumed in the original OCS formulation. As a result, the standard constraint must be replaced by a FedPLT-aware communication budget that accounts for heterogeneous training fractions:
\[
\sum_{k=1}^K r_k p_k^t = \kappa.
\]
This constraint ensures that the expected total ratios of trained parameter per round is bounded by the communication budget~$\kappa$.

Solving the associated optimization problem with Lagrangian methods yields a modified rule for client selection probabilities:
{
\footnotesize
\begin{equation}
p_k^t =
\begin{cases}
\displaystyle \frac{\kappa - \sum_{j=1}^{K}r_j + \sum_{j\in\mathcal{O}} r_j}{\sqrt{r_k}} \cdot \frac{n_k \|U_k^t\|}{\sum_{j\in\mathcal{O}} \sqrt{r_j} n_j \|U_j^t\|}, & \text{if } k \in \mathcal{O} \\[1ex]
1, & \text{if } k \in \mathcal{K} - \mathcal{O}
\end{cases}
\end{equation}
}
where the set \( \mathcal{O} \) is defined by:
\small
\begin{equation}
\mathcal{O} = \left\{ k \in \mathbb{K} \Big/  \sqrt{r_k} n_k \|U_k^t\| < \frac{\sum_{j\in\mathcal{O}} \sqrt{r_j} n_j \|U_j^t\|}{\kappa - \sum_{j=1}^{K}r_j + \sum_{j\in\mathcal{O}} r_j} \right\}.
\end{equation}
\normalsize

This formulation preserves the variance-minimization objective while adapting to the heterogeneous communication costs introduced by partial layer training. Clients with larger datasets, larger gradient norms, and smaller $r_k$ (i.e., cheaper partial participation) receive higher sampling probabilities. Full mathematical details are provided in Appendix~\ref{app:OCS-PLT}.

This extension enables client sampling to explicitly account for heterogeneous partial training costs, allowing FedPLT to jointly exploit parameter-level adaptivity (through partial layer training) and client-level adaptivity (through variance-optimal sampling). While FedPLT balances layer-wise contributions across heterogeneous devices, optimal client sampling further controls which clients participate at each round under a fixed communication budget. In the experimental section, we evaluate the combined impact of these two mechanisms and study how variance-aware client selection interacts with partial layer training under heterogeneous data and system conditions.

\section{Efficiency Analysis}
\label{sec:FedPLT_efficiency}

In this section, we analyze the efficiency of FedPLT relative to FedAvg along three axes: computation cost, communication cost, and round time, under a simple system model.

We consider a system with $K$ clients, each characterized by a computation speed $\gamma_k$ (in FLOPs/s), an uplink bandwidth $B_k^{\uparrow}$ (in B/s), and a downlink bandwidth $B_k^{\downarrow}$ (in B/s). We consider a global model containing $P$ parameters, each occupying $s$ bytes. For each parameter, the forward computation requires $\alpha$ FLOPs and the backward computation requires $\beta$ FLOPs, following the standard relation $\beta \approx 2\alpha$ in typical neural networks. Each training round consists of a fixed number of local iterations, denoted by $\tau$, and incurs a fixed latency term $\delta$ accounting for connection setup, synchronization delay, or other non-payload overhead. The server is assumed to act only as an aggregator with unbounded computational resources.

\subsection{Computation Cost}
Under full-model training, the per-round computation cost of client $k$ is
\[
\Xi_k^{\text{full}} = (\alpha+\beta) \tau P.
\]
Under FedPLT, only the backward part scales with the trained fraction $r_k$, hence
\[
\Xi_k^{\text{plt}} = (\alpha+\beta r_k) \tau P.
\]
Therefore, the relative computation reduction is
\[
\Delta_k^{\text{comp}}
= 1 - \frac{\Xi_k^{\text{plt}}}{\Xi_k^{\text{full}}}
= 1 - \frac{\alpha+\beta r_k}{\alpha+\beta}.
\]
Thus, FedPLT reduces computation through the backward stage, with gains increasing as $r_k$ decreases.

\subsection{Communication Cost}
In full-model FL, each client downloads and uploads the entire model. Hence, the bidirectional communication volume for each client $k$ per round is
\[
\Theta_k^{\text{full}} = 2Ps.
\]
In FedPLT, the full model is still downloaded, but only the updated fraction is uploaded:
\[
\Theta_k^{\text{plt}} = (1+r_k)Ps.
\]
Hence, the relative uplink reduction is
\[
\Delta_k^{\text{comm,up}} = 1-r_k,
\]
while the total bidirectional communication reduction is
\[
\Delta_k^{\text{comm,tot}} = 1-\frac{\Theta_k^{\text{plt}}}{\Theta_k^{\text{full}}}
= \frac{1-r_k}{2}.
\]
Therefore, FedPLT preserves the downlink cost while significantly reducing uplink communication.

\subsection{Round Time and Stragglers}
In synchronous federated learning, the duration of each round is determined by the slowest participating client. Under full-model training, the round time is
\[
\begin{aligned}
T_{\text{round}}^{\text{full}}
=&\ \delta
+ \max_{k}\!\left(
    \underbrace{\tfrac{\alpha\tau P}{\gamma_k}}_{\text{forward}}
    + \underbrace{\tfrac{\beta\tau P}{\gamma_k}}_{\text{backward}}
    + \underbrace{\tfrac{Ps}{B_k^{\downarrow}}}_{\text{downlink}}
    + \underbrace{\tfrac{Ps}{B_k^{\uparrow}}}_{\text{uplink}}
\right) \\
=&\ \delta + \left(
    \tfrac{(\alpha+\beta)\tau P}{\hat{\gamma}}
    + \tfrac{Ps}{\hat{B}^{\downarrow}}
    + \tfrac{Ps}{\hat{B}^{\uparrow}}
\right),
\end{aligned}
\]
where $\hat{k}$ denotes the straggler client attaining the maximum, with $\hat{\gamma}=\gamma_{\hat{k}}$, $\hat{B}^{\downarrow}=B_{\hat{k}}^{\downarrow}$, and $\hat{B}^{\uparrow}=B_{\hat{k}}^{\uparrow}$.

FedPLT can mitigate the straggler effect by selecting the fractions $\{r_k\}$ so that clients finish their local computation and communication in approximately the same time:
\[
T_1^{\text{plt}}
\;\approx\;
T_2^{\text{plt}}
\;\approx\;
\dots
\;\approx\;
T_K^{\text{plt}}
\;\approx\;
T_{\text{round}}^{\text{plt}}.
\]

Hence, the round time under FedPLT can be expressed as
\[
T_{\text{round}}^{\text{plt}}
= \delta
+ \tfrac{\alpha\tau P}{\hat{\gamma}}
+ \tfrac{Ps}{\hat{B}^{\downarrow}}
+ \hat{r}\left(
    \tfrac{\beta\tau P}{\hat{\gamma}}
    + \tfrac{Ps}{\hat{B}^{\uparrow}}
\right),
\]
where $\hat{r}$ denotes the fraction assigned to the limiting client under the balanced FedPLT allocation.

The relative round-time efficiency is
\[
\Delta_{\text{time}}
= 1 - \frac{T_{\text{round}}^{\text{plt}}}{T_{\text{round}}^{\text{full}}}
= \frac{
    (1-\hat{r})
    \left(
        \tfrac{\beta\tau P}{\hat{\gamma}}
        + \tfrac{Ps}{\hat{B}^{\uparrow}}
    \right)
}{
    \tfrac{(\alpha+\beta)\tau P}{\hat{\gamma}}
    + \tfrac{Ps}{\hat{B}^{\downarrow}}
    + \tfrac{Ps}{\hat{B}^{\uparrow}}
}.
\]

Thus, since synchronous FL is governed by the slowest participating client, assigning smaller fractions $r_k$ to slower clients directly shortens the overall round duration compared with full-model training and mitigates the straggler effect.

Detailed derivations and a concrete numerical example are provided in Appendix~\ref{app:eff-Gain}.

\section{Experimentation and Results}
\label{sec:ER}

We evaluate FedPLT through three experiments covering different settings:

\begin{itemize}
    \item \textbf{Homogeneous low-resource system:} 
    We consider a system of resource-constrained devices, with limited communication and computation capabilities, such as IoT devices or drone networks, where no client can realistically train the full model locally. In this setting, all clients use the same small training ratio \(r_k\). We evaluate whether FedPLT can enable effective training under such constraints and compare it with full-model training (FedAvg).

    \item \textbf{Highly heterogeneous system:} 
    We consider a system with devices of varying computation and communication capacities, leading to different feasible training ratios \(r_k\). We compare FedPLT with existing sub-model training and partial parameter training methods under this heterogeneous setting.

    \item \textbf{FedPLT with optimal client sampling:} 
    We investigate the impact of integrating FedPLT with optimal client sampling. Specifically, we compare the original OCS formulation with our FedPLT-aware extension and analyze the resulting performance gains and trade-offs.

\end{itemize}

\subsection{Experimental Setup}

\begin{itemize}

    \item \textbf{Datasets and Data Distribution:} 
    We use two benchmark datasets: Fashion-MNIST~\cite{Fashion} and CIFAR-10~\cite{CIFAR10}, each split into 50{,}000 training samples and 10{,}000 validation samples. Data are distributed across clients using a Dirichlet partitioning strategy~\cite{Dirichlet} to simulate non-IID settings. For Experiments~1 and~2, data are partitioned among 50 clients with Dirichlet concentration parameter $\alpha=0.2$. For Experiment~3, data are partitioned among 100 clients with a more heterogeneous split using $\alpha=0.1$

    \item \textbf{Model Architectures:} 
    For Fashion-MNIST, we use a fully connected network (FCN) with four hidden layers following the architecture, $(\text{Input} \rightarrow 512 \rightarrow 256 \rightarrow 128 \rightarrow \text{Output})$, with ReLU activations in the hidden layers and Softmax at the output. For CIFAR-10, we evaluate two models: the same FCN model and ResNet-8~\cite{ResNet}.

    \item \textbf{Training Configuration:} 
    In Experiments~1 and~2, all 50 clients participate in every communication round. For (Fashion-MNIST, FCN), we train for 300 rounds, while for (CIFAR-10, FCN), we train for 1{,}000 rounds. In both cases, each client performs one local epoch per round with a batch size of 64. For the (CIFAR-10, ResNet-8), each client performs 16 local iterations per round,  each on a randomly sampled mini-batch of 64 images. In Experiment~3, we use the (CIFAR-10, FCN) configuration with client sampling, so the number of participating clients varies by strategy. These experiments run for 500 rounds with three local epochs per round.

    \item \textbf{Optimization and Loss Function:} 
    Across all experiments, we use stochastic gradient descent (SGD) with a fixed learning rate of 0.01 and optimize the categorical cross-entropy loss. Performance is measured using classification accuracy.

    \item \textbf{Federated Aggregation and Settings:} 
    Client updates are aggregated using a weighted average based on local sample counts, each time adapted to the used scheme. For our FedPLT scheme, we use the aggregation rule defined in Eq.~\ref{eq:masked_aggregation}. For the (CIFAR-10, ResNet-8) setting, we additionally adopt FedBN~\cite{FedBN}, keeping batch-normalization layers local to preserve client-specific feature statistics. 

    \item \textbf{Reproducibility and Randomness Control:} 
    To ensure robustness and reproducibility, each experiment is repeated multiple times with different random seeds, and reported results are averaged across runs. We maintain a consistent experimental setup across all runs.

\end{itemize}

\subsection{Evaluating FedPLT in a homogeneous system}

In this experiment, we consider a homogeneous low-resource federated system in which all clients have limited computation and communication capabilities, so that full-model local training is impractical. We therefore evaluate FedPLT in this setting, where each client trains only a small fraction of the model.

For each experimental setting, all clients share the same training ratio $r_k$ and the same layer-wise allocation vector $Q_k$. To study the effect of parameter allocation, we repeat the experiment with several choices of $Q_k$ corresponding to different balancing levels of the induced contribution vector $X_k$. We conduct this evaluation on three dataset-model pairs: Fashion-MNIST with a custom fully connected network, CIFAR-10 with the same fully connected model, and CIFAR-10 with ResNet-8.

Recall that the most balanced contribution vector $X_k^*$ is obtained by solving~\eqref{eq:proj_obj}--\eqref{eq:proj_bounds}. To quantify how far a given allocation deviates from this balanced solution, we define the relative error
\[
E(X)=\frac{J(X)-J(X^*)}{J(X^*)},
\]
which measures the imbalance of $X_k$ relative to the optimal balanced vector $X_k^*$. For each dataset-model pair, we therefore consider the default allocation $Q_k^*$ associated with $X_k^*$ together with alternative allocations of increasing imbalance.

The corresponding configurations are reported in Table~\ref{tab:exp_1_setup}.

\begin{table}[!h]
\centering
\caption{Experimental settings used in Exp.~1. For each setting, full-model training with $r_k=100\%$ is also simulated as a benchmark.}
\label{tab:exp_1_setup}
\renewcommand{\arraystretch}{1.3}
\setlength{\tabcolsep}{6pt}

\begin{tabular}{|p{0.31\columnwidth} p{0.42\columnwidth} p{0.1\columnwidth}|}
\multicolumn{3}{c}{\textbf{Fashion-MNIST + custom FCN} \quad ($r_k=29\%$)} \\
\hline
Config. &
\parbox[c]{0.42\columnwidth}{\centering $Q_k$} &
\parbox[c]{0.1\columnwidth}{\centering $E$} \\
\hline
Most-balanced ($Q_k^*$) &
\parbox[c]{0.42\columnwidth}{\centering $(0.16,\,0.50,\,1.00,\,1.00)$} &
\parbox[c]{0.1\columnwidth}{\centering $0$} \\
Mildly-unbalanced &
\parbox[c]{0.42\columnwidth}{\centering $(0.20,\,0.45,\,0.76,\,0.80)$} &
\parbox[c]{0.1\columnwidth}{\centering $30.53\%$} \\
Moderately-unbalanced &
\parbox[c]{0.42\columnwidth}{\centering $(0.10,\,0.73,\,0.86,\,0.90)$} &
\parbox[c]{0.1\columnwidth}{\centering $66.5\%$} \\
Largely-unbalanced &
\parbox[c]{0.42\columnwidth}{\centering $(0.25,\,0.29,\,0.78,\,1.00)$} &
\parbox[c]{0.1\columnwidth}{\centering $87.43\%$} \\
\hline
\end{tabular}

\vspace{0.25cm}

\begin{tabular}{|p{0.31\columnwidth} p{0.42\columnwidth} p{0.1\columnwidth}|}
\multicolumn{3}{c}{\textbf{CIFAR-10 + custom FCN} \quad ($r_k=23\%$)} \\
\hline
Config. &
\parbox[c]{0.42\columnwidth}{\centering $Q_k$} &
\parbox[c]{0.1\columnwidth}{\centering $E$} \\
\hline
Most-balanced ($Q_k^*$) &
\parbox[c]{0.42\columnwidth}{\centering $(0.15,\,1.00,\,1.00,\,1.00)$} &
\parbox[c]{0.1\columnwidth}{\centering $0$} \\
Mildly-unbalanced &
\parbox[c]{0.42\columnwidth}{\centering $(0.17,\,0.81,\,1.00,\,1.00)$} &
\parbox[c]{0.1\columnwidth}{\centering $18.79\%$} \\
Largely-unbalanced &
\parbox[c]{0.42\columnwidth}{\centering $(0.20,\,0.45,\,0.76,\,0.80)$} &
\parbox[c]{0.1\columnwidth}{\centering $92.15\%$} \\
\hline
\end{tabular}

\vspace{0.25cm}

\begin{tabular}{|p{0.31\columnwidth} p{0.42\columnwidth} p{0.1\columnwidth}|}
\multicolumn{3}{c}{\textbf{CIFAR-10 + ResNet-8} \quad ($r_k=18\%$)} \\
\hline
Config. &
\parbox[c]{0.42\columnwidth}{\centering $Q_k$} &
\parbox[c]{0.1\columnwidth}{\centering $E$} \\
\hline
Most-balanced ($Q_k^*$) &
\parbox[c]{0.42\columnwidth}{\centering $(1.00,\,0.88,\,0.31,\,0.08,\,1.00)$} &
\parbox[c]{0.1\columnwidth}{\centering $0$} \\
Mildly-unbalanced &
\parbox[c]{0.42\columnwidth}{\centering $(0.75,\,0.75,\,0.25,\,0.10,\,1.00)$} &
\parbox[c]{0.1\columnwidth}{\centering $25.25\%$} \\
Largely-unbalanced &
\parbox[c]{0.42\columnwidth}{\centering $(0.50,\,0.50,\,0.25,\,0.12,\,1.00)$} &
\parbox[c]{0.1\columnwidth}{\centering $85.94\%$} \\
\hline
\end{tabular}

\vspace{2pt}
\footnotesize
For ResNet-8, each entry of $Q_k$ corresponds to a block: stem convolution, block~1, block~2, block~3, and the final fully connected layer.
\end{table}

\begin{figure*}[t]
  \centering
  \begin{minipage}[t]{0.32\linewidth}
    \centering
    \includegraphics[width=\linewidth]{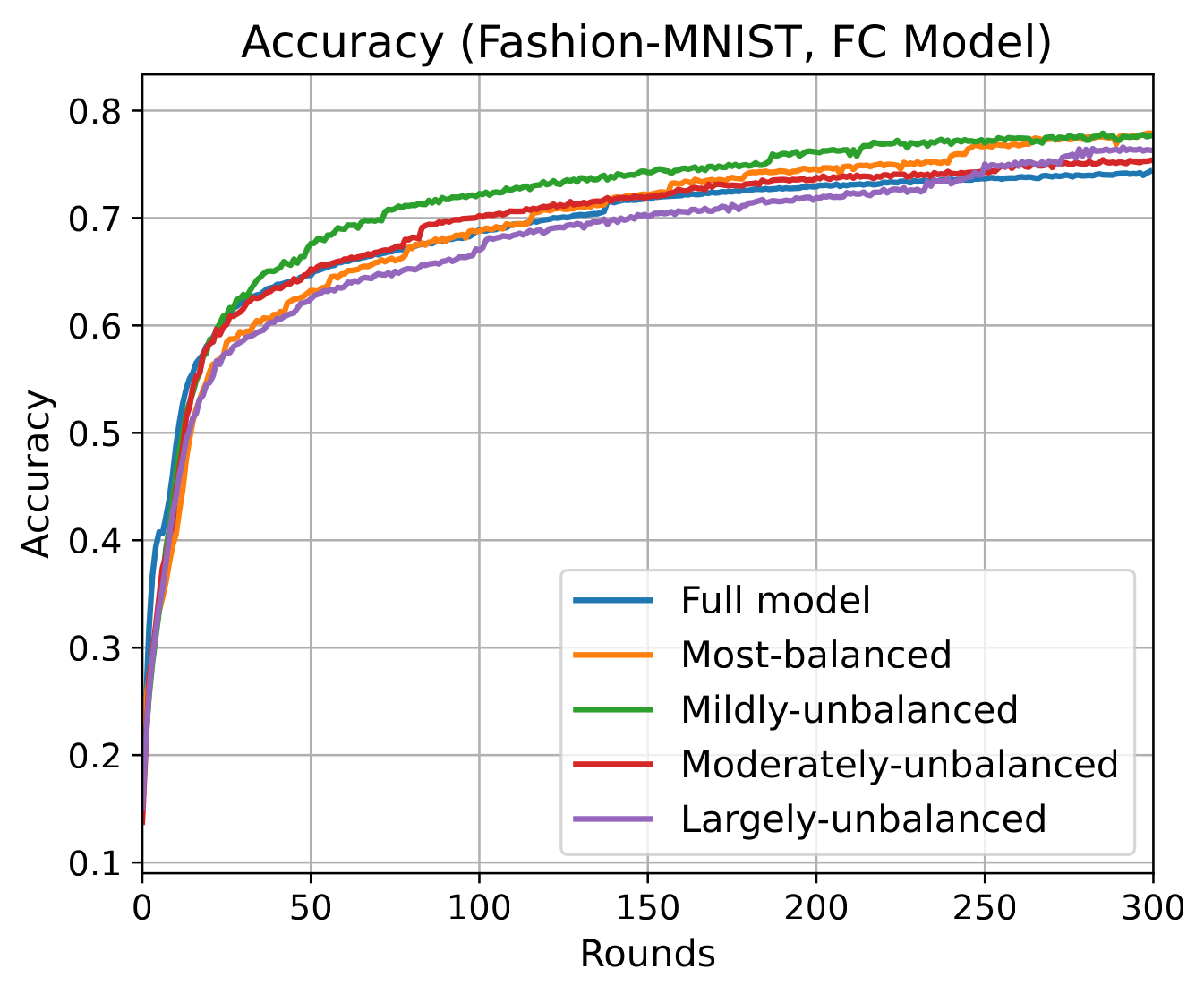}
  \end{minipage}
  \hfill
  \begin{minipage}[t]{0.32\linewidth}
    \centering
    \includegraphics[width=\linewidth]{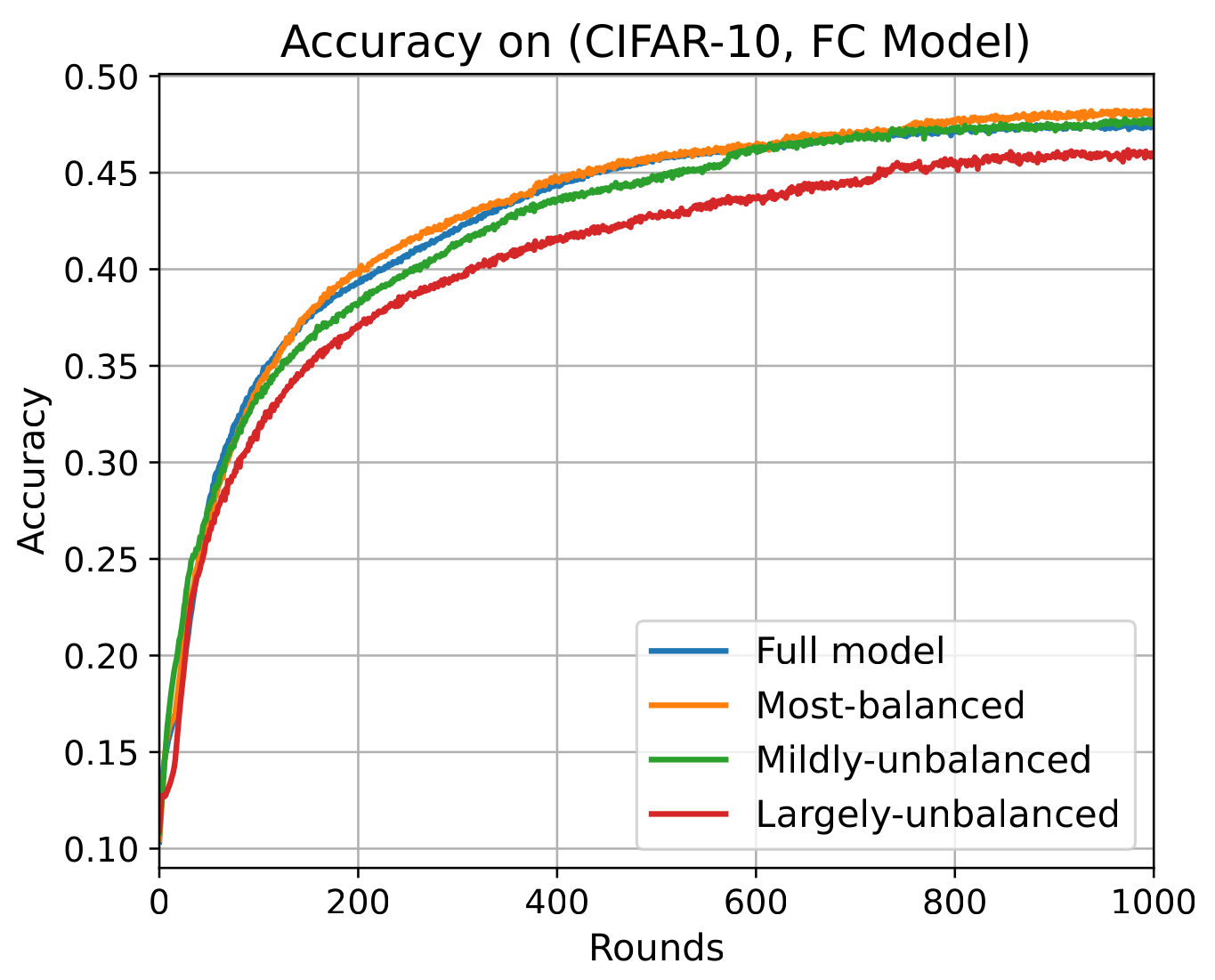}
  \end{minipage}
  \hfill
  \begin{minipage}[t]{0.32\linewidth}
    \centering
    \includegraphics[width=\linewidth]{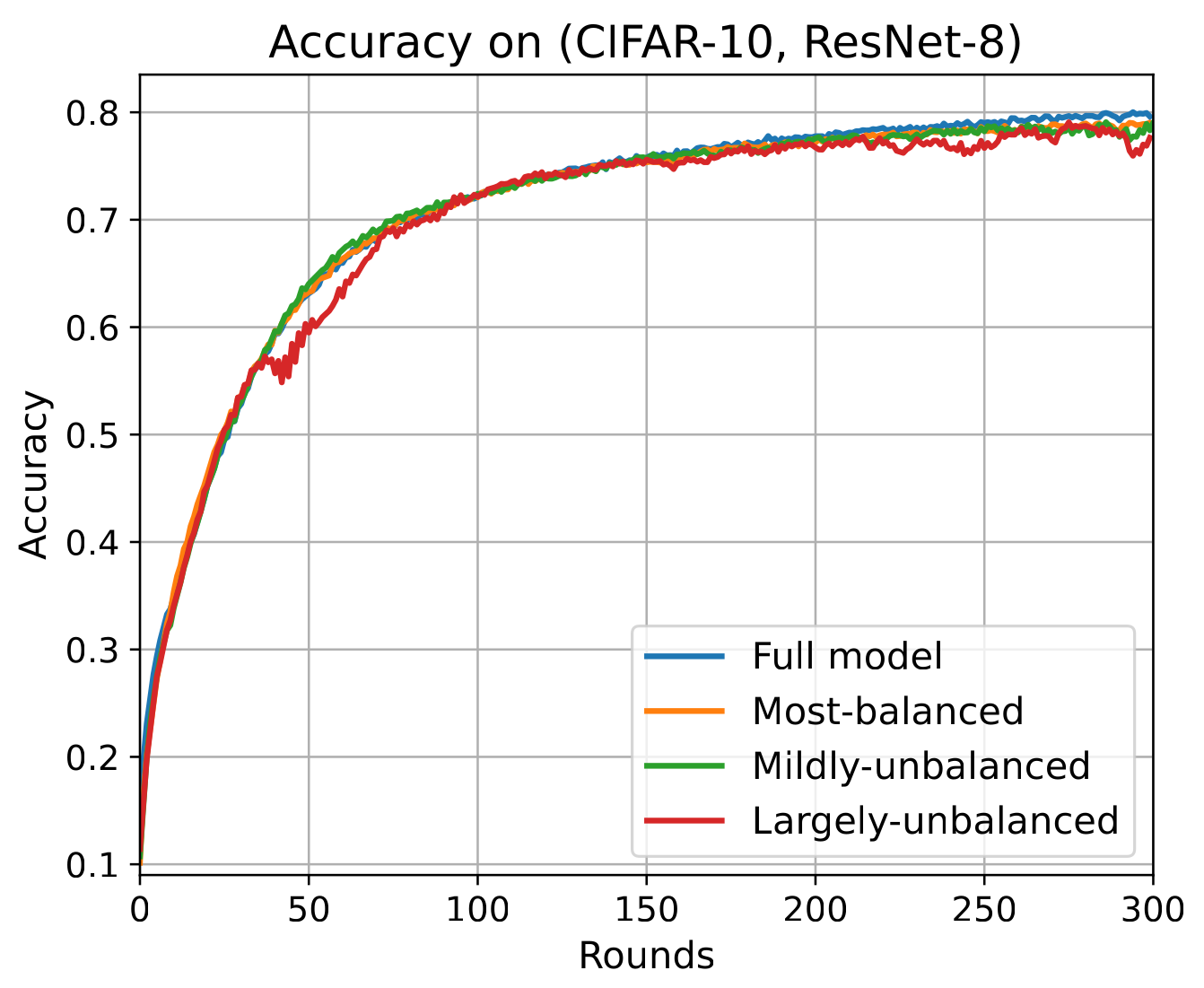}
  \end{minipage}
  \caption{Validation accuracy curves of FedPLT in homogeneous low-resource systems under different layer-wise allocation vectors $Q_k$, compared with full-model training, for (Fashion-MNIST, FCN), (CIFAR-10, FCN), and (CIFAR-10, ResNet-8).}
  \label{fig:results_exp1}
\end{figure*}

For each dataset-model pair, full-model training with $r_k=100\%$ is also included as a benchmark. Each experiment is repeated over multiple random seeds. We report the validation accuracy curves in Figure~\ref{fig:results_exp1} and the final validation accuracy (mean $\pm$ standard deviation) in Table~\ref{tab:results_exp1}.

\begin{table*}[!h]
\centering
\caption{Final validation accuracy (mean $\pm$ standard deviation) over multiple runs of FedPLT and FedAvg in homogeneous low-resource systems across different datasets, models, and allocation balance levels.}
\label{tab:results_exp1}

\renewcommand{\arraystretch}{1.15}

\begin{tabular}{l cc cc cc}
\toprule

\textbf{Methods}
& \multicolumn{2}{c}{\textbf{FashionMNIST + FCN}}
& \multicolumn{2}{c}{\textbf{CIFAR-10 + FCN}}
& \multicolumn{2}{c}{\textbf{CIFAR-10 + ResNet-8}} \\

\midrule

& Allocation & \textbf{Accuracy (\%)}
& Allocation & \textbf{Accuracy (\%)}
& Allocation & \textbf{Accuracy (\%)} \\

\cmidrule(lr){2-3}
\cmidrule(lr){4-5}
\cmidrule(lr){6-7}

\textbf{FedAvg}
& Full model  & $74.33 \pm 9.53$
& Full model & $47.42 \pm 4.12$
& Full model & $\mathbf{79.61 \pm 1.10}$ \\[3pt]

\multirow{4}{*}{\textbf{FedPLT}}

& Most-balanced & $\mathbf{77.91 \pm 5.13}$
& Most-balanced & $\mathbf{48.18 \pm 0.43}$
& Most-balanced & $79.04 \pm 1.22$ \\

& Mildly-unbalanced & $77.65 \pm 4.51$
& Mildly-unbalanced & $47.58 \pm 0.56$
& Mildly-unbalanced & $78.32 \pm 1.87$ \\

& Moderately-unbalanced & $75.35 \pm 5.35$
& Largely-unbalanced & $45.98 \pm 2.01$
& Largely-unbalanced & $77.63 \pm 1.95$ \\

& Largely-unbalanced & $76.32 \pm 5.50$
&  &  
&  &  \\

\bottomrule
\end{tabular}

\end{table*}

Both the table and the figure show that FedPLT remains competitive with full-model training across all three homogeneous low-resource settings, despite each client training only a small fraction of the model. On Fashion-MNIST with the FCN model, all FedPLT configurations outperform FedAvg, with gains ranging from $1.02\%$ to $3.58\%$ while training only $29\%$ of the parameters. On CIFAR-10 with the same FCN model, the most-balanced and mildly-unbalanced configurations improve upon FedAvg by $0.76\%$ and $0.16\%$, respectively, whereas the largely-unbalanced configuration incurs a modest drop of $1.44\%$ while reducing the trained parameters by $77\%$. On CIFAR-10 with ResNet-8, where only $18\%$ of the model is trained, FedPLT remains close to full-model training, with accuracy drops of only $0.57\%$, $1.29\%$, and $1.98\%$ for the most-balanced, mildly-unbalanced, and largely-unbalanced configurations, respectively.

These results indicate that partial-layer training does not inherently compromise accuracy and may even improve generalization in some cases. A possible explanation lies in how FedPLT distributes parameter updates across clients. In FedAvg, all clients update all parameters, so each parameter receives heterogeneous gradients from the whole population. Under non-IID data, this can create strong inter-client interference and allow dominant client distributions to bias the entire model. In contrast, in FedPLT, each parameter block is updated by fewer clients, and different blocks receive gradients from different subsets of the population. This reduces per-parameter gradient conflict and prevents any single client group from dominating the whole model. In this sense, FedPLT distributes client influence across the parameter space, which may reduce client drift, encourage block-wise specialization, and improve convergence and generalization in heterogeneous settings.

Table~\ref{tab:results_exp1} also highlights the variance in accuracy across runs. On Fashion-MNIST with FCN, FedAvg exhibits a standard deviation of $9.53$ points, which is markedly higher than those of the FedPLT configurations. Similarly, on CIFAR-10 with FCN, FedAvg attains a standard deviation of $4.12$, compared to $0.43$-$2.01$ for FedPLT configurations, all of which remain below half of FedAvg’s value. On CIFAR-10 with ResNet-8, FedAvg and FedPLT have similarly low variability, with standard deviations below $2$, indicating stable training. Overall, these results suggest that FedPLT exhibits greater robustness and provides more consistent and reliable training behavior, especially for the fully connected models.













\begin{table*}[t]
\centering
\caption{FedPLT gains over FedAvg under three target budgets (or goals) in homogeneous low-resource systems: accuracy, communication, and computation.}
\label{tab:effective_FedPLT}
\renewcommand{\arraystretch}{1.15}
\setlength{\tabcolsep}{4.2pt}

\begin{tabular}{l ccc ccc ccc}
\toprule

& \multicolumn{3}{c}{\textbf{FashionMNIST + FCN}}
& \multicolumn{3}{c}{\textbf{CIFAR-10 + FCN}}
& \multicolumn{3}{c}{\textbf{CIFAR-10 + ResNet-8}} \\

\cmidrule(lr){2-4}
\cmidrule(lr){5-7}
\cmidrule(lr){8-10}

& \textbf{Eq-Acc.} & \textbf{Eq-Comm.} & \textbf{Eq-Comp.}
& \textbf{Eq-Acc.} & \textbf{Eq-Comm.} & \textbf{Eq-Comp.}
& \textbf{Eq-Acc.} & \textbf{Eq-Comm.} & \textbf{Eq-Comp.} \\

\textbf{Fixed Target}
& $74\%$ & $43.75$ GB & $27.50$ TFLOPs
& $47\%$ & $425$ GB & $260$ TFLOPs
& $78\%$ & $5.25$ GB & $512.50$ TFLOPs \\

\midrule

\textbf{Accuracy Gain (\%)}
& -- & $+5.17$ & $+5.91$
& -- & $+1.75$ & $+2.61$
& -- & $+2.09$ & $+4.02$ \\

\textbf{System Comm. Gain (GB)}
& $+36.08$ & -- & $-8.27$
& $+208.39$ & -- & $-90.04$
& $+2.23$ & -- & $-1.29$ \\

\textbf{System Comp. Gain (TFLOPs)}
& $+31.43$ & $+6.13$ & --
& $+204.44$ & $+68.20$ & --
& $+396.32$ & $+151.15$ & -- \\

\bottomrule
\end{tabular}
\end{table*}

Across FedPLT configurations, the most-balanced allocation consistently provides the best performance. As allocations become increasingly unbalanced, accuracy degrades across all three settings. Comparing the most-balanced and largely-unbalanced configurations, accuracy drops by $1.59\%$ on Fashion-MNIST + FCN, $2.20\%$ on CIFAR-10 + FCN, and $1.41\%$ on CIFAR-10 + ResNet-8. The same trend is reflected in the standard deviations, which generally increase with imbalance. These results confirm that balanced parameter allocation improves both performance and training stability.



In terms of convergence speed, FedPLT and FedAvg appear to exhibit similar learning curves when plotted against communication rounds. However, the horizontal axis in Fig.~\ref{fig:results_exp1} does not reflect the actual communication or computation cost. To account for this, Table~\ref{tab:effective_FedPLT} reports target-based comparisons under three scenarios: equal accuracy, equal communication budget, and equal computation budget.




As discussed in Section~\ref{sec:FedPLT_efficiency}, FedPLT preserves the full forward pass and downlink cost, while the backward pass and uplink communication scale with the training ratio $r_k$. Consequently, its lower per-round cost allows more training rounds under the same total resource budget. Table~\ref{tab:effective_FedPLT} shows that, across all datasets and models, FedPLT consistently achieves higher accuracy than FedAvg under the same total communication or computation budget. Conversely, for a fixed target accuracy, FedPLT requires substantially less total system communication and computation than FedAvg. Under the equal-computation comparison, the communication gain may become negative, since FedPLT can require more rounds to match the same computation budget despite its lower per-round backward cost.


Overall, these results indicate that FedPLT is well suited to homogeneous low-resource federated settings in which no client can train the full model locally. Although each client updates only a small portion of the network (29\%, 23\%, and 18\% in our scenarios), the global model is still effectively optimized through complementary sub-layer updates across clients. This makes FedPLT a practical and scalable alternative to full-model training, particularly for large models, where full local optimization is often infeasible on resource-constrained edge devices such as IoT devices.

\subsection{Benchmarking FedPLT against Existing Methods in a Heterogeneous Setting}



In this experiment, we evaluate FedPLT in a heterogeneous federated system where clients differ significantly in computation and communication capacity. Such heterogeneity creates both resource imbalance and straggler effects, making this setting particularly challenging for federated training. Our objective is to compare FedPLT with existing federated learning methods under these realistic system constraints.

As in the previous subsection, we conduct three sets of experiments using different dataset-model pairs: Fashion-MNIST with an FCN model, CIFAR-10 with an FCN model, and CIFAR-10 with ResNet-8.

For all three cases, we consider a realistic heterogeneous system with three levels of computational and communication capability:
\begin{itemize}
    \item \textbf{Highly capable devices}: 10\% of clients, capable of training the full global model (\(r_k=1.00\)).
    \item \textbf{Moderately capable devices}: 30\% of clients, capable of training 29\% of the model for Fashion-MNIST + FCN, 23\% for CIFAR-10 + FCN, and 18\% for CIFAR-10 + ResNet-8.
    \item \textbf{Low-capability devices}: 60\% of clients, capable of training only 6\% of the model.
\end{itemize}



For each training ratio \(r_k\), we use the default layer-wise allocation vector \(Q_k^*\) whose induced contribution vector \(X_k\) is as balanced as possible. The resulting FedPLT configurations are reported in Table~\ref{tab:Exp2_setups}.

\begin{table}[!h]
\centering
\caption{FedPLT configurations used across different dataset-model pairs in the heterogeneous setting.}
\label{tab:Exp2_setups}

\begin{tabular}{l c c}
\toprule

Dataset/Model & $r_k$ & $Q_k^*$\\
\midrule

\multirow{3}{*}{Fashion-MNIST + FCN}
& $1.00$ & $(1.0, 1.0, 1.0, 1.0)$ \\
& $0.29$ & $(0.16, 0.5, 1.0, 1.0)$ \\
& $0.06$ & $(0.03, 0.08, 0.31, 1.0)$ \\

\midrule

\multirow{3}{*}{CIFAR-10 + FCN}
& $1.00$ &  $(1.0, 1.0, 1.0, 1.0)$ \\
& $0.23$ & $(0.15, 1.0, 1.0, 1.0)$ \\
& $0.06$ & $(0.02, 0.27, 1.0, 1.0)$ \\

\midrule

\multirow{3}{*}{CIFAR-10 + ResNet-8}

& $1.00$ &  $(1.0, 1.0, 1.0, 1.0, 1.0)$ \\
& $0.18$ & $(1.0, 0.88, 0.31, 0.08, 1.0)$ \\
& $0.06$ & $(1.0, 0.25, 0.06, 0.03, 1.0)$ \\

\bottomrule
\end{tabular}
\end{table}



\begin{figure*}[t]
  \centering
  \begin{minipage}[t]{0.32\linewidth}
    \centering
    \includegraphics[width=\linewidth]{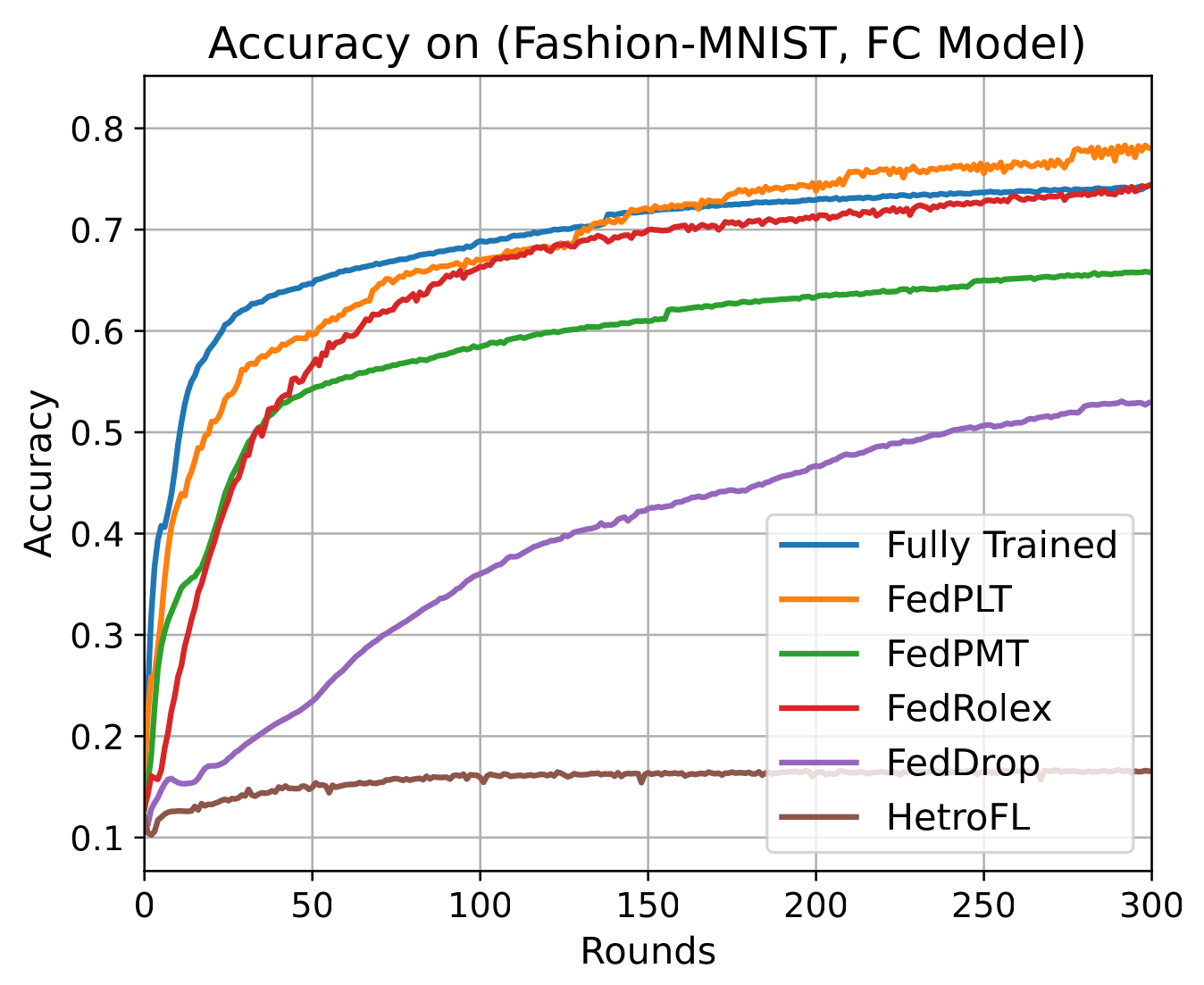}
  \end{minipage}
  \hfill
  \begin{minipage}[t]{0.32\linewidth}
    \centering
    \includegraphics[width=\linewidth]{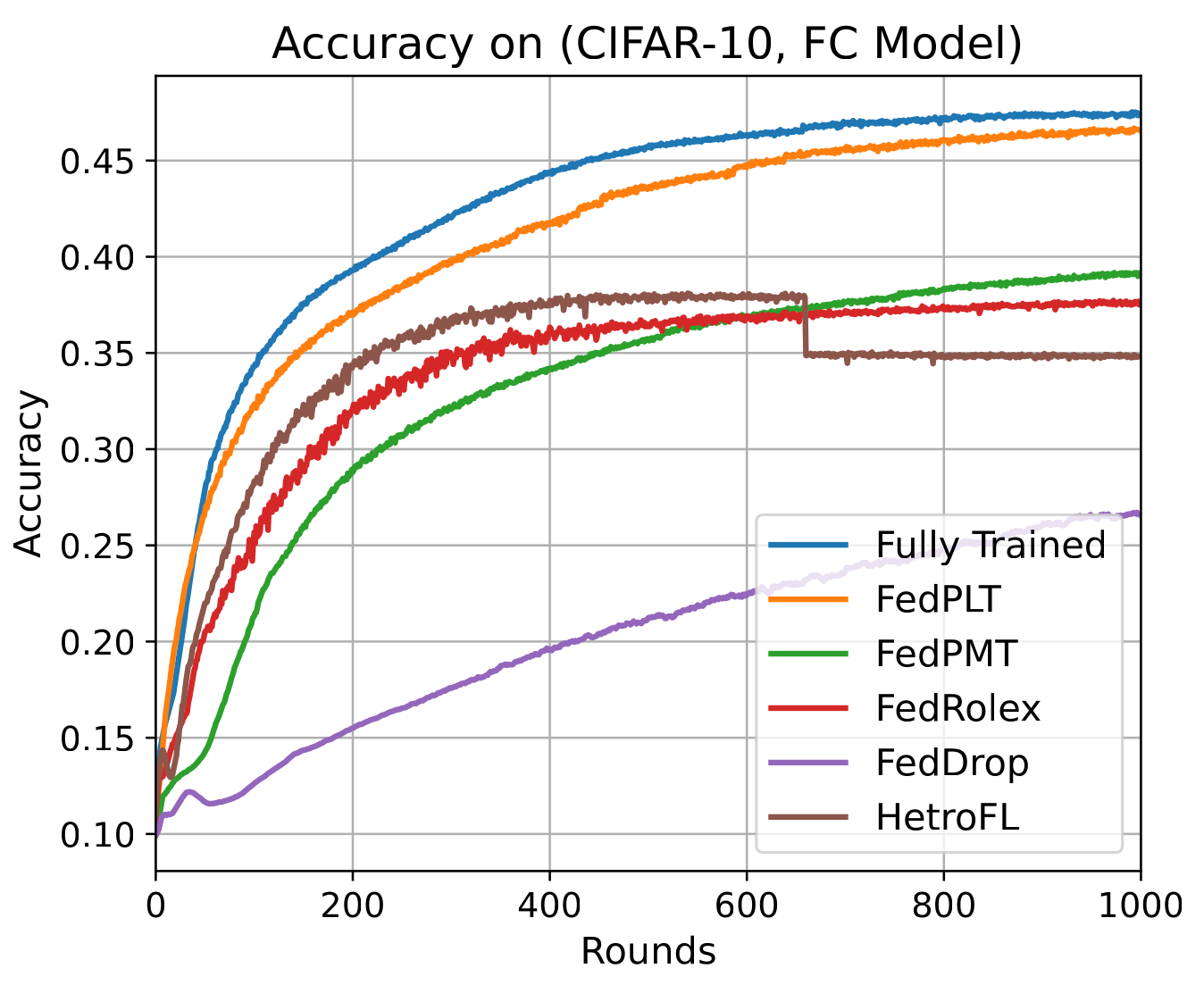}
  \end{minipage}
  \hfill
  \begin{minipage}[t]{0.32\linewidth}
    \centering
    \includegraphics[width=\linewidth]{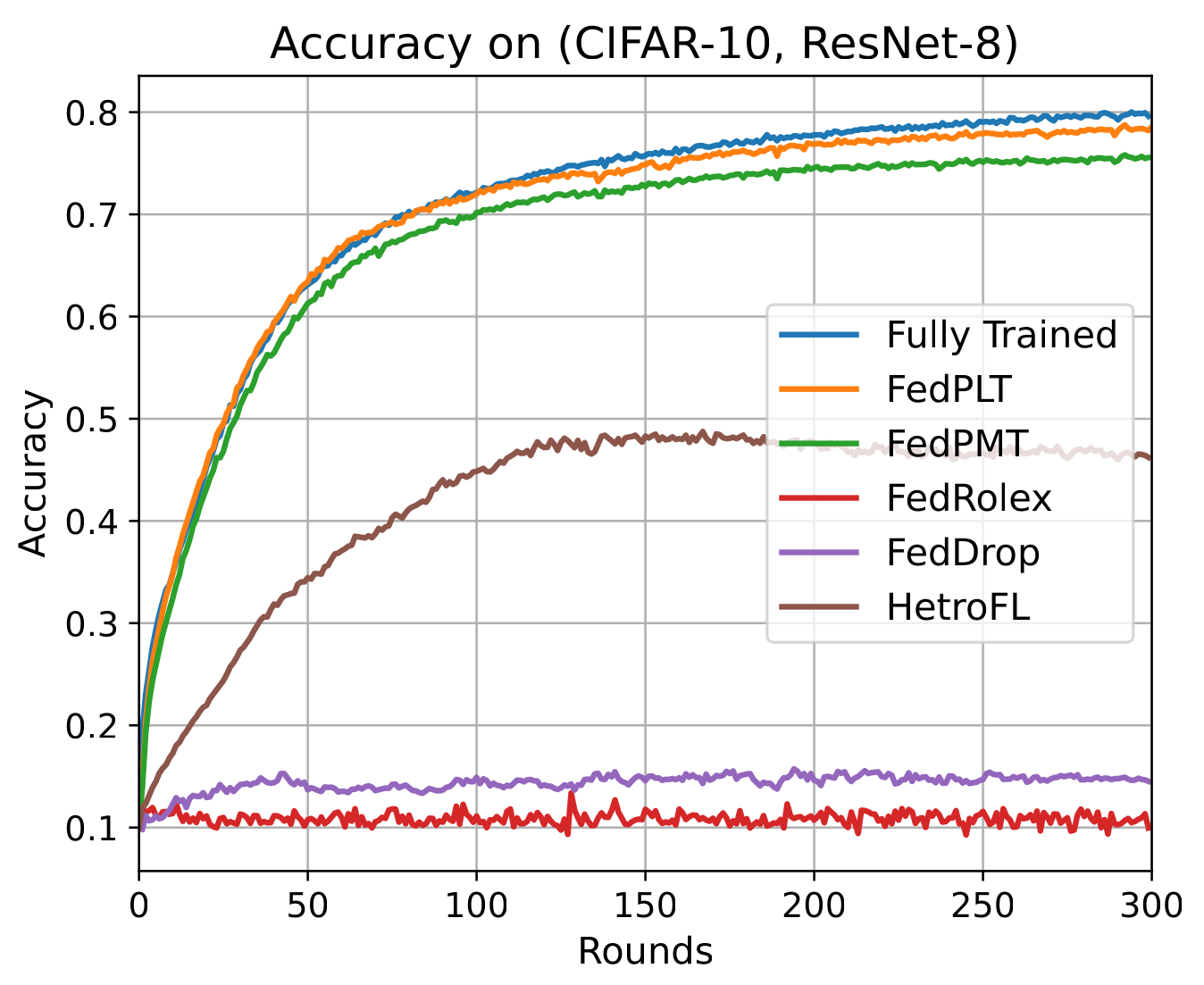}
  \end{minipage}
  \caption{Validation accuracy curves comparing FedPLT with existing methods in a heterogeneous and low-resource FL system, for (Fashion-MNIST, FCN), (CIFAR-10, FCN), and (CIFAR-10, ResNet-8).}
  \label{fig:results_exp2}
\end{figure*}

We benchmark FedPLT against FedAvg, FedPMT, FedRolex, FedDrop, and HeteroFL. For fairness, each baseline is simulated under a configuration that matches, as closely as possible, the corresponding FedPLT trained parameter count. Each setting was repeated multiple times with different random seeds. We report the validation accuracy curves in Figure~\ref{fig:results_exp2} and the final validation accuracy (mean \(\pm\) standard deviation) in Table~\ref{tab:results_exp2}.

\begin{table}[!h]
\centering
\caption{Final validation accuracy (mean $\pm$ standard deviation) compared with existing methods in a heterogeneous and low-resource FL system across different datasets and models.}
\label{tab:results_exp2}

\renewcommand{\arraystretch}{1.15}

\begin{tabular}{l c c c}
\toprule

\textbf{Methods}
& \makecell{\textbf{Fashion MNIST}\\\textbf{+ FCN}}
& \makecell{\textbf{CIFAR-10}\\\textbf{+ FCN}}
& \makecell{\textbf{CIFAR-10}\\\textbf{+ ResNet-8}} \\

\midrule

\textbf{FedAvg}
& $ 74.33 \pm 9.53 $
& $\mathbf{47.42 \pm 4.12}$
& $\mathbf{79.61 \pm 1.10}$ \\

\textbf{FedPLT}
& $\mathbf{78.08 \pm 4.65}$
& $46.56 \pm 0.71$
& $78.21 \pm 1.57$ \\

\textbf{FedPMT}
& $65.81 \pm 5.8$
& $39.16 \pm 3.47$
& $75.57 \pm 3.04$ \\

\textbf{FedRolex}
& $74.32 \pm 3.85$
& $37.65 \pm 1.32$
& $9.96 \pm 1.48$  \\

\textbf{FedDrop}
& $52.88 \pm 8.35$
& $26.60 \pm 2.94$
& $14.53 \pm 0.81$ \\

\textbf{HeteroFL}
& $16.56 \pm 19.70$
& $34.81 \pm 8.31$
& $46.22 \pm 1.58$ \\

\bottomrule
\end{tabular}

\end{table}


Figure~\ref{fig:results_exp2} and Table~\ref{tab:results_exp2} show that FedPLT consistently outperforms existing partial-training and submodeling approaches across all heterogeneous settings. On Fashion-MNIST with the FCN model, FedPLT achieves the highest final accuracy among all methods, exceeding FedAvg by $3.75\%$ and substantially outperforming FedPMT, FedRolex, FedDrop, and HeteroFL. On CIFAR-10 with the FCN model, FedPLT remains close to FedAvg (46.56\% vs.\ 47.42\%) while outperforming FedPMT by $7.40\%$, FedRolex by $8.91\%$, HeteroFL by $11.75\%$, and FedDrop by $19.96\%$. A similar trend is observed for CIFAR-10 with ResNet-8, where FedPLT remains within $1.4\%$ of FedAvg while improving upon FedPMT by $2.64\%$ and HeteroFL by $31.00\%$. In this setting, FedRolex and FedDrop fail and remain below $15\%$ accuracy.


These results help explain why FedPLT performs better than existing heterogeneity-aware methods. FedPMT restricts low-capability clients to training the deeper layers, leading to highly imbalanced contribution vectors and leaving the shallow layers undertrained across the heterogeneous client datasets. HeteroFL also trains the full model unevenly, since clients operate on reduced-width local models, which results in biased training. FedDrop and FedRolex use temporal masking, where the set of trained parameters changes from one round to another, leading to inconsistent training. In contrast, FedPLT assigns trainable model parts in a fixed and balanced manner, while rotating sublayers across clients so that all parameters are trained more uniformly across the heterogeneous client population. This leads to more stable and less biased training and helps explain the stronger accuracy observed in Fig.~\ref{fig:results_exp2} and Table~\ref{tab:results_exp2}.

Beyond mean accuracy, FedPLT also exhibits markedly reduced variance across runs. For example, on CIFAR-10 with the FCN model, FedPLT achieves a standard deviation of $0.71$, compared to $4.12$ for FedAvg and larger values for the competing methods. This improved stability reflects the structured and deterministic nature of FedPLT’s partial updates, which avoids both the rigidity of suffix-layer training and the randomness of sub-model training schemes.

Table~\ref{tab:exp2_target_budget} complements the final-accuracy comparison by reporting equal-accuracy and equal-communication evaluations on CIFAR-10. Across both models, FedPLT reaches the target accuracy with substantially lower total system communication and computation than FedAvg, while several competing baselines fail to reach the target. For example, with ResNet-8, FedPLT reaches the $75\%$ target using roughly half the communication and computation required by FedAvg. Under the same total communication budget, FedPLT also achieves higher accuracy than FedAvg for both models and remains clearly stronger than the other heterogeneity-aware baselines in the CIFAR-10 settings. These results show that, even in heterogeneous systems, FedPLT remains considerably more resource-efficient and converges faster under a limited budget than FedAvg and the other baselines.

\begin{table*}[t]
\centering
\caption{Comparison under equal-accuracy and equal-communication evaluations in the heterogeneous setting. For equal-accuracy, we report the total system communication and computation required to reach the target accuracy. For equal-communication, we report the accuracy achieved under the common total system communication budget.}
\label{tab:exp2_target_budget}
\renewcommand{\arraystretch}{1.15}
\setlength{\tabcolsep}{4.5pt}

\begin{tabular}{l ccc ccc}
\toprule
\multirow{3}{*}{\textbf{Method}}
& \multicolumn{3}{c}{\textbf{CIFAR-10 + FCN}}
& \multicolumn{3}{c}{\textbf{CIFAR-10 + ResNet-8}} \\

\cmidrule(lr){2-4}
\cmidrule(lr){5-7}

& \multicolumn{2}{c}{\textbf{Target Acc.} (\(45\%\))}
& \textbf{Comm. Budget} (\(139.26\) GB)
& \multicolumn{2}{c}{\textbf{Target Acc.} (\(75\%\))}
& \textbf{Comm. Budget} (\(1.77\) GB) \\

\cmidrule(lr){2-3}
\cmidrule(lr){4-4}
\cmidrule(lr){5-6}
\cmidrule(lr){7-7}

& \textbf{Comm.} & \textbf{Comp.} & \textbf{Acc.}
& \textbf{Comm.} & \textbf{Comp.} & \textbf{Acc.} \\

\midrule
\textbf{FedAvg}
& $304.67$ GB & $233.96$ TFLOPs
& $39.37\%$
& $4.05$ GB & $511.31$ TFLOPs
& $64.98\%$ \\

\textbf{FedPLT}
& $260.26$ GB & $155.90$ TFLOPs
& $40.52\%$ 
& $2.72$ GB & $264.82$ TFLOPs
& $71.26\%$ \\

\textbf{FedPMT}
& $\infty$ & $\infty$
& $33.45\%$
& $3.82$ GB & $353.00$ TFLOPs
& $70.20\%$ \\

\textbf{FedRolex}
& $\infty$ & $\infty$
& $37.65\%$
& $\infty$ & $\infty$
& $9.96\%$ \\

\textbf{FedDrop}
& $\infty$ & $\infty$
& $26.60\%$
& $\infty$ & $\infty$
& $14.53\%$ \\

\textbf{HeteroFL}
& $\infty$ & $\infty$
& $34.81\%$
& $\infty$ & $\infty$
& $46.22\%$ \\
\bottomrule

\end{tabular}
\end{table*}

Overall, these experiments demonstrate that, despite severe system heterogeneity, where 60\% of clients are limited to training only 6\% of the model, FedPLT consistently achieves accuracy comparable to FedAvg while outperforming the other baselines across all configurations. The results also highlight its stability, robustness, and resource efficiency under highly imbalanced client capabilities. Together, these findings suggest that FedPLT effectively reduces optimization bias and variance through fixed and balanced layer-wise training and more uniform parameter exposure across the client population, making it a high-performing and resource-efficient solution for heterogeneous federated systems with severe device constraints.

\subsection{FedPLT and Optimal Client Sampling}

This experiment evaluates how FedPLT interacts with variance-optimal client sampling under severe communication constraints. In particular, we compare the original Optimal Client Sampling (OCS) formulation with our FedPLT-aware OCS to assess whether accounting for partial training ratios improves convergence and generalization in heterogeneous systems.

We consider two heterogeneous systems with 100 clients:

\begin{itemize}
    \item \textbf{Low-resource system:} 10\% of clients train the full model, while 90\% are restricted to training only 20\% of the model.
    \item \textbf{Moderate-resource system:} 10\% of clients train the full model, while 90\% train 50\% of the model.
\end{itemize}

In both settings, we fix the communication budget to $\kappa=5$ out of a maximum of 100, corresponding to extremely limited resource availability. We compare the original OCS formulation, where the budget constrains only the number of selected clients, against our FedPLT-aware OCS, where the budget accounts for the partial training ratios $r_k$.

\begin{figure}[h]
  \centering
  \begin{minipage}[t]{0.49\linewidth}
    \centering
    \includegraphics[width=\linewidth]{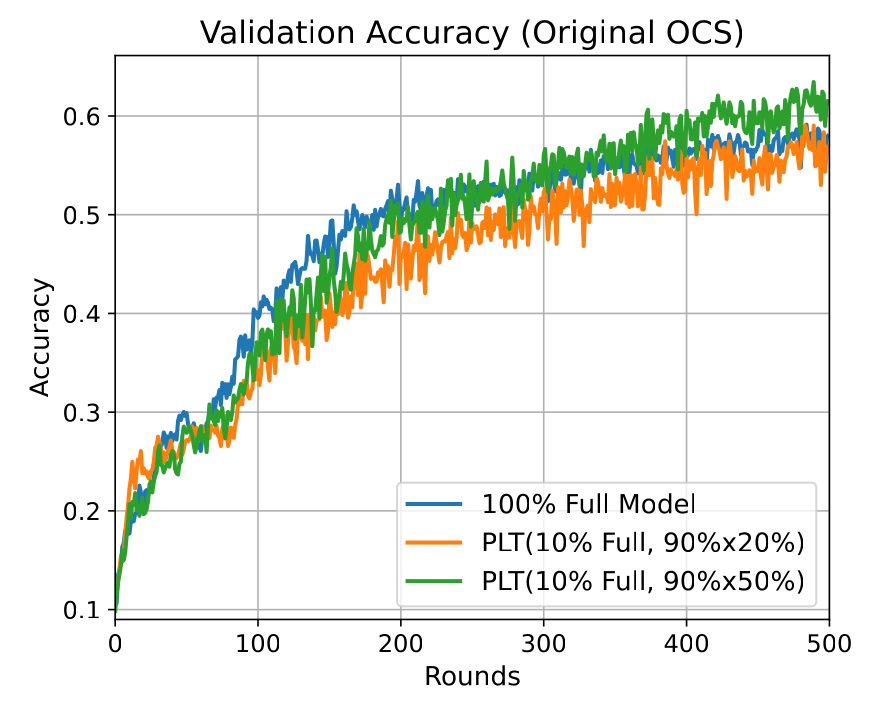}
  \end{minipage}
  \hfill
  \begin{minipage}[t]{0.49\linewidth}
    \centering
    \includegraphics[width=\linewidth]{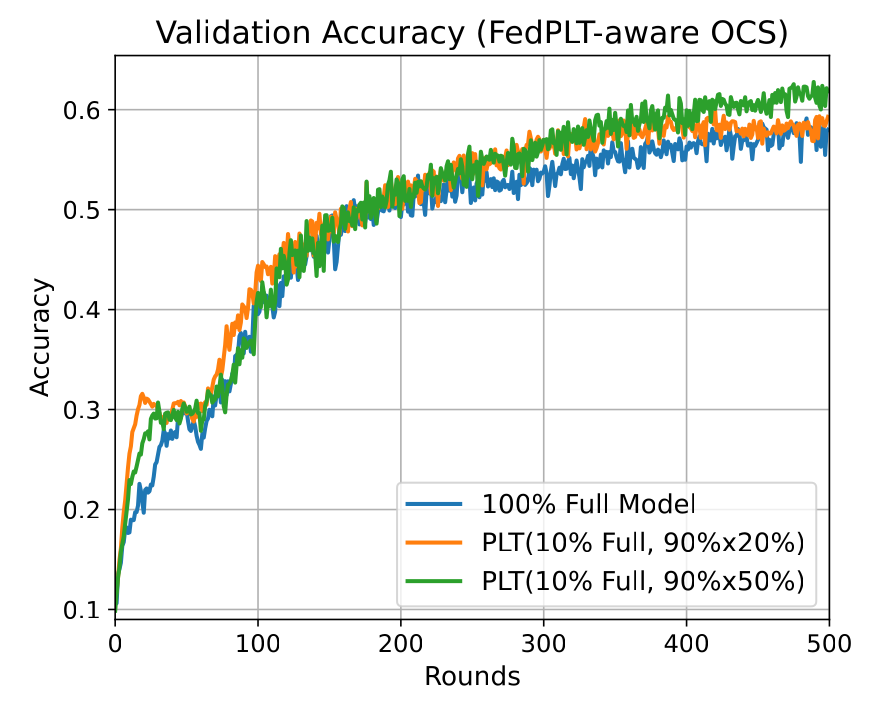}
  \end{minipage}
  \caption{Validation accuracy curves comparing OCS with and without FedPLT. Left: original OCS formulation. Right: FedPLT-aware OCS.}
  \label{fig:results_56}
\end{figure}

With the original OCS formulation, where the communication budget counts only the number of clients but ignores their layer ratios $r_k$, we observe mixed behavior. In the moderate-resource case (50\% training), FedPLT combined with OCS improves validation accuracy by +3.51\% relative to full-model training while reducing communication by 36\% (6,375 vs.\ 10,000 units). However, in the low-resource case (20\% training), accuracy drops slightly by -1.11\%, despite achieving 43\% lower communication (5,700 vs.\ 10,000 units). This occurs because the original constraint limits the expected number of selected clients to $\kappa$, regardless of whether they train the full model or only a fraction, thereby restricting data coverage per round and biasing sampling toward less informative updates under strong heterogeneity.



By contrast, the FedPLT-aware OCS consistently improves performance under the same normalized communication budget of 10,000. Accuracy gains of +3.84\% in the moderate-resource system and +1.29\% in the low-resource system confirm that incorporating the $r_k$ constraint yields a more efficient allocation of the communication budget. Since the budget now accounts for trained fractions rather than client counts, the expected number of sampled clients can exceed $\kappa$ when $r_k<1$, allowing multiple partial trainers to be included instead of a single full trainer.

Importantly, this improvement is not merely due to selecting more clients per round, but also to the interaction between communication-aware sampling and FedPLT’s balanced partial training. By allowing multiple low-$r_k$ clients to participate simultaneously, FedPLT-aware OCS exposes a larger portion of the global model to more diverse data, leading to improved generalization and more stable convergence.

Overall, these experiments highlight that FedPLT benefits significantly from being integrated with communication-aware sampling schemes. While standard OCS already reduces communication cost, our FedPLT-aware extension achieves both higher accuracy and fairer client participation under heterogeneous system and data constraints, further reinforcing FedPLT’s scalability and robustness.

\section{Conclusion}

In this paper, we proposed FedPLT, a federated learning framework that enables flexible and fine-grained partial parameter training in heterogeneous systems. Unlike existing sub-modeling and partial-training approaches that rely on coarse layer-wise freezing or stochastic parameter selection, FedPLT decomposes each layer into sub-layers and assigns them deterministically across clients, ensuring balanced parameter exposure and more consistent optimization of the global model despite highly unequal device capabilities.

We introduced a principled formulation of partial training through the training ratio $r_k$, layer-wise allocation vector $Q_k$, and contribution vector $X_k$, and showed empirically that balanced contribution configurations yield stable optimization and strong performance across both FCN and CNN architectures. Building on this insight, FedPLT distributes training effort across layers in a structured manner while preserving the full forward pass, allowing all model components to be optimized collaboratively even when individual clients can train only small fractions of the model.

Extensive experiments demonstrate that FedPLT achieves accuracy comparable to or exceeding FedAvg while substantially outperforming existing heterogeneity-aware baselines under severe system constraints, including scenarios where up to 60\% of clients train only 6\% of the model. Moreover, FedPLT consistently exhibits reduced variance across runs, indicating improved stability and robustness in non-IID environments. We further extended Optimal Client Sampling to account for partial training ratios and showed that FedPLT-aware sampling yields additional gains in communication efficiency and generalization.


Collectively, these results establish FedPLT as a scalable and resource-efficient alternative to full-model federated learning, capable of maintaining near full-model federated performance while accommodating extreme device heterogeneity. By enabling flexible partial participation without sacrificing model coherence, FedPLT is well suited to IoT environments, where edge devices such as sensors, wearables, smart cameras, drones, and gateways often operate under strict computation and communication constraints. It thus opens new opportunities for deploying large-scale federated learning systems on resource-constrained edge devices and heterogeneous real-world infrastructures.

\appendices
\onecolumn

\section{Geometric Interpretation of the Masked Gradient Alignment Assumption}
\label{app:masked_alignment}

In the convergence analysis, we introduce the following assumption specific to static partial training schemes. For each client \(k\) and model iterate \(W\), define the masking attenuation ratio
\[
\rho_k(W)
:=
\frac{\left\langle m_k \odot \nabla F_k(W),\, W-W^* \right\rangle}
{\left\langle \nabla F_k(W),\, W-W^* \right\rangle},
\]
whenever the denominator is nonzero, where \(W^*\) denotes a global minimizer. The assumption requires that
\[
\rho_k(W) > 0.
\]

This ratio measures how masking affects the alignment between the local gradient and the direction \(W-W^*\). To interpret it geometrically, let
\begin{itemize}
    \item \(\theta_1\) be the angle between \(\nabla F_k(W)\) and \(W-W^*\),
    \item \(\theta_2\) be the angle between \(m_k \odot \nabla F_k(W)\) and \(W-W^*\),
    \item \(\theta_3\) be the angle between \(m_k \odot \nabla F_k(W)\) and \(\nabla F_k(W)\).
\end{itemize}
Then
\[
\left\langle \nabla F_k(W),\, W-W^* \right\rangle
=
\|\nabla F_k(W)\|\,\|W-W^*\|\,\cos(\theta_1),
\]
\[
\left\langle m_k \odot \nabla F_k(W),\, W-W^* \right\rangle
=
\|m_k \odot \nabla F_k(W)\|\,\|W-W^*\|\,\cos(\theta_2),
\]
and, since masking corresponds to a coordinate-wise projection onto the active coordinates,
\[
\|m_k \odot \nabla F_k(W)\|
=
\|\nabla F_k(W)\|\,\cos(\theta_3).
\]
Substituting into the definition of \(\rho_k(W)\) gives
\[
\rho_k(W)
=
\frac{\cos(\theta_3)\cos(\theta_2)}{\cos(\theta_1)}.
\]

\noindent \textbf{Interpretation:} \\
Since \(\cos(\theta_3)\ge 0\), the condition \(\rho_k(W)>0\) requires \(\cos(\theta_1)\) and \(\cos(\theta_2)\) to have the same sign. In other words, masking must preserve the overall descent orientation of the local gradient relative to the direction \(W-W^*\). Thus, if the full gradient is aligned with progress toward \(W^*\), then the masked gradient must remain aligned in the same general direction.

Moreover:
\begin{itemize}
    \item If masking improves alignment with the direction \(W-W^*\), that is, \(\cos(\theta_2)>\cos(\theta_1)\), then \(\rho_k(W)>\cos(\theta_3)\).
    \item If masking degrades alignment, that is, \(\cos(\theta_2)<\cos(\theta_1)\), then \(\rho_k(W)<\cos(\theta_3)\).
\end{itemize}

Therefore, \(\rho_k(W)\) quantifies how much masking preserves or degrades the directional usefulness of the local gradient.

\noindent \textbf{Why the assumption is reasonable:} \\
Under balanced participation and mild heterogeneity, local gradients are typically expected to contribute to global progress. The assumption \(\rho_k(W)>0\) rules out degenerate masking patterns that would reverse this contribution. It is therefore a natural condition ensuring that static partial training preserves a meaningful descent direction at the client level.
\newpage
\section{Convergence Proof of Static Partial Parameter Update Strategies}
\label{app:convergence}

\subsection{Introduction}

\subsubsection{Objective}

In this subsection, we present a detailed convergence analysis of any \textbf{static partial model training scheme}, including our proposed method \textbf{FedPLT}. The objective is to show that, under standard assumptions, such methods converge.

We consider the case where each client's objective is smooth and strongly convex, and the global objective is defined as a weighted sum of local losses. Our analysis takes into account the key conditions:
\begin{itemize}
    \item \textbf{Static Partial Layer Training:} Clients update only a fixed subset of model parameters using predefined binary masks over all training rounds.
    \item \textbf{Stochastic Gradient Descent (SGD):} Clients train locally using mini-batch SGD, leading to stochastic updates.
\end{itemize}

The convergence analysis aims to:
\begin{itemize}
    \item Quantify how these factors (partial updates and stochastic gradients) influence the global model update and convergence.
    \item Derive a bound on the expected squared distance between the global model at any round and its optimal value over $t$ rounds.
    \item Establish a convergence rate to stationary points in the non-convex setting.
\end{itemize}

The overall goal is to demonstrate that despite the constraint imposed by restricted update masks, FedPLT, as well as other static partial model update methods, remains an effective and theoretically grounded algorithm with provable convergence guarantees.

\subsubsection{Problem Formulation}

We consider the standard federated learning optimization problem, where the goal is to minimize a global objective function defined as the weighted average of local losses across $K$ clients:
\begin{equation}
\min_{W \in \mathbb{R}^d} F(W) := \sum_{k=1}^{K} \frac{n_k}{n} F_k(W), \quad \text{with } n = \sum_{k=1}^K n_k,
\end{equation}
where $F_k(W)$ is the empirical risk on client $k$ given by:
\begin{equation}
F_k(W) := \frac{1}{n_k} \sum_{i=1}^{n_k} \ell(W; x_i^{(k)}, y_i^{(k)}),
\end{equation}
and $\ell(\cdot; x, y)$ is the per-sample loss function evaluated on a data point $(x_i^{(k)}, y_i^{(k)})$ from client $k$.

\subsubsection{Partial Update Rule}

Let $W^t \in \mathbb{R}^d$ denote the global model at communication round $t$. Each client $k$ is associated with a binary mask vector $m_k \in \{0,1\}^d$, which specifies the subset of model parameters that client $k$ is responsible for updating. Specifically, for each coordinate $i \in \{1, \dots, d\}$:
\begin{equation}
(m_k)_i = 
\begin{cases}
1 & \text{if client } k \text{ is assigned to update coordinate } i, \\
0 & \text{otherwise.}
\end{cases}
\end{equation}

The mask $m_k$ reflects the client's computational budget and its assigned sub-layers in the model. Let $r_k = \|m_k\|_0 / d$ denote the fraction of coordinates that client $k$ is allowed to update.

At the beginning of round $t$, each selected client $k$ receives the current global model $W^t$ and initializes its local model:
\[
W_k^{t,0} := W^t.
\]

Each client performs $\tau$ local update steps, only modifying the coordinates specified by $m_k$.

\noindent We distinguish between two possible settings:

\begin{enumerate}
    \item \textbf{Local Stochastic Gradient Descent (SGD):}
    In the default and practical setting, each client uses mini-batch SGD. At each local step $s$, a random mini-batch $\xi_k^{(s)}$ is sampled from client $k$'s local dataset, and the update rule is:
    \begin{equation}
    W_k^{t,s+1} = W_k^{t,s} - \eta_k \left(m_k \odot \nabla \ell(W_k^{t,s}; \xi_k^{(s)})\right), \quad s = 0, \dots, \tau-1,
    \end{equation}
    where $\eta_k > 0$ is the local learning rate, and $\nabla \ell(W; \xi)$ is the stochastic gradient over the mini-batch $\xi$.

    \item \textbf{Local Full Gradient Descent (FGD):}
    In the idealized setting, each client uses full-batch gradient descent on its local data. The update rule becomes:
    \begin{equation}
    W_k^{t,s+1} = W_k^{t,s} - \eta_k \left(m_k \odot \nabla F_k(W_k^{t,s})\right), \quad s = 0, \dots, \tau-1,
    \end{equation}
    where $\nabla F_k(W)$ is the full local gradient:
    \[
    \nabla F_k(W) := \frac{1}{n_k} \sum_{i=1}^{n_k} \nabla \ell(W; x_i^{(k)}, y_i^{(k)}).
    \]
\end{enumerate}

After completing local training, the client returns only its local updates: $m_k \odot U_k^t$ in case of SGD or $m_k \odot \bar{U}_k^t$ for FGD, where:

\begin{equation}
U_k^t = W_k^{t,\tau} - W_k^{t,0}  = - \eta_k \sum_{s=0}^{\tau - 1} \nabla \ell(W_k^{t,s}; \xi_k^{(s)})
\end{equation}

and 

\begin{equation}
\bar{U}_k^t = W_k^{t,\tau} - W_k^{t,0}  = - \eta_k \sum_{s=0}^{\tau - 1} \nabla F_k(W_k^{t,s})
\end{equation}

The server aggregates the received partial local updates using coordinate-wise weighted averaging.


We define the normalized weight of client \(k\) as:
\[
c_k := \frac{n_k}{\sum_{j=1}^K n_j}, \quad \text{so that } \sum_{k=1}^K c_k = 1.
\]

Then the global model update at round \(t\) can be written as:
\[
\Delta^t := \left( \sum_{k=1}^K c_k \cdot (m_k \odot U_k^t) \right),
\]



\subsection{Convergence Analysis}

In this subsection, we present the convergence analysis of static partial training schemes, including our proposed method, FedPLT.

We begin by introducing a set of standard assumptions commonly adopted in the federated learning literature to ensure theoretical rigor. Additionally, we propose a new assumption specifically tailored to the partial update setting, which captures the alignment properties between masked and full gradients.

Our proof follows the general structure of the analysis in \cite{li2019convergence}, which we extend to accommodate the static partial update mechanisms used in schemes like FedPLT. We detail each step of the convergence proof, highlighting where our contributions diverge from or generalize existing results.

In the cited paper, virtual aggregation is applied at each local step $s$, whereas in a real setting, aggregation occurs only once every $\tau$ steps. We will be using their proving flow by using this notation and then, when necessary in the analysis, we revert to the per-step formulation by explicitly indexing local iterations $s$.

Thus, we use the simplified update notation:
\[
U_k^t = -\eta^t \nabla \ell(W_k^t; \xi_k), \quad \text{and} \quad \bar{U}_k^t = -\eta^t \nabla F_k(W_k^t),
\]

\subsubsection{Finding the Expected Distance Toward the Optimal Global Model}

We aim to analyze the expected squared distance to the optimal global model \( W^* \) after round \( t+1 \) of federated training with partial updates. Specifically, our goal is to bound the following quantity:
\[
\mathbb{E} \left[ \| W^{t+1} - W^* \|^2 \right],
\]
where the expectation is taken over the randomness in local mini-batch selection. This quantity captures the expected gap between the global model and the optimum across rounds.

\paragraph*{Step 1: Expanding the Distance to the Optimum}

We begin by expanding the global model's parameters using the global update as follow:
\[
W^{t+1} = W^t + \Delta^t,
\]
where the global update is defined as:
\[
\Delta^t :=  \sum_{k=1}^K c_k \left( m_k \odot U_k^t \right),
\]
and each client’s stochastic local update \( U_k^t \) is given by:
\[
U_k^t = -\eta^t \nabla \ell(W_k^t; \xi_k),
\]

thus we have:
\[
\| W^{t+1} - W^* \|^2 = \| W^t + \Delta^t - W^* \|^2.
\]

We then introduce the full (non-stochastic) update:
\[
\bar{\Delta}^t :=\sum_{k=1}^K c_k \left( m_k \odot \bar{U}_k^t \right),
\quad \text{where} \quad
\bar{U}_k^t = -\eta^t \nabla F_k(W_k^t).
\]

by add and subtract it inside the norm:
\[
\| W^{t+1} - W^* \|^2
= \left\| (W^t - W^* + \bar{\Delta}^t) + (\Delta^t - \bar{\Delta}^t) \right\|^2.
\]

Applying the Euclidean norm expansion (a.k.a. the cosine or parallelogram identity), we obtain:
\begin{equation}
\| W^{t+1} - W^* \|^2 = 
\underbrace{\| \Delta^t - \bar{\Delta}^t \|^2}_{A_1}
+ \underbrace{\| W^t - W^* + \bar{\Delta}^t \|^2}_{A_2}
+ \underbrace{2 \left\langle \Delta^t - \bar{\Delta}^t, \; W^t - W^* + \bar{\Delta}^t \right\rangle}_{A_3}.
\end{equation}

\paragraph*{Step 2: Expectation of the Cross-Term \( A_3 \)}

We now analyze the expectation of \( A_3 \). The key idea is that the stochastic error term \( \Delta^t - \bar{\Delta}^t \) arises from the randomness due to mini-batch sampling at round \( t \), while the other term in the inner product is deterministic.

Specifically, we compute:
\[
\mathbb{E}_{\zeta^t}[A_3]
= \mathbb{E}_{\zeta^t} \left[
2 \langle W^t - W^* + \bar{\Delta}^t, \Delta^t - \bar{\Delta}^t \rangle
\right].
\]

Note that:
\begin{itemize}
    \item \( W^t \), \( W^* \), and \( \bar{\Delta}^t \) are deterministic with respect to the mini-batch sampling at round \( t \),
    \item \( \Delta^t - \bar{\Delta}^t \) is a zero-mean random variable:
    \[
    \mathbb{E}_{\zeta^t}[\Delta^t - \bar{\Delta}^t] = 0.
    \]
\end{itemize}

Therefore, the expectation of the inner product vanishes:
\[
\mathbb{E}_{\zeta^t} \left[
\langle W^t - W^* + \bar{\Delta}^t,\; \Delta^t - \bar{\Delta}^t \rangle
\right] = 0,
\]
which implies that:
\[
\mathbb{E}[A_3] = 0.
\]

The term \( A_3 \) contributes nothing in expectation. Hence, the expected squared distance simplifies to:
\[
\mathbb{E}\left[\|W^{t+1} - W^*\|^2\right]
= \mathbb{E}\left[A_1\right]
+ \mathbb{E}\left[A_2\right].
\]

\paragraph*{Step 3: Bounding the Stochastic Gradient Variance Term (\(A_1\))}

We now provide a concise upper bound for the variance of the global update due to stochastic gradient noise. Recall that the actual update is:
\[
\Delta^t := W^{t+1} - W^t =  \sum_{k=1}^K c_k (m_k \odot U_k^t),
\]
and the full gradient update is:
\[
\bar{\Delta}^t :=  \sum_{k=1}^K c_k (m_k \odot \bar{U}_k^t),
\]
where \(U_k^t\) and \(\bar{U}_k^t\) are the stochastic and full local updates, respectively.

Letting \(\delta_k^t := U_k^t - \bar{U}_k^t\), the update difference becomes:
\[
\Delta^t - \bar{\Delta}^t = \sum_{k=1}^K c_k (m_k \odot \delta_k^t).
\]

We now bound the squared norm:
\[
\|\Delta^t - \bar{\Delta}^t\|^2
= \left\|  \sum_{k=1}^K c_k (m_k \odot \delta_k^t) \right\|^2.
\]


We take the expectation of the squared norm:
\[
\mathbb{E}\left[ \|\Delta^t - \bar{\Delta}^t\|^2 \right]
=   \mathbb{E}\left[ \left\| \sum_{k=1}^K c_k (m_k \odot \delta_k^t) \right\|^2 \right]
\leq  \cdot \sum_{k=1}^K c_k^2 \mathbb{E}\left[\| m_k \odot \delta_k^t \|^2\right].
\]

Now applying the masking inequality $\| m_k \odot \delta_k^t \| \leq \alpha_{r_k} \| \delta_k^t \|$, we get:

\[
\mathbb{E}\left[ \|\Delta^t - \bar{\Delta}^t\|^2 \right]
\leq  \cdot \sum_{k=1}^K c_k^2 \alpha_{r_k}^2 \cdot \mathbb{E}[\| \delta_k^t \|^2].
\]

Finally, applying Assumption 3 (bounded variance):
\[
\mathbb{E}[\| \delta_k^t \|^2] = \mathbb{E}[\| U_k^t - \bar{U}_k^t \|^2] \leq (\eta^t)^2 \sigma^2,
\]

We conclude:
\[
\mathbb{E}[\|\Delta^t - \bar{\Delta}^t\|^2]
\leq (\eta^t)^2 \sigma^2 \cdot  \sum_{k=1}^K c_k^2 \alpha_{r_k}^2.
\]

\paragraph*{Step 4: Decomposition of \(A_2\) and Finding its expectation}
Recall the second term in our main inequality:
\[
A_2 := \|W^t - W^* + \bar{\Delta}^t\|^2,
\]

We expand this term as:
\[
A_2 = 
\|W^t - W^*\|^2
+ \underbrace{\|\bar{\Delta}^t\|^2}_{A_{2.1}} 
+ \underbrace{2 \langle W^t - W^*, \bar{\Delta}^t \rangle}_{A_{2.2}}.
\]

We define the following components for further analysis:
\begin{align*}
A_{2.1} &:= \|\bar{\Delta}^t\|^2, \\
A_{2.2} &:= 2 \langle W^t - W^*, \bar{\Delta}^t \rangle.
\end{align*}

\paragraph*{Step 4.1: Bounding \(A_{2.1}\) (Ideal Update Norm)}

Recall the definition of the ideal global update:
\[
\bar{\Delta}^t :=\sum_{k=1}^K c_k \left( m_k \odot \bar{U}_k^t \right),
\]
where \( \bar{U}_k^t = -\eta^t \nabla F_k(W_k^t) \) is the full local gradient descent.

Using the masking inequality \( \|m_k \odot v\| \leq \alpha_{r_k} \|v\| \), and applying the convexity of the squared norm \( \| \cdot \|^2 \), we get:
\[
\|\bar{\Delta}^t\|^2
= \left\|  \sum_{k=1}^K c_k (m_k \odot \bar{U}_k^t) \right\|^2
\leq  \left\| \sum_{k=1}^K c_k (m_k \odot \bar{U}_k^t) \right\|^2
\leq  \sum_{k=1}^K c_k \|m_k \odot \bar{U}_k^t\|^2.
\]

Applying the masking bound:
\[
\|m_k \odot \bar{U}_k^t\| \leq \alpha_{r_k} \|\bar{U}_k^t\|,
\quad \text{so} \quad
\|m_k \odot \bar{U}_k^t\|^2 \leq \alpha_{r_k}^2 \|\bar{U}_k^t\|^2,
\]
we obtain:
\[
\|\bar{\Delta}^t\|^2 \leq  \sum_{k=1}^K c_k \alpha_{r_k}^2 \|\bar{U}_k^t\|^2.
\]

We now bound \( \|\bar{U}_k^t\|^2 \):
\[
\|\bar{U}_k^t\|^2 = (\eta^t)^2 \|\nabla F_k(W_k^t)\|^2.
\]

Using the \(L\)-smoothness of \(F_k\) and the local optimal value \(F_k^* := \min_W F_k(W)\), we apply:
\[
\|\nabla F_k(W_k^t)\|^2 \leq 2L \left( F_k(W_k^t) - F_k^* \right).
\]

Hence,
\[
\|\bar{U}_k^t\|^2 \leq 2 (\eta^t)^2 L \left( F_k(W_k^t) - F_k^* \right).
\]

Substituting into the bound on \( \|\bar{\Delta}^t\|^2 \), we obtain:
\[
A_{2.1} = \|\bar{\Delta}^t\|^2
\leq 2 L (\eta^t)^2  \sum_{k=1}^K \alpha_{r_k}^2\ c_k \left( F_k(W_k^t) - F_k^* \right).
\]

\paragraph*{Step 4.2: Bounding the Inner Product Term \(A_{2.2}\)}

We now focus on bounding the term:
\[
A_{2.2} := 2 \left\langle W^t - W^*, \bar{\Delta}^t \right\rangle.
\]

We start by expanding \(\bar{\Delta}^t\) using the aggregation rule:
\[
\bar{\Delta}^t :=  \sum_{k=1}^K c_k (m_k \odot \bar{U}_k^t), \quad \text{where} \quad \bar{U}_k^t := -\eta \nabla F_k(W_k^t),
\]

We decompose \(W^t - W^*\) as:
\[
W^t - W^* = (W^t - W_k^t) + (W_k^t - W^*).
\]

Substituting this into the inner product and splitting the sum:
\begin{align*}
A_{2.2} &= 2 \left\langle W^t - W^*,\;  \sum_{k=1}^K c_k (m_k \odot \bar{U}_k^t) \right\rangle \\
&= 2 \sum_{k=1}^K \left[
\left\langle W^t - W_k^t,\; \left(c_k m_k \odot \bar{U}_k^t \right) \right\rangle
+
\left\langle W_k^t - W^*,\; c_k m_k \odot \bar{U}_k^t \right\rangle
\right].
\end{align*}

\subparagraph{Part 1:}

Applying the Cauchy–Schwarz inequality, we get:
\begin{align*}
\left\langle W^t - W_k^t,\; \left(c_k m_k \odot \bar{U}_k^t\right) \right\rangle
&\leq \|\left(c_k m_k \odot \bar{U}_k^t\right)\| \cdot \|W^t - W_k^t\| \\
&\leq  \cdot c_k \cdot \|m_k \odot \bar{U}_k^t\| \cdot \|W^t - W_k^t\|,\\
&= \cdot \eta^t \cdot c_k \cdot \|m_k \odot \nabla F_k(W_k^t)\| \cdot \|W^t - W_k^t\|,
\end{align*}

Then, applying the masking bound \(\|m_k \odot v\| \leq \alpha_{r_k} \|v\|\), where \( \alpha_{r_k} := \min(1, \sqrt{d r_k}) \), we obtain:
\[
\left\langle W^t - W_k^t,\; \left(c_k m_k \odot \bar{U}_k^t\right) \right\rangle
\leq   \eta^t \cdot c_k \cdot \alpha_{r_k} \cdot \|\nabla F_k(W_k^t)\| \cdot \|W^t - W_k^t\|.
\]

Now applying the AM--GM inequality with parameter \( \eta^t > 0 \), we get:
\[
\|W^t - W_k^t\| \cdot \left( \cdot \alpha_{r_k} \cdot \|\nabla F_k(W_k^t)\|\right)
\leq \frac{1}{\eta^t}\ \|W^t - W_k^t\|^2 
+ \eta^t\ (\ \alpha_{r_k})^2\ \|\nabla F_k(W_k^t)\|^2.
\]

Multiplying both sides by the positive scalar \( c_k \cdot \eta^t \), we obtain:
\[
\left\langle W^t - W_k^t,\;  \left(c_k m_k \odot \bar{U}_k^t\right) \right\rangle
\leq c_k \left( 
\|W^t - W_k^t\|^2 
+ (\eta^t\ \ \alpha_{r_k})^2\ \|\nabla F_k(W_k^t)\|^2 
\right).
\]

Recall the \(L\)-smoothness of \( F_k \), which implies
\[
\|\nabla F_k(W_k^t)\|^2 \leq 2L \left( F_k(W_k^t) - F_k^* \right),
\]

Substituting it into the previous inequality:
\[
\left\langle W^t - W_k^t,\;\left(c_k m_k \odot \bar{U}_k^t\right) \right\rangle
\leq c_k \left( 
\|W^t - W_k^t\|^2 
+ (\eta^t\  \ \alpha_{r_k})^2\ L \left( F_k(W_k^t) - F_k^* \right)
\right).
\]

\subparagraph{Part 2:}

We aim to bound the inner product:
\[
\left\langle W_k^t - W^*,\;   c_k m_k \odot \bar{U}_k^t  \right\rangle,
\quad \text{where } \bar{U}_k^t := -\eta^t \nabla F_k(W_k^t).
\]

Expanding, we obtain:
\[
\left\langle W_k^t - W^*,\; \left( c_k m_k \odot \bar{U}_k^t \right) \right\rangle
= -\eta^t \left\langle W_k^t - W^*,\; c_k\;  m_k \odot \nabla F_k(W_k^t) \right\rangle.
\]


Then we apply the client-wise masked alignment ratio:
\[
\rho_k^t := \frac{
\left\langle m_k \odot \nabla F_k(W_k^t),\; W_k^t - W^* \right\rangle
}{
\left\langle \nabla F_k(W_k^t),\; W_k^t - W^* \right\rangle
},
\]
which measures how well the masked gradient aligns with the true gradient. Under Assumption 5, we assume the alignment is uniformly lower bounded across all rounds and clients:
\[
\underline{\rho} := \min_{k \in [K],\; t \in \mathbb{N}} \rho_k^t > 0.
\]

Using the strong convexity of \( F_k \), we have:
\[
\left\langle \nabla F_k(W_k^t),\; W_k^t - W^* \right\rangle
\geq F_k(W_k^t) - F_k(W^*) + \frac{\mu}{2} \|W_k^t - W^*\|^2.
\]

Combining the above, we obtain:
\begin{align*}
\left\langle W_k^t - W^*,\; (c_k m_k \odot \bar{U}_k^t) \right\rangle
&= -\eta^t c_k \left\langle W_k^t - W^*,\;  m_k \odot \nabla F_k(W_k^t) \right\rangle \\
&\leq -\eta^t  c_k \left\langle W_k^t - W^*,\; m_k \odot \nabla F_k(W_k^t) \right\rangle \\
&= -\eta^t  \rho_k^t c_k \left\langle \nabla F_k(W_k^t),\; W_k^t - W^* \right\rangle \\
&\leq -\eta^t  \rho_k^t c_k \left[ F_k(W_k^t) - F_k(W^*) + \frac{\mu}{2} \|W_k^t - W^*\|^2 \right].
\end{align*}

Finally, using \( \rho_k^t \geq \underline{\rho} \), we get the uniform lower bound:
\[
\left\langle W_k^t - W^*,\; (c_k m_k \odot \bar{U}_k^t) \right\rangle
\leq -\eta^t\ \ \underline{\rho}\ c_k
\left[ F_k(W_k^t) - F_k(W^*) + \frac{\mu}{2} \|W_k^t - W^*\|^2 \right].
\]

\subparagraph{Part 3: Final Bound on \(A_{2.2}\)}

Combining the two parts, we obtain:
\begin{align*}
A_{2.2} = 2 \left\langle W^t - W^*,\ \bar{\Delta}^t \right\rangle
\leq \sum_{k=1}^K \bigg[
& c_k \|W^t - W_k^t\|^2 \\
& + 2 (\eta^t\ \alpha_{r_k})^2\ L \left(F_k(W_k^t) - F_k^*\right) \\
& - 2 \eta^t\  \underline{\rho}\ c_k
\left( F_k(W_k^t) - F_k(W^*) + \frac{\mu}{2} \|W_k^t - W^*\|^2 \right)
\bigg],
\end{align*}

\paragraph*{Step 4.3: Grouping and Rearranging the Terms of \(A_2\)}

Recall the decomposition:
\[
A_2 := \|W^t - W^* + \bar{\Delta}^t\|^2
= \underbrace{\|W^t - W^*\|^2}_{\text{Initial distance}} 
+ \underbrace{\|\bar{\Delta}^t\|^2}_{A_{2.1}} 
+ \underbrace{2 \langle W^t - W^*, \bar{\Delta}^t \rangle}_{A_{2.2}}.
\]

Substituting the bounds from previous steps, we obtain:
\begin{align*}
A_2 \leq 
&\underbrace{\|W^t - W^*\|^2}_{\text{Initial distance}} \\
&+ \underbrace{
\sum_{k=1}^K 
\left[
2 (\eta^t)^2  \alpha_{r_k}^2 L\ c_k
(F_k(W_k^t) - F_k^*) 
\right]
}_{\text{Ideal update norm } (A_{2.1})} \\
&+ \underbrace{
\sum_{k=1}^K 
\left[
c_k \|W^t - W_k^t\|^2 
- \mu \eta^t\  \underline{\rho}\ c_k \|W_k^t - W^*\|^2
\right]
}_{\text{Model difference terms from } A_{2.2}} \\
&+ \underbrace{
\sum_{k=1}^K 
\left[
2 (\eta^t\ \ \alpha_{r_k})^2\ L\ c_k (F_k(W_k^t) - F_k^*) 
- 2 \eta^t \ \underline{\rho}\ c_k\ (F_k(W_k^t) - F_k(W^*))
\right]
}_{\text{Loss terms from } A_{2.2}}.
\end{align*}

Combining the result and regrouping all components, we obtain:
\begin{align*}
A_2 \leq \|W^t - W^*\|^2 
+ \sum_{k=1}^K \bigg[
& c_k \|W^t - W_k^t\|^2
- \mu \eta^t\ \ \underline{\rho}\ c_k \|W_k^t - W^*\|^2 \\
& + 4 (\eta^t\  \alpha_{r_k})^2\ L\ c_k (F_k(W_k^t) - F_k^*) 
- 2 \eta^t\  \underline{\rho}\ c_k(F_k(W_k^t) - F_k(W^*))
\bigg].
\end{align*}

We recall that the global model at round \(t\) can be written as a weighted average of local models:
\[
W^t = \sum_{k=1}^K c_k W_k^t, \quad \text{with} \quad c_k := \frac{n_k}{\sum_{j=1}^K n_j}.
\]
Using the convexity of the squared norm, we apply:
\[
\|W^t - W^*\|^2 \leq \sum_{k=1}^K c_k \|W_k^t - W^*\|^2.
\]

Substituting into the earlier bound, we obtain:
\begin{align*}
A_2 \leq
& (1 - \mu \eta^t\  \underline{\rho}) \cdot \|W^t - W^*\|^2 \\
& + \sum_{k=1}^K c_k \|W^t - W_k^t\|^2 \\
& + \underbrace{\sum_{k=1}^K \bigg[
4 (\eta^t\  \alpha_{r_k})^2 L\ c_k (F_k(W_k^t) - F_k^*)
- 2 \eta^t\  \underline{\rho}\ c_k (F_k(W_k^t) - F_k(W^*))
\bigg]}_{A_{2.4}}.
\end{align*}

\paragraph*{Step 4.4: Bounding \(A_{2.4}\)}

We define \(A_{2.4}\) to capture all the loss-related contributions arising in the expansion of \(A_2\):
\begin{align*}
A_{2.4} := 
\sum_{k = 1}^K \bigg[
4 (\eta^t\ \ \alpha_{r_k})^2 L \cdot c_k (F_k(W_k^t) - F_k^*)
- 2 \eta^t\  \underline{\rho} \cdot c_k (F_k(W_k^t) - F_k(W^*))
\bigg].
\end{align*}

We regroup the above term by adding and subtracting \(F_k^*\) in the expression \((F_k(W_k^t) - F_k(W^*))\):
\begin{align*}
A_{2.4} := 
\sum_{k = 1}^K \bigg[
\underbrace{\left(4 (\eta^t \ \alpha_{r_k})^2 L
- 2 \eta^t \ \underline{\rho} \right)}_{:=\ -\gamma_k^t}
\cdot c_k (F_k(W_k^t) - F_k^*) 
+ \underbrace{2 \eta^t \ \underline{\rho}}_{:=\ \beta^t} \cdot c_k (F_k(W^*) - F_k^*)
\bigg].
\end{align*}

We insert and subtract \(F^*\) inside both gap terms accordingly:
\begin{align*}
A_{2.4} = 
\sum_{k = 1}^K \bigg[
&- \gamma_k^t \cdot c_k \left( F_k(W_k^t) - F^* + F^* - F_k^* \right)
+ \beta^t \cdot c_k \left( F_k(W^*) - F^* + F^* - F_k^* \right)
\bigg].
\end{align*}

We regroup and simplify to highlight the three key components:
\begin{align*}
A_{2.4} =
\sum_{k = 1}^K \bigg[
& - \gamma_k^t \cdot c_k \left( F_k(W_k^t) - F^* \right) \\
& + (\beta^t - \gamma_k^t) \cdot c_k \left( F^* - F_k^* \right) \\
& + \beta^t \cdot c_k \left( F_k(W^*) - F^* \right)
\bigg].
\end{align*}

Noting that \( F^* = \sum_{k=1}^K c_k F_k(W^*) \), this implies:
\[
\sum_{k=1}^K c_k (F_k(W^*) - F^*) = 0.
\]

We thus eliminate the final term and rewrite the expression as:
\begin{align*}
A_{2.4} =
\sum_{k = 1}^K \bigg[
- \gamma_k^t \cdot c_k \left( F_k(W_k^t) - F^* \right)
+ (\beta^t - \gamma_k^t) \cdot c_k \left( F^* - F_k^* \right)
\bigg].
\end{align*}

\subparagraph{Bounding \( (F_k(W_k^t) - F^*) \) gap term}

By adding and subtracting \( F_k(W^t) \), we write:
\begin{align*}
F_k(W_k^t) - F_k^* 
&= \big(F_k(W_k^t) - F_k(W^t)\big) + \big(F_k(W^t) - F^*\big).
\end{align*}

Using the convexity of \( F_k \), we have:
\begin{align*}
F_k(W_k^t) - F_k(W^t)
&\geq \left\langle \nabla F_k(W^t),\ W_k^t - W^t \right\rangle.
\end{align*}

Applying the AM--GM inequality with parameter \( \eta^t \), we get:
\begin{align*}
\left\langle \nabla F_k(W^t),\ W_k^t - W^t \right\rangle
&\geq -\frac{\eta^t}{2} \|\nabla F_k(W^t)\|^2 
- \frac{1}{2\eta^t} \|W_k^t - W^t\|^2.
\end{align*}

Using the \(L\)-smoothness of \(F_k\):
\[
\|\nabla F_k(W^t)\|^2 \leq 2L \left( F_k(W^t) - F_k^* \right),
\]

we obtain the following bound:
\begin{align*}
F_k(W_k^t) - F^* 
&\geq F_k(W^t) - F^* - \eta^t L \left(F_k(W^t) - F_k^* \right)
- \frac{1}{2\eta^t} \|W_k^t - W^t\|^2.
\end{align*}

Now we decompose the bound by adding and subtracting \( F^* \) in $(F_k(W^t) - F_k^*)$:
\begin{align*}
F_k(W_k^t) - F_k^* 
\geq\ 
& \left(1 - \eta^t L \right)(F_k(W^t) - F^*) \\
& - \eta^t L (F^* - F_k^*) 
- \frac{1}{2\eta^t} \|W_k^t - W^t\|^2.
\end{align*}

We substitute the lower bound on \( F_k(W_k^t) - F^* \) into the original upper bound for \( A_{2.4} \). This gives:
\begin{align*}
A_{2.4} \leq
\sum_{k = 1}^K
\bigg[
& \gamma_k^t \cdot c_k\ \bigg(
(\eta^t L - 1)(F_k(W^t) - F^*) 
+ \eta^t L (F^* - F_k^*) 
+ \frac{1}{2\eta^t} \|W_k^t - W^t\|^2 \bigg) \\
& + (\beta^t - \gamma_k^t) \cdot c_k (F^* - F_k^*)
\bigg].
\end{align*}

We now regroup the terms in the corrected upper bound of \( A_{2.4} \):
\begin{align*}
A_{2.4} \leq
\sum_{k = 1}^K \bigg[ 
& \gamma_k^t (\eta^t L - 1) \cdot c_k\ (F_k(W^t) - F^*) \\
& + \left[\gamma_k^t (\eta^t L - 1) + \beta^t \right] \cdot c_k\ (F^* - F_k^*) \\
& + \gamma_k^t \cdot \frac{1}{2\eta^t} \cdot c_k\ \|W_k^t - W^t\|^2
\bigg].
\end{align*}

We now simplify the bound on \( A_{2.4} \) under the following conditions:
\begin{itemize}
    \item \( \eta^t < \frac{1}{L} \) \( \Rightarrow (\eta^t L - 1) < 0 \),
    \item \( \gamma_k^t > 0 \), so \( \gamma_k^t (\eta^t L - 1) < 0 \),
    \item \( \frac{\gamma_k^t}{2 \eta^t} < 1 \), so \( \gamma_k^t \cdot \frac{1}{2\eta^t} \cdot c_k\ \|W_k^t - W^t\|^2 < c_k\ \|W_k^t - W^t\|^2 \),
    \item The combined coefficient: 
    \[
    \gamma_k^t (\eta^t L - 1) + \beta^t 
    = 2 (\eta^t)^2 L \cdot \left[2  \alpha_{r_k}^2 (1 - \eta^t L) +  \underline{\rho} \right] 
    \leq 2 (\eta^t)^2 L \nu,
    \]
    where \( \nu := \max_{t} \left[2 \alpha_{r_k}^2 (1 - \eta^t L) + \underline{\rho} \right] > 0 \).

\end{itemize}

Applying these observations to the previous bound:
\begin{align*}
A_{2.4} \leq
\sum_{k = 1}^K \bigg[ 
& 2 (\eta^t)^2 L \nu \cdot c_k (F^* - F_k^*) \\
& + c_k \|W_k^t - W^t\|^2
\bigg].
\end{align*}

We define the aggregate heterogeneity gap:
\[
\Lambda := \sum_{k=1}^K c_k (F^* - F_k^*),
\]

thus,
Applying these observations to the previous bound:
\begin{align*}
A_{2.4} \leq
\sum_{k = 1}^K \bigg[ 
& 2 (\eta^t)^2 L \nu \cdot c_k\ \Lambda \\
& + c_k \|W_k^t - W^t\|^2
\bigg].
\end{align*}

\paragraph*{Step 4.5: Regrouping the Final Bound on \(A_2\)}

We now regroup and simplify the final upper bound on \(A_2\), using the decompositions derived in previous steps.

Recall that:

\begin{align*}
A_2 \leq
& (1 - \mu \eta^t \ \underline{\rho}) \cdot \|W^t - W^*\|^2 \\
& + \sum_{k=1}^K c_k \|W^t - W_k^t\|^2 \\
& + \underbrace{\sum_{k=1}^K \bigg[
4 (\eta^t \ \alpha_{r_k})^2 L\ c_k (F_k(W_k^t) - F_k^*)
- 2 \eta^t \ \underline{\rho}\ c_k (F_k(W_k^t) - F_k(W^*))
\bigg]}_{A_{2.4}}.
\end{align*}

We substitute the simplified bound for the loss-related component $A_{2.4}$ derived in step 6.4, yielding:
\begin{align*}
A_2 \leq
& (1 - \mu \eta^t \ \underline{\rho}) \|W^t - W^*\|^2 \\
& + \sum_{k=1}^K 2 c_k \|W_k^t - W^t\|^2 \\
& + 2 (\eta^t)^2 L \nu \cdot \Lambda.
\end{align*}

\paragraph*{Step 4.6: Taking the Expectation of \(A_2\)}

To compute \( \mathbb{E}[A_2] \), we note that:
\[
\mathbb{E}[A_2] \leq
(1 - \mu \eta^t \ \underline{\rho}) \|W^t - W^*\|^2 
+ 2\ \mathbb{E} \left[ \sum_{k=1}^K c_k \|W_k^t - W^t\|^2 \right]
+ 2 (\eta^t)^2 L \nu \cdot \Lambda.
\]

We now bound the expected local divergence term:
\[
\mathbb{E} \left[ \sum_{k=1}^K c_k \|W_k^t - W^t\|^2 \right].
\]

Recall that:
\begin{itemize}
    \item \( W_k^t = W_k^{t,\tau} \) is the local model of client \(k\) after \(\tau\) local steps,
    \item \( W^t = W_k^{t,0} \) is the global model at round \(t\), identical to the local initialization,
    \item Local updates are computed with coordinate-wise masks: 
    \[
    W_k^{t,s+1} = W_k^{t,s} - \eta^t \left( m_k \odot \nabla \ell(W_k^{t,s}; \xi_k^{(s)}) \right).
    \]
\end{itemize}

By unrolling the update over \(\tau\) local steps, we obtain:
\[
W_k^t - W^t = W_k^{t,\tau} - W_k^{t,0} = - \eta^t \sum_{s=0}^{\tau - 1} \left( m_k \odot \nabla \ell(W_k^{t,s}; \xi_k^{(s)}) \right).
\]

Taking the squared norm and applying Jensen's inequality:
\begin{align*}
\mathbb{E} \left[ \|W_k^t - W^t\|^2 \right]
&= (\eta^t)^2 \cdot \mathbb{E} \left[ \left\| \sum_{s=0}^{\tau - 1} m_k \odot \nabla \ell(W_k^{t,s}; \xi_k^{(s)}) \right\|^2 \right] \\
&\leq (\eta^t)^2 \cdot \tau \sum_{s=0}^{\tau - 1} \mathbb{E} \left[ \| m_k \odot \nabla \ell(W_k^{t,s}; \xi_k^{(s)}) \|^2 \right].
\end{align*}

Applying the masking bound \( \| m_k \odot v \| \leq \alpha_{r_k} \cdot \|v\| \) and the variance assumption:
\[
\mathbb{E} \left[ \| \nabla \ell(W_k^{t,s}; \xi_k^{(s)}) \|^2 \right] \leq G^2 + \sigma^2,
\]
We obtain:
\[
\mathbb{E} \left[ \|W_k^t - W^t\|^2 \right] \leq (\eta^t)^2 \cdot \tau^2 \cdot \alpha_{r_k}^2 \cdot (G^2 + \sigma^2).
\]

Finally, summing over clients with weights \( c_k \), we get:
\[
\mathbb{E} \left[ \sum_{k=1}^K c_k \|W_k^t - W^t\|^2 \right]
\leq (\eta^t)^2 \cdot \tau^2 \cdot (G^2 + \sigma^2) \sum_{k=1}^K c_k \alpha_{r_k}^2.
\]

We denote this term by:
\[
\Gamma := \sum_{k=1}^K c_k \alpha_{r_k}^2,
\]
and thus conclude:
\[
\mathbb{E} \left[ \sum_{k=1}^K c_k \|W_k^t - W^t\|^2 \right] \leq (\eta^t)^2 \cdot \tau^2 \cdot (G^2 + \sigma^2) \cdot \Gamma.
\]

Combining this with the earlier result for \( A_2 \), we obtain the following upper bound in expectation:
\[
\mathbb{E}[A_2] \leq (1 - \mu \eta^t \ \underline{\rho}) \cdot \|W^t - W^*\|^2
+ 2\ (\eta^t)^2 \cdot \tau^2 \cdot (G^2 + \sigma^2) \cdot \Gamma
+ 2\ (\eta^t)^2 L \nu \cdot \Lambda,
\]

\paragraph*{Step 5: Recursive bound over the expected optimal gap}

We recall the Euclidean expansion:
\[
\| W^{t+1} - W^* \|^2 = 
\underbrace{\| \Delta^t - \bar{\Delta}^t \|^2}_{A_1}
+ \underbrace{\| W^t - W^* + \bar{\Delta}^t \|^2}_{A_2}
+ \underbrace{2 \left\langle \Delta^t - \bar{\Delta}^t,\; W^t - W^* + \bar{\Delta}^t \right\rangle}_{A_3}.
\]

Taking expectations over the stochasticity at round \(t\), and applying the derived bounds, we get:
\begin{align*}
\mathbb{E} \left[ \| W^{t+1} - W^* \|^2 \right]
&= \mathbb{E}[A_1] + \mathbb{E}[A_2] + \mathbb{E}[A_3] \\
&\leq (\eta^t)^2 \sigma^2 \cdot  \Gamma \\
&\quad + (1 - \mu \eta^t \ \underline{\rho}) \cdot \mathbb{E}\left[\|W^t - W^*\|^2\right] \\
&\quad + 2\ (\eta^t)^2 \tau^2 (G^2 + \sigma^2) \cdot \Gamma \\
&\quad + 2 (\eta^t)^2 L \nu \cdot \Lambda.
\end{align*}

We define the sequence:
\[
D^t := \mathbb{E} \left[ \|W^t - W^*\|^2 \right],
\]
and obtain the recursive inequality:
\[
D^{t+1} \leq \left(1 - z\ \eta^t \right) \cdot D^t + (\eta^t)^2\ B,
\]
where:
\[
z := \mu\  \underline{\rho},
\quad
B := 2\ \tau^2 (G^2 + \sigma^2)\ \Gamma + 2 L \nu\ \Lambda
+ \sigma^2 \Gamma.
\]

\subsubsection*{Convergence Analysis of the Recursive Inequality}

We recall the recursive inequality derived in Step 5:
\[
D^{t+1} \leq (1 - z \eta^t) \cdot D^t + (\eta^t)^2 B,
\]

We now analyze the convergence behavior under two scenarios: (1) constant step size, and (2) decaying step size.

\paragraph*{Case 1: Constant Learning Rate}

Assume a constant step size \( \eta^t = \eta \in (0, \frac{1}{z}) \). Then the recursion becomes:
\[
D^{t+1} \leq (1 - z\eta)\, D^t + \eta^2 B.
\]

Let \( \omega := 1 - z\eta < 1 \). Then by recursion:
\[
D^{t+1} \leq \omega^{t+1} D^0 + \eta^2 B \sum_{j=0}^{t} \omega^j = \omega^{t+1} D^0 + \eta^2 B \cdot \frac{1 - \omega^{t+1}}{1 - \omega}.
\]

Since \( \frac{1 - \omega^{t+1}}{1 - \omega} \leq \frac{1}{1 - \omega} = \frac{1}{z\eta} \), we obtain the final bound:
\[
D^{t+1} \leq \omega^{t+1} D^0 + \frac{\eta B}{z}.
\]

\subparagraph{Interpretation:} The first term decays geometrically, while the second term is a residual error that depends on \( \eta \). Therefore, the convergence is fast, but with a gap of size: $\frac{\eta B}{z}$ around the optimal point \( W^* \).

\subparagraph{Implication:} Choosing a smaller \( \eta \) reduces the residual error, but slows down convergence. This is a speed-error trade-off: a smaller learning rate implies lower asymptotic error but slower descent.

\paragraph*{Case 2: Decaying Learning Rate}

We now analyze the convergence behavior under a decaying learning rate of the form:
\[
\eta^t = \frac{1}{z(t+1)},
\]

Recall the recursive inequality:
\[
D^{t+1} \leq (1 - z \eta^t) D^t + (\eta^t)^2 B.
\]
Substituting the decaying step size into the recursion, we get:
\[
D^{t+1} \leq \left(1 - \frac{1}{t+1} \right) D^t + \frac{B}{z^2(t+1)^2}
= \frac{t}{t+1} D^t + \frac{B}{z^2(t+1)^2}.
\]

We now prove by mathematical induction that for all \( t \geq 0 \), there exists a constant \( C \geq \frac{B}{z^2} \) such that:
\[
D^t \leq \frac{C}{t+1}.
\]

\subparagraph{Base Case:} At \( t = 0 \),
\[
D^1 \leq D^0 + \frac{B}{z^2} \leq \frac{C}{0 + 1},
\]
provided that:
\[
C \geq D^0 + \frac{B}{z^2} \geq \frac{B}{z^2}.
\]

\subparagraph{Inductive Step:} Assume the hypothesis holds at step \( t \), i.e.,
\[
D^t \leq \frac{C}{t+1}.
\]
Then:
\begin{align*}
D^{t+1}
&\leq \frac{t}{t+1} \cdot \frac{C}{t+1} + \frac{B}{z^2(t+1)^2} \\
&= \frac{C t + \frac{B}{z^2}}{(t+1)^2}.
\end{align*}

We now compare this to \( \frac{C}{t+2} \) by defining the function:
\[
f(t) := \frac{C t + \frac{B}{z^2}}{(t+1)^2} - \frac{C}{t+2}.
\]
It can be proved through typical methods of calculus that \( f(t) \leq 0 \) if $C \geq \frac{B}{z^2}$.

Thus, for any \( C \geq \frac{B}{z^2} \), the inequality \( D^{t+1} \leq \frac{C}{t+2} \) holds. This completes the inductive step.

\subparagraph{Conclusion:} By induction, for all \( t \geq 0 \), we have:
\[
D^t \leq \frac{C}{t+1}, \quad \text{for some constant } C \geq \frac{B}{z^2}.
\]

Therefore, the expected squared distance to the optimum decays at the rate:
\[
\boxed{D^t = \mathcal{O}\left( \frac{1}{t} \right)}.
\]

\subparagraph{Interpretation:} This gives asymptotic convergence to the optimum \( W^* \), but at a sublinear rate. The error shrinks slowly over time, especially during early rounds.

\paragraph*{Summary and Comparison}

\begin{itemize}
    \item With a constant step size, we converge fast to a small neighborhood around \( W^* \), with asymptotic error \( \frac{\eta B}{z} \).
    \item With a decaying step size, we converge exactly to \( W^* \), but at a slower \( \mathcal{O}(1/t) \) rate.
\end{itemize}

\subparagraph{Practical Guideline:} In practice, we can use a hybrid strategy, starting with a large constant step size to ensure fast initial progress, then decaying it slowly to improve final accuracy.

\newpage
\section{Static Partial Parameter Training with OCS}
\label{app:OCS-PLT}

\subsection{Problem Formulation}

We now consider the optimization problem:
\[
\begin{aligned}
\min_{\{p_k^t\}} \quad & \mathbb{E}_{A^t} \left[ \left\| \sum_{k \in A^t} \frac{n_k}{p_k^t}\, U_k^t - \sum_{k=1}^{K} n_k\, U_k^t \right\|^2 \right] \\
\text{such that} \quad & 0 \leq p_k^t \leq 1, \quad \forall\, k=1,\dots,N, \\
& \sum_{k=1}^{K} r_k\, p_k^t = m.
\end{aligned}
\]
where:
\begin{itemize}
    \item \(p_k^t\) is the probability of sampling client \(k\) at round \(t\),
    \item \(A^t\) is the set of clients selected at round \(t\),
    \item \(n_k\) is the number of data samples for client \(k\),
    \item \(U_k^t\) is the gradient transferred from client \(k\) ar round \(t\),
    \item \(r_k\) is a ratio in \([0,1]\) representing the training proportion of client's \(k\) model.
\end{itemize}

\subsection{Problem Simplification}

Observe that
\[
\mathbb{E}_{A^t}\left[ \sum_{k \in A^t} \frac{n_k}{p_k^t}\, U_k^t \right] = \sum_{k=1}^{K} n_k\, U_k^t.
\]
Thus, the objective function
\[
\mathbb{E}_{A^t}\left[ \left\| \sum_{k \in A^t} \frac{n_k}{p_k^t}\, U_k^t - \sum_{k=1}^{K} n_k\, U_k^t \right\|^2 \right]
\]
is the variance of the first term. By the variance identity,
\[
\mathbb{E}_{A^t}\left[ \left\| \sum_{k \in A^t} \frac{n_k}{p_k^t}\, U_k^t - \sum_{k=1}^{K} n_k\, U_k^t \right\|^2 \right] 
= \mathbb{E}_{A^t}\left[ \left\| \sum_{k \in A^t} \frac{n_k}{p_k^t}\, U_k^t \right\|^2 \right] - \left\|\sum_{k=1}^{K} n_k\, U_k^t\right\|^2.
\]

By expanding the squared norms into inner products and distributing the transpose over the sum, we obtain
\[
\left\| \sum_{k \in A^t} \frac{n_k}{p_k^t}\, U_k^t \right\|^2 
= \left( \sum_{k \in A^t} \frac{n_k}{p_k^t}\, U_k^t \right)^T \left( \sum_{j \in A^t} \frac{n_j}{p_j^t}\, U_j^t \right)
= \left( \sum_{k \in A^t} \left( \frac{n_k}{p_k^t}\, U_k^t \right)^T \right) \left( \sum_{j \in A^t} \frac{n_j}{p_j^t}\, U_j^t \right).
\]
Thus, the squared norm can be written as
\[
\left\| \sum_{k \in A^t} \frac{n_k}{p_k^t}\, U_k^t \right\|^2 
= \sum_{k \in A^t} \sum_{j \in A^t} \left( \frac{n_k}{p_k^t}\, U_k^t \right)^T \left( \frac{n_j}{p_j^t}\, U_j^t \right).
\]

Next, we split the double summation into two terms: one for the case \(k = j\) and one for \(k \neq j\):
\[
\sum_{k \in A^t} \sum_{j \in A^t} \left( \frac{n_k}{p_k^t}\, U_k^t \right)^T \left( \frac{n_j}{p_j^t}\, U_j^t \right)
= \underbrace{\sum_{k \in A^t} \left( \frac{n_k}{p_k^t}\, U_k^t \right)^T \left( \frac{n_k}{p_k^t}\, U_k^t \right)}_{\text{terms with } k = j}
+ \underbrace{\sum_{k \in A^t} \sum_{\substack{j \in A^t \\ j \neq k}} \left( \frac{n_k}{p_k^t}\, U_k^t \right)^T \left( \frac{n_j}{p_j^t}\, U_j^t \right)}_{\text{terms with } k \neq j}.
\]

Taking the expectation over \(A^t\) and noting that each client \(k\) is selected with probability \(p_k^t\), we have
\[
\begin{aligned}
\mathbb{E}_{A^t} \Biggl[ \sum_{k \in A^t} \sum_{j \in A^t} &\left( \frac{n_k}{p_k^t}\, U_k^t \right)^T \left( \frac{n_j}{p_j^t}\, U_j^t \right) \Biggr] \\
&= \sum_{k=1}^{K} \mathbb{E}[I_k]\, \left( \frac{n_k}{p_k^t}\, U_k^t \right)^T \left( \frac{n_k}{p_k^t}\, U_k^t \right)
+ \sum_{k=1}^{K} \sum_{\substack{j=1 \\ j\neq k}}^{K} \mathbb{E}[I_k I_j]\, \left( \frac{n_k}{p_k^t}\, U_k^t \right)^T \left( \frac{n_j}{p_j^t}\, U_j^t \right) \\
&= \sum_{k=1}^{K} p_k^t \left( \frac{n_k}{p_k^t}\, U_k^t \right)^T \left( \frac{n_k}{p_k^t}\, U_k^t \right)
+ \sum_{k=1}^{K} \sum_{\substack{j=1 \\ j\neq k}}^{K} \left(p_k^t p_j^t\right) \left( \frac{n_k}{p_k^t}\, U_k^t \right)^T \left( \frac{n_j}{p_j^t}\, U_j^t \right) \\
&= \sum_{k=1}^{K} \frac{n_k^2}{p_k^t}\, \|U_k^t\|^2
+ \sum_{k=1}^{K} \sum_{\substack{j=1 \\ j\neq k}}^{K} n_k\, K_j \left( U_k^t \right)^T \left( U_j^t \right).
\end{aligned}
\]

Similarly, for the second term we have
\[
\left\|\sum_{k=1}^{K} n_k\, U_k^t\right\|^2 
= \left( \sum_{k=1}^{K} n_k\, U_k^t \right)^T \left( \sum_{j=1}^{K} n_j\, U_j^t \right)
= \sum_{k=1}^{K} \sum_{j=1}^{K} \left( n_k\, U_k^t \right)^T \left( n_j\, U_j^t \right).
\]
Dividing the double summation into the diagonal and off-diagonal parts, we obtain
\[
\sum_{k=1}^{K} \sum_{j=1}^{K} \left( n_k\, U_k^t \right)^T \left( n_j\, U_j^t \right)
= \underbrace{\sum_{k=1}^{K} \left( n_k\, U_k^t \right)^T \left( n_k\, U_k^t \right)}_{k=j}
+ \underbrace{\sum_{k=1}^{K} \sum_{\substack{j=1 \\ j \neq k}}^{K} \left( n_k\, U_k^t \right)^T \left( n_j\, U_j^t \right)}_{k \neq j}.
\]

Thus, from the previous reduction, we have
\[
\begin{aligned}
\mathbb{E}_{A^t}\left[ \left\| \sum_{k \in A^t} \frac{n_k}{p_k^t}\, U_k^t \right\|^2 \right] 
&= \sum_{k=1}^{K} \frac{n_k^2}{p_k^t}\, \|U_k^t\|^2 
+ \sum_{\substack{k,j=1 \\ i\neq j}}^{K} n_k\, K_j\, \left(U_k^t\right)^T \left(U_j^t\right), \\
\left\|\sum_{k=1}^{K} n_k\, U_k^t\right\|^2 
&= \sum_{k=1}^{K} n_k^2\, \|U_k^t\|^2 
+ \sum_{\substack{k,j=1 \\ i\neq j}}^{K} n_k\, K_j\, \left(U_k^t\right)^T \left(U_j^t\right).
\end{aligned}
\]
By rewriting the variance in terms of the separated summations, the total variance is given by
\[
\begin{aligned}
\text{Var} 
&= \mathbb{E}_{A^t}\left[ \left\| \sum_{k \in A^t} \frac{n_k}{p_k^t}\, U_k^t \right\|^2 \right]
- \left\|\sum_{k=1}^{K} n_k\, U_k^t\right\|^2 \\
&= \left[ \sum_{k=1}^{K} \frac{n_k^2}{p_k^t}\, \|U_k^t\|^2 
+ \sum_{\substack{k,j=1 \\ i\neq j}}^{K} n_k\, K_j\, \left(U_k^t\right)^T \left(U_j^t\right) \right]
- \left[ \sum_{k=1}^{K} n_k^2\, \|U_k^t\|^2 
+ \sum_{\substack{k,j=1 \\ i\neq j}}^{K} n_k\, K_j\, \left(U_k^t\right)^T \left(U_j^t\right) \right] \\
&= \sum_{k=1}^{K} \left(\frac{n_k^2}{p_k^t} - n_k^2\right) \|U_k^t\|^2 \\
&= \sum_{k=1}^{K} n_k^2\, \|U_k^t\|^2 \left(\frac{1}{p_k^t} - 1\right).
\end{aligned}
\]

Thus, the original optimization problem can be rewritten as
\[
\min_{\{p_k^t\}} \; \sum_{k=1}^{K} n_k^2\, \|U_k^t\|^2 \left(\frac{1}{p_k^t} - 1\right)
\]
subject to
\[
0 \leq p_k^t \leq 1, \quad \forall\, i, \quad \text{and} \quad \sum_{k=1}^{K} r_k\, p_k^t = m.
\]

\subsection{Variable Substitution}

We perform the substitutions:
\[
q_k = \frac{1}{p_k^t} - 1 \quad \Longrightarrow \quad p_k^t = \frac{1}{q_k+1},
\]
and define
\[
x_k = n_k^2\, \|U_k^t\|^2.
\]
Then, the objective becomes
\[
\min_{\{q_k\}} \; \sum_{k=1}^{K} x_k\, q_k,
\]
with the constraints:
\[
q_k \geq 0, \quad \forall\, i,
\]
and the modified equality constraint is now
\[
\sum_{k=1}^{K} \frac{r_k}{q_k+1} = m.
\]

\subsection{Lagrangian Formulation and KKT Conditions}

Define the Lagrangian with Lagrange multiplier \(\lambda\) for the equality constraint and multipliers \(\mu_k\ge0\) for the non-negativity constraints on \(q_k\):
\[
\mathcal{L}(\{q_k\}, \lambda, \{\mu_k\}) = \sum_{k=1}^{K} x_k\, q_k + \lambda \left( \sum_{k=1}^{K} \frac{r_k}{q_k+1} - m \right) - \sum_{k=1}^{K} \mu_k\, q_k.
\]
The KKT conditions are:
\begin{enumerate}
    \item \textbf{Stationarity:} For each \(k=1,\dots,K\),
    \[
    \frac{\partial \mathcal{L}}{\partial q_k} = x_k - \lambda\, \frac{r_k}{(q_k+1)^2} - \mu_k = 0.
    \]
    \item \textbf{Primal Feasibility:}
    \[
    q_k \geq 0, \quad \forall\, i, \quad \text{and} \quad \sum_{k=1}^{K} \frac{r_k}{q_k+1} = m.
    \]
    \item \textbf{Dual Feasibility:}
    \[
    \mu_k \geq 0, \quad \forall\, i.
    \]
    \item \textbf{Complementary Slackness:}
    \[
    \mu_k\, q_k = 0, \quad \forall\, i.
    \]
\end{enumerate}
These conditions will be used to solve for the optimal \(\{q_k\}\), from which the original probabilities are recovered via
\[
p_k^t = \frac{1}{q_k+1}.
\]

\subsection{Solving the KKT Conditions}

The stationarity condition is
\[
x_k - \frac{\lambda\, r_k}{(q_k+1)^2} - \mu_k = 0.
\]
Isolating \(\mu_k\) yields
\[
x_k - \frac{\lambda\, r_k}{(q_k+1)^2} = \mu_k.
\]
We now consider two cases:

\paragraph*{Case 1: \(q_k > 0\).}  
Then complementary slackness implies \(\mu_k = 0\), so that
\[
x_k = \frac{\lambda\, r_k}{(q_k+1)^2}.
\]
Solving for \(q_k\) gives
\[
q_k = \sqrt{\frac{\lambda\, r_k}{x_k}} - 1.
\]

\paragraph*{Case 2: \(q_k = 0\).}  
In this case, the stationarity condition becomes
\[
x_k - \lambda\, r_k = \mu_k \ge 0,
\]
which implies
\[
x_k \ge \lambda\, r_k.
\]

Thus, we have:
\begin{itemize}
    \item For \(q_k > 0\): \(q_k = \sqrt{\frac{\lambda\, r_k}{x_k}} - 1\).
    \item For \(q_k = 0\): \(x_k \ge \lambda\, r_k\).
\end{itemize}

\subsection{Determining \(\lambda\) via the Equality Constraint}

The modified equality constraint is
\[
\sum_{k=1}^{K} \frac{r_k}{q_k+1} = m.
\]
Define the index set
\[
\mathcal{O} = \{ k \in \{1, \dots, K\} : x_k < \lambda\, r_k\ \},
\]
so that for \(k\in\mathcal{O}\) (by Case 1) we have
\[
\quad \frac{1}{q_k+1} = \sqrt{\frac{x_k}{\lambda\, r_k}},
\]
and for indices \(k\in\mathcal{O}^c=\mathcal{K}-\mathcal{O}\) (where \(q_k=0\) by Case 2) we have
\[
\frac{1}{q_k+1} = 1.
\]
Thus, the equality constraint becomes
\[
\sum_{k\in\mathcal{O}} r_k\, \sqrt{\frac{x_k}{\lambda\, r_k}} + \sum_{k\in\mathcal{O}^c} r_k = m.
\]
note that
\[
\sum_{k\in\mathcal{O}^c} r_k = \sum_{k=1}^{K} r_k - \sum_{k\in\mathcal{O}} r_k.
\]
Thus, the equality constraint becomes
\[
\frac{1}{\sqrt{\lambda}} \sum_{k\in\mathcal{O}} \sqrt{r_k\, x_k} + \left(\sum_{k=1}^{K} r_k - \sum_{k\in\mathcal{O}} r_k \right) = m.
\]
Isolating \(\lambda\) yields
\[
\lambda = \left(\frac{\sum_{k\in\mathcal{O}} \sqrt{r_k\, x_k}}{m - \sum_{k=1}^{K} r_k + \sum_{k\in\mathcal{O}} r_k}\right)^2.
\]

\subsection{Finding the Explicit Solution for \(q_k\)}

For \(k \in \mathcal{O}\) (i.e. where \(q_k > 0\)), substituting the expression for \(\lambda\) into
\[
q_k = \sqrt{\frac{\lambda\, r_k}{x_k}} - 1,
\]
yields
\[
q_k = \frac{\sqrt{r_k}}{m - \sum_{j=1}^{K}r_j + \sum_{j\in\mathcal{O}} r_j} \cdot \frac{\sum_{j\in\mathcal{O}} \sqrt{r_j\, x_j}}{\sqrt{x_k}} - 1.
\]
For \(k \in \mathcal{O}^c\), we set \(q_k = 0\).

Thus, the piecewise definition is:
\[
q_k =
\begin{cases}
\displaystyle \frac{\sqrt{r_k}}{m - \sum_{j=1}^{K}r_j + \sum_{j\in\mathcal{O}} r_j} \cdot \frac{\sum_{j\in\mathcal{O}} \sqrt{r_j\, x_j}}{\sqrt{x_k}} - 1, & \text{if } k \in \mathcal{O}, \\
0, & \text{if } k \in \mathcal{O}^c.
\end{cases}
\]

\subsection{Returning to the Original Variable \(p_k^t\)}

Recall that
\[
p_k^t = \frac{1}{q_k+1}.
\]
Thus, for \(k \in \mathcal{O}\) (where \(q_k>0\)) we have
\[
p_k^t = \frac{m - \sum_{j=1}^{K}r_j + \sum_{j\in\mathcal{O}} r_j}{\sqrt{r_k}} \cdot \frac{\sqrt{x_k}}{\sum_{j\in\mathcal{O}} \sqrt{r_j\, x_j}},
\]
and for \(k \in \mathcal{O}^c\) (where \(q_k = 0\)) we obtain
\[
p_k^t = 1.
\]
Substituting the expression and recalling that \(x_k = n_k^2\, \|U_k^t\|^2\), the final solution is
\[
p_k^t =
\begin{cases}
\displaystyle \frac{m - \sum_{j=1}^{K}r_j + \sum_{j\in\mathcal{O}} r_j}{\sqrt{r_k}} \cdot \frac{n_k\, \|U_k^t\|}{\sum_{j\in\mathcal{O}} \sqrt{r_j}\, K_j\, \|U_j^t\|}, & \text{if } k \in \mathcal{O}, \\[1ex]
1, & \text{if } k \in \mathcal{O}^c.
\end{cases}
\]

\noindent{Where} 
\[
\mathcal{O} = \left\{ k \in \{1, \dots, K\} : \sqrt{r_k}\, K_k\, \|U_k^t\| < \frac{\sum_{j\in\mathcal{O}} \sqrt{r_j}\, K_j\, \|U_j^t\|}{m - \sum_{j=1}^{K}r_j + \sum_{j\in\mathcal{O}} r_j} \right\}.
\]

\subsection{Determining the Index Set \(\mathcal{O}\)}

Since for indices in \(\mathcal{O}^c\) we have \(p_k^t = 1\), the equality constraint
\[
\sum_{k=1}^{K} r_k\, p_k^t = m
\]
becomes
\[
\sum_{k\in\mathcal{O}} r_k\, p_k^t + \sum_{k\in\mathcal{O}^c} r_k = m.
\]
and by splitting the second term, we got:
\[
\sum_{k\in\mathcal{O}} r_k\, p_k^t + \sum_{k=1}^{K} r_k - \sum_{k\in\mathcal{O}} r_k = m.
\]
equivalent to
\[
\sum_{k\in\mathcal{O}} r_k\, \left(1-p_k^t \right) = \sum_{k=1}^{K} r_k - m.
\]
Thus, the indices in \(\mathcal{O}\) must satisfy
\[
\sum_{k\in\mathcal{O}} r_k > \sum_{k=1}^{K} r_k - m.
\]
That is, the total \(r\)-mass in \(\mathcal{O}\) must exceed \(\sum_{k=1}^{K} r_k - m\).

The procedure to determine \(\mathcal{O}\) and its complement \(\mathcal{O}^c\) is as follows:
\begin{enumerate}
    \item \textbf{Initialization:}  
    Order the indices in increasing order of \(\sqrt{r_k}\, K_k^t\, \|U_k^t\|\). Begin by selecting the indices corresponding to the smallest values until the cumulative sum \(\sum_{k\in\mathcal{O}} r_k\) exceeds \(\sum_{k=1}^{K}r_k - m\).

    \item \textbf{Iteration:}  
    Let \(\mathcal{O}^c = \{1,\dots,N\} \setminus \mathcal{O}\). For the next candidate index \(k\) in \(\mathcal{O}^c\) (with the next smallest \(\sqrt{r_k}\,n_k^t\, \|U_k^t\|\)), check whether adding \(k\) to \(\mathcal{O}\) maintains the inequality
    \[
    \sqrt{r_k}\, n_k^t\, \|U_k^t\| < \frac{\sum_{j\in \mathcal{O}\cup\{k\}} \sqrt{r_j}\, K_j^t\, \|U_j^t\|}{m - \sum_{j=1}^{K}r_j + \sum_{j\in\mathcal{O}} r_j}
    \]
    
    \item \textbf{Update:}  
    \begin{itemize}
        \item If the inequality holds, update \(\mathcal{O} \leftarrow \mathcal{O} \cup \{k\}\) and \(\mathcal{O}^c \leftarrow \mathcal{O}^c \setminus \{k\}\), and then repeat the iteration.
        \item If the inequality fails for the candidate \(k\), NO further indices can be added to \(\mathcal{O}\); then \(\mathcal{O}\) and \(\mathcal{O}^c\) are finalized.
    \end{itemize}
\end{enumerate}

\newpage
\section{Detailed Derivations and Numerical Analysis of Efficiency Gain Analysis}
\label{app:eff-Gain}

\subsection{Derivations of Efficiency Gains}

In this appendix, we provide the detailed derivations underlying the efficiency analysis presented in Section~\ref{sec:FedPLT_efficiency}.

\subsubsection{Detailed Derivation of Computation Cost}
Under full-model training, each client performs both forward and backward computation over all $P$ parameters during each of the $\tau$  local iterations.

The forward and the backward workloads are, respectively:
\[
\Xi_{k,\mathrm{fwd}}^{\mathrm{full}} = \alpha \tau P,
\qquad
\Xi_{k,\mathrm{bwd}}^{\mathrm{full}} = \beta \tau P
\]

Hence, the total per-round computation cost under full-model training is
\[
\Xi_k^{\mathrm{full}} =
\Xi_{k,\mathrm{fwd}}^{\mathrm{full}} + \Xi_{k,\mathrm{bwd}}^{\mathrm{full}} =
(\alpha+\beta) \tau P.
\]

Under FedPLT, the forward pass still involves the entire model. However, only the assigned fraction $r_k$ of the model participates in backward computation. Therefore, the forward backward remains unchanged, but backward workload scales with $r_k$.
\[
\Xi_{k,\mathrm{fwd}}^{\mathrm{plt}} = \alpha \tau P,
\qquad
\Xi_{k,\mathrm{bwd}}^{\mathrm{plt}} = \beta r_k \tau P.
\]
The total per-round computation cost under FedPLT becomes then
\[
\Xi_k^{\mathrm{plt}} =
\Xi_{k,\mathrm{fwd}}^{\mathrm{plt}} + \Xi_{k,\mathrm{bwd}}^{\mathrm{plt}} =
\alpha \tau P + \beta r_k \tau P =
(\alpha+\beta r_k) \tau P.
\]

The absolute computation reduction is
\[
\Delta \Xi_k =
\Xi_k^{\mathrm{full}}-\Xi_k^{\mathrm{plt}} =
(\alpha+\beta)\tau P-(\alpha+\beta r_k) \tau P =
\beta(1-r_k) \tau P.
\]
Dividing by $\Xi_k^{\mathrm{full}}$ yields the relative computation reduction:
\[
\Delta_k^{\mathrm{comp}} =
1-\frac{\Xi_k^{\mathrm{plt}}}{\Xi_k^{\mathrm{full}}} =
1-\frac{(\alpha+\beta r_k) \tau P}{(\alpha+\beta) \tau P} =
1-\frac{\alpha+\beta r_k}{\alpha+\beta}.
\]

Therefore, the computation gain comes entirely from the backward stage, and it increases as $r_k$ decreases.

\subsubsection{Derivation of Communication Cost}
Under full-model training, each client downloads the full global model and uploads the full local update. Since the model size is $Ps$ bytes, the per-round communication volume is
\[
\Theta_k^{\mathrm{full}} = Ps + Ps = 2Ps.
\]

Under FedPLT, the full model is still downloaded, but only the updated fraction $r_k$ is uploaded. Thus,
\[
\Theta_k^{\mathrm{plt}} = Ps + r_k Ps = (1+r_k)Ps.
\]

The absolute communication reduction is
\[
\Delta \Theta_k =
\Theta_k^{\mathrm{full}}-\Theta_k^{\mathrm{plt}} =
2Ps-(1+r_k)Ps =
(1-r_k)Ps.
\]

and the elative communication reduction is
\[
\Delta_k^{\mathrm{comm}} =
1-\frac{\Theta_k^{\mathrm{plt}}}{\Theta_k^{\mathrm{full}}} =
1-\frac{(1+r_k)Ps}{2Ps} =
\frac{1-r_k}{2}.
\]

It is also useful to separate downlink and uplink:
\[
\Theta_{k,\downarrow}^{\mathrm{full}} = \Theta_{k,\downarrow}^{\mathrm{plt}} = Ps,
\]
\[
\Theta_{k,\uparrow}^{\mathrm{full}} = Ps,
\qquad
\Theta_{k,\uparrow}^{\mathrm{plt}} = r_k Ps.
\]
Hence, the relative uplink reduction is
\[
\Delta_k^{\mathrm{comm,up}} =
1-\frac{\Theta_{k,\uparrow}^{\mathrm{plt}}}{\Theta_{k,\uparrow}^{\mathrm{full}}} =
1-\frac{r_k Ps}{Ps} =
1-r_k.
\]

Therefore, FedPLT preserves the downlink cost while reducing the uplink proportionally to the assigned fraction $r_k$.

\subsubsection{Derivation of Round Time and Straggler Mitigation}
In synchronous FL, the duration of a round is determined by the slowest participating client.

For a client $k$, under full-model training, the forward  and backward computation time are respectively,
\[
\frac{\alpha \tau P}{\gamma_k},
\qquad
\frac{\beta \tau P}{\gamma_k}.
\]
and the download and upload times are respectively
\[
\frac{Ps}{B_k^{\downarrow}},
\qquad
\frac{Ps}{B_k^{\uparrow}}.
\]

Including the fixed overhead $\delta$, the total per-round time of client $k$ is
\[
\begin{aligned}
T_k^{\mathrm{full}}
&=
\delta + \frac{\alpha \tau P}{\gamma_k} + \frac{\beta \tau P}{\gamma_k} + \frac{Ps}{B_k^{\downarrow}} + \frac{Ps}{B_k^{\uparrow}} \\
&=
\delta + \frac{(\alpha+\beta)\tau P}{\gamma_k} + \frac{Ps}{B_k^{\downarrow}} + \frac{Ps}{B_k^{\uparrow}}.
\end{aligned}
\]

Hence, the full-model round time is
\[
T_{\mathrm{round}}^{\mathrm{full}} =
\max_k T_k^{\mathrm{full}} =
\delta + \max_k \left( \frac{(\alpha+\beta) \tau P}{\gamma_k} + \frac{Ps}{B_k^{\downarrow}} + \frac{Ps}{B_k^{\uparrow}} \right).
\]

Let $\hat{k}$ the straggler client be the one attaining this maximum. Defining
\[
\hat{\gamma}=\gamma_{\hat{k}},
\qquad
\hat{B}^{\downarrow}=B_{\hat{k}}^{\downarrow},
\qquad
\hat{B}^{\uparrow}=B_{\hat{k}}^{\uparrow},
\]
we obtain
\[
T_{\mathrm{round}}^{\mathrm{full}} = 
\delta + \frac{(\alpha+\beta) \tau P}{\hat{\gamma}} + \frac{Ps}{\hat{B}^{\downarrow}} + \frac{Ps}{\hat{B}^{\uparrow}}.
\]

Under FedPLT, the forward computation and model download remain unchanged, whereas the backward computation and upload scale with $r_k$. Therefore, the per-round time of client $k$ becomes
\[
\begin{aligned}
T_k^{\mathrm{plt}} =
\delta
+ \underbrace{\left(\frac{\alpha \tau P}{\gamma_k} + \frac{Ps}{B_k^{\downarrow}}
\right)}_{\text{fixed part}}
+ r_k \underbrace{\left(\frac{\beta \tau P}{\gamma_k} + \frac{Ps} {B_k^{\uparrow}}\right)}_{\text{scaled part}}.
\end{aligned}
\]

FedPLT mitigates the straggler effect by adapting $r_k$ to clients' capacity so that client completion times become approximately balanced:
\[
T_1^{\mathrm{plt}}
\approx
T_2^{\mathrm{plt}}
\approx
\dots
\approx
T_K^{\mathrm{plt}}
\approx
T_{\mathrm{round}}^{\mathrm{plt}}.
\]

Choosing a target round duration $T$ imposes
\[
T_k^{\mathrm{plt}} = 
\delta + \frac{\alpha \tau P}{\gamma_k} + \frac{Ps}{B_k^{\downarrow}} + r_k\left(
\frac{\beta \tau P}{\gamma_k} + \frac{Ps}{B_k^{\uparrow}} \right) \approx T.
\]
Hence, to mitigate stragglers, the training ration to be applied by each client $k$ is
\[
r_k
\approx
\frac{
T-\delta-\left(\frac{\alpha \tau P}{\gamma_k}+\frac{Ps}{B_k^{\downarrow}}\right)
}{
\left(\frac{\beta \tau P}{\gamma_k}+\frac{Ps}{B_k^{\uparrow}}\right)
}.
\]
In practice, this expression is meaningful when the resulting value lies in $[0,1]$; otherwise, the target $T$ must be relaxed or the value clipped to the feasible range.

Under approximate equalization, the FedPLT round time can be written in terms of the limiting client as
\[
T_{\mathrm{round}}^{\mathrm{plt}}
= \delta + \frac{\alpha \tau P}{\hat{\gamma}} + \frac{Ps}{\hat{B}^{\downarrow}} + \hat{r}\left(\frac{\beta \tau P}{\hat{\gamma}} + \frac{Ps}{\hat{B}^{\uparrow}}\right),
\]
where $\hat{r}$ denotes the fraction assigned to the limiting client.

The absolute round-time reduction is then
\[
\begin{aligned}
\Delta T
&=
T_{\mathrm{round}}^{\mathrm{full}}
-
T_{\mathrm{round}}^{\mathrm{plt}} \\
&=
\left[
\delta
+
\frac{(\alpha+\beta) \tau P}{\hat{\gamma}}
+
\frac{Ps}{\hat{B}^{\downarrow}}
+
\frac{Ps}{\hat{B}^{\uparrow}}
\right]
-
\left[
\delta
+
\frac{\alpha \tau P}{\hat{\gamma}}
+
\frac{Ps}{\hat{B}^{\downarrow}}
+
\hat{r}\left(
\frac{\beta \tau P}{\hat{\gamma}}
+
\frac{Ps}{\hat{B}^{\uparrow}}
\right)
\right] \\
&=
\frac{\beta \tau P}{\hat{\gamma}}
+
\frac{Ps}{\hat{B}^{\uparrow}}
-
\hat{r}\left(
\frac{\beta \tau P}{\hat{\gamma}}
+
\frac{Ps}{\hat{B}^{\uparrow}}
\right) \\
&=
(1-\hat{r})
\left(
\frac{\beta \tau P}{\hat{\gamma}}
+
\frac{Ps}{\hat{B}^{\uparrow}}
\right).
\end{aligned}
\]

Dividing by $T_{\mathrm{round}}^{\mathrm{full}}$ gives the relative round-time efficiency:
\[
\Delta_{\mathrm{time}}
=
1-\frac{T_{\mathrm{round}}^{\mathrm{plt}}}{T_{\mathrm{round}}^{\mathrm{full}}}
=
\frac{
(1-\hat{r})
\left(
\frac{\beta \tau P}{\hat{\gamma}}
+
\frac{Ps}{\hat{B}^{\uparrow}}
\right)
}{
\frac{(\alpha+\beta)iP}{\hat{\gamma}}
+
\frac{Ps}{\hat{B}^{\downarrow}}
+
\frac{Ps}{\hat{B}^{\uparrow}}
}.
\]

Therefore, the round-time gain comes from reducing the variable part of the slowest client's workload, namely backward computation and uplink transmission, which directly mitigates the straggler effect.

\subsection{Numerical Example}

We consider a system with $K=5$ heterogeneous clients. The global model contains $P=5\times10^6$ parameters, stored using $s=4$ bytes per parameter, so that
\[
Ps = 20\times10^6 \text{ bytes} \approx 20\text{ MB}.
\]
Each client performs $\tau=150$ local iterations per round. The per-parameter forward and backward costs are $\alpha=2$ and $\beta=4$, respectively, with $\beta\approx 2\alpha$. The fixed latency is $\delta=0.2$ s.

Client $k$ has computation speed $\gamma_k$ (in GFLOPs/s), downlink bandwidth $B_k^{\downarrow}$, and uplink bandwidth $B_k^{\uparrow}$ (in Mb/s).%
\footnote{If bandwidth is expressed in Mb/s, the corresponding transmission rate in MB/s is $B/8$. Hence, the transmission time of a payload of size $Ps$ is $\tfrac{Ps}{B/8}$.}

\subsubsection{Full-Training Per-Client Round Time and FedPLT Equalization}
Under full-model training, the per-round time of client $k$ is
\[
T_k^{\mathrm{full}}
=
\delta
+
\left(
\frac{\alpha \tau P}{\gamma_k}
+
\frac{Ps}{B_k^{\downarrow}/8}
\right)
+
\left(
\frac{\beta \tau P}{\gamma_k}
+
\frac{Ps}{B_k^{\uparrow}/8}
\right).
\]

We first compute the workload terms:
\[
\alpha \tau P = 2\times150\times5\times10^6 = 1.5\times10^9 \text{ FLOPs} = 1.5 \text{ GFLOPs},
\]
\[
\beta \tau P = 4\times150\times5\times10^6 = 3\times10^9 \text{ FLOPs} = 3 \text{ GFLOPs}.
\]
Thus, under full-model training, each client performs
\[
(\alpha+\beta) \tau P= 4.5 \text{ GFLOPs}
\]
per round.

Let
\[
T^* := \min_k T_k^{\mathrm{full}}
\]
denote the target round duration used for equalization. In this example, the fastest full-training client determines the target. FedPLT selects $r_k$ so that
\[
T_k^{\mathrm{plt}}
=
\delta
+
\left(
\frac{\alpha \tau P}{\gamma_k}
+
\frac{Ps}{B_k^{\downarrow}/8}
\right)
+
r_k
\left(
\frac{\beta \tau P}{\gamma_k}
+
\frac{Ps}{B_k^{\uparrow}/8}
\right)
\approx T^*.
\]
Solving for $r_k$ gives
\[
r_k
=
\frac{
T^*-\delta-\left(\frac{\alpha \tau P}{\gamma_k}+\frac{Ps}{B_k^{\downarrow}/8}\right)
}{
\left(\frac{\beta \tau P}{\gamma_k}+\frac{Ps}{B_k^{\uparrow}/8}\right)
}.
\]

Table~\ref{tab:perclient_times} reports the client parameters, the full-training round times, and the resulting FedPLT fractions. The value of $T^*$ is highlighted in bold.

\begin{table}[H]
\centering
\caption{Per-client parameters, full-training round time $T_k^{\mathrm{full}}$, and FedPLT allocation $r_k$. Units: $\gamma_k$ in GFLOPs/s, $B_k^{\downarrow},B_k^{\uparrow}$ in Mb/s, and $T_k^{\mathrm{full}}$ in seconds. Bold indicates $T^*=\min_k T_k^{\mathrm{full}}$.}
\label{tab:perclient_times}
\begin{tabular}{lccccc}
\toprule
Client & $\gamma_k$ & $B_k^{\downarrow}$ & $B_k^{\uparrow}$ & $T_k^{\mathrm{full}}$ & $r_k$ \\
\midrule
$C_1$ (smartphone, WiFi)   & $80$ & $120$ & $40$ & $5.59$ & $0.69$ \\
$C_2$ (smart TV, Ethernet) & $30$ & $200$ & $50$ & \textbf{4.35} & $1.00$ \\
$C_3$ (drone, 5G)          & $18$ & $60$  & $25$ & $9.52$ & $0.21$ \\
$C_4$ (smart AC, WiFi)     & $10$ & $50$  & $16$ & $13.85$ & $0.078$ \\
$C_5$ (IoT device, LTE/4G) & $8$  & $45$  & $12$ & $17.65$ & $0.03$ \\
\bottomrule
\end{tabular}
\end{table}

\subsubsection{Computation Efficiency}
Under full-model training,
\[
\Xi_k^{\mathrm{full}} = (\alpha+\beta) \tau P= 4.5 \text{ GFLOPs}.
\]
Under FedPLT,
\[
\Xi_k^{\mathrm{plt}} = \alpha \tau P + \beta r_k \tau P = 1.5 + 3r_k \quad \text{(GFLOPs)}.
\]
Therefore,
\[
\Delta_{k,\mathrm{comp}}
=
1-\frac{\Xi_k^{\mathrm{plt}}}{\Xi_k^{\mathrm{full}}}
=
1-\frac{\alpha+\beta r_k}{\alpha+\beta}.
\]

For interpretability, we also report the derived computation time
\[
T_{k,\mathrm{comp}}^{(\cdot)} = \frac{\Xi_k^{(\cdot)}}{\gamma_k},
\]
with $\gamma_k$ in GFLOPs/s.

For example, for client $C_1$,
\[
\Xi_{1}^{\mathrm{plt}} = 1.5 + 3(0.69)=3.57 \text{ GFLOPs},
\]
\[
T_{1,\mathrm{comp}}^{\mathrm{full}} = \frac{4.5}{80}=0.05625\text{ s}=56.3\text{ ms},
\qquad
T_{1,\mathrm{comp}}^{\mathrm{plt}} = \frac{3.57}{80}=0.0446\text{ s}=44.6\text{ ms}.
\]

Table~\ref{tab:comp_saving} summarizes the resulting values for all clients.

\begin{table}[H]
\centering
\caption{Per-client computation workload and derived computation time. $\Xi$ is reported in GFLOPs and $T_{k,\mathrm{comp}}=\Xi/\gamma_k$ in milliseconds.}
\label{tab:comp_saving}
\begin{tabular}{lccccc}
\toprule
 & $C_1$ & $C_2$ & $C_3$ & $C_4$ & $C_5$ \\
\midrule
$\Xi_k^{\mathrm{full}}$ (GFLOPs) & $4.50$ & $4.50$ & $4.50$ & $4.50$ & $4.50$ \\
$T_{k,\mathrm{comp}}^{\mathrm{full}}$ (ms) & $56.3$ & $150.0$ & $250.0$ & $450.0$ & $562.5$ \\
$\Xi_k^{\mathrm{plt}}$ (GFLOPs) & $3.57$ & $4.50$ & $2.13$ & $1.73$ & $1.59$ \\
$T_{k,\mathrm{comp}}^{\mathrm{plt}}$ (ms) & $44.6$ & $150.0$ & $118.3$ & $173.4$ & $198.8$ \\
$\Delta_{k,\mathrm{comp}}$ (\%) & $20.7$ & $0.0$ & $52.7$ & $61.5$ & $64.7$ \\
\bottomrule
\end{tabular}
\end{table}

As expected, clients with weaker computational capabilities ($C_3$--$C_5$) receive smaller fractions and thus achieve the largest computation savings.

\subsubsection{Communication Efficiency}
In full-model training, each client exchanges the full model:
\[
\Theta_k^{\mathrm{full}} = 2Ps = 40 \text{ MB}.
\]
Under FedPLT,
\[
\Theta_k^{\mathrm{plt}} = (1+r_k)Ps = 20(1+r_k)\text{ MB}.
\]
The relative uplink and total communication savings are
\[
\Delta_k^{\mathrm{comm,up}} = 1-r_k,
\qquad
\Delta_k^{\mathrm{comm,tot}} = \frac{1-r_k}{2}.
\]

For example, for client $C_4$ with $r_4=0.078$,
\[
\Theta_4^{\mathrm{plt}} = 20(1+0.078)=21.56 \text{ MB},
\]
\[
\Delta_4^{\mathrm{comm,up}} = 1-0.078 = 0.922 = 92.2\%,
\]
\[
\Delta_4^{\mathrm{comm,tot}} = \frac{1-0.078}{2} = 0.461 = 46.1\%.
\]

Table~\ref{tab:comm_saving} reports the per-client communication costs and the corresponding efficiency gains.

\begin{table}[H]
\centering
\caption{Per-client communication cost and efficiency gains. Values are reported in MB, assuming $Ps=20$ MB.}
\label{tab:comm_saving}
\begin{tabular}{lccccc}
\toprule
 & $C_1$ & $C_2$ & $C_3$ & $C_4$ & $C_5$ \\
\midrule
$\Theta_k^{\mathrm{full}}$ (MB) & $40.0$ & $40.0$ & $40.0$ & $40.0$ & $40.0$ \\
$\Theta_k^{\mathrm{plt}}$ (MB) & $33.8$ & $40.0$ & $24.2$ & $21.56$ & $20.6$ \\
$\Delta_k^{\mathrm{comm,up}}$ (\%) & $31.0$ & $0.0$ & $79.0$ & $92.2$ & $97.0$ \\
$\Delta_k^{\mathrm{comm,tot}}$ (\%) & $15.5$ & $0.0$ & $39.5$ & $46.1$ & $48.5$ \\
\bottomrule
\end{tabular}
\end{table}

These results highlight that the communication gain comes entirely from the reduced uplink, while the model download remains unchanged.

\subsection{Round Time and Straggler Mitigation}
Under full-model synchronous training, the global round time is determined by the slowest client:
\[
T_{\mathrm{round}}^{\mathrm{full}} = \max_k T_k^{\mathrm{full}} = 17.65 \text{ s}.
\]

Under the FedPLT equalization target, all clients are adjusted to finish in approximately
\[
T_{\mathrm{round}}^{\mathrm{plt}} = T^* = 4.35 \text{ s}.
\]
Therefore, the relative round-time efficiency is
\[
\Delta_{\mathrm{time}}
=
1-\frac{T_{\mathrm{round}}^{\mathrm{plt}}}{T_{\mathrm{round}}^{\mathrm{full}}}
=
1-\frac{4.35}{17.65}
=
75.35\%.
\]

Table~\ref{tab:time_global} summarizes the reduction in overall round duration.

\begin{table}[H]
\centering
\caption{Global round time before and after FedPLT (seconds).}
\label{tab:time_global}
\begin{tabular}{lcc}
\toprule
$T_{\mathrm{round}}^{\mathrm{full}}$ & $T_{\mathrm{round}}^{\mathrm{plt}}$ & $\Delta_{\mathrm{time}}$ \\
\midrule
$17.65$ & $4.35$ & $75.35\%$ \\
\bottomrule
\end{tabular}
\end{table}

Under full training, each client may finish early and then remain idle while waiting for the slowest client. The idle time of client $k$ is
\[
I_k^{\mathrm{full}} = T_{\mathrm{round}}^{\mathrm{full}} - T_k^{\mathrm{full}} \ge 0.
\]
Under ideal FedPLT equalization, this idle time is removed. For example, for client $C_1$,
\[
I_1^{\mathrm{full}} = 17.65 - 5.59 = 12.06 \text{ s},
\]
which corresponds to
\[
\frac{12.06}{17.65} = 68.33\%.
\]

Table~\ref{tab:idle_time} reports the idle time avoided for all clients.

\begin{table}[H]
\centering
\caption{Idle time avoided by FedPLT. Absolute values are in seconds and relative values are with respect to $T_{\mathrm{round}}^{\mathrm{full}}$.}
\label{tab:idle_time}
\begin{tabular}{lccccc}
\toprule
 & $C_1$ & $C_2$ & $C_3$ & $C_4$ & $C_5$ \\
\midrule
$T_k^{\mathrm{full}}$ (s) & $5.59$ & $4.35$ & $9.52$ & $13.85$ & \textbf{17.65} \\
Idle avoided (s) & $12.06$ & $13.30$ & $8.13$ & $3.80$ & $0.00$ \\
Idle avoided (\%) & $68.33$ & $75.35$ & $46.06$ & $21.53$ & $0.00$ \\
\bottomrule
\end{tabular}
\end{table}

Overall, FedPLT reduces the global round time from $17.65$ s to $4.35$ s, while also yielding substantial savings in backward computation and uplink communication. This numerical example supports the analytical results by showing how partial-layer training improves efficiency while mitigating the straggler effect in heterogeneous systems.

\newpage
\twocolumn

\bibliographystyle{IEEEtran}
\bibliography{References}

\end{document}